\documentclass[12pt]{report}   
\usepackage{graphicx}  
\usepackage[letterpaper, left=1.5in, right=1in, top=1in, bottom=1in]{geometry}
\usepackage{setspace}  
\usepackage{times}  
\usepackage[explicit]{titlesec}  
\usepackage[titles]{tocloft}  
\usepackage[bookmarks=true, hidelinks,breaklinks = true]{hyperref}
\usepackage[page]{appendix}  
\usepackage{rotating}  
\usepackage[normalem]{ulem}  
\usepackage{subcaption}
\usepackage[utf8]{inputenc}
\usepackage{pdfpages}

\usepackage{tabularx}
 \usepackage{multirow}
\usepackage{bigstrut}
\usepackage{adjustbox}

\usepackage{enumerate}
\usepackage{multirow}

\usepackage{hyperref}

\usepackage{longtable}

\usepackage{color}

\newcommand{\ignore}[1]{}
\usepackage{alltt}
\usepackage{epsfig}
\usepackage{fancyhdr}
\usepackage{amssymb,amsmath}
\usepackage{float}
\usepackage{flushend}
\usepackage{tikz}
\usepackage{calc}
\usepackage{xcolor, colortbl}
\usepackage{multirow}
\usepackage{pifont}
\usepackage{adjustbox}
\usepackage{array} 
\newcolumntype{L}[1]{>{\raggedright\let\newline\\\arraybackslash\hspace{0pt}}m{#1}} 
\newcolumntype{C}[1]{>{\centering\let\newline\\\arraybackslash\hspace{0pt}}m{#1}} 
\newcolumntype{R}[1]{>{\raggedleft\let\newline\\\arraybackslash\hspace{0pt}}m{#1}} 
\definecolor{Green}{rgb}{0, 1, 0}
\definecolor{Red}{rgb}{1, 0, 0}
\newcolumntype{R}[2]{%
    >{\adjustbox{angle=#1,lap=\width-(#2)}\bgroup}%
    l%
    <{\egroup}%
}
\newcommand*\circled[1]{\tikz[baseline=(char.base)]{%
            \node[shape=circle,fill=black,draw,inner sep=0.5pt] (char) {#1};}}
\newcolumntype{g}{>{\columncolor{Green}}c}
\newcolumntype{R}{>{\columncolor{Red}}c}
\begin{document}
\doublespacing  


\newcommand{\thesisTitle}{Architectural Techniques to Enable Reliable and Scalable Memory Systems}
\newcommand{\yourName}{Prashant Jayaprakash Nair}
\newcommand{\yourSchool}{Electrical and Computer Engineering}
\newcommand{\yourMonth}{May}
\newcommand{\yourYear}{2017}


\begin{titlepage}
\begin{center}

\begin{singlespacing}

\textbf{\MakeUppercase{\thesisTitle}}\\
\vspace{10\baselineskip}
A Dissertation\\
Presented to\\
The Academic Faculty\\
\vspace{3\baselineskip}
By\\
\vspace{3\baselineskip}
\yourName\\
\vspace{3\baselineskip}
In Partial Fulfillment\\
of the Requirements for the Degree\\
Doctor of Philosophy in the\\
School of \yourSchool\\
\vspace{3\baselineskip}
Georgia Institute of Technology\\
\vspace{\baselineskip}
\yourMonth{} \yourYear{}

\vfill
Copyright \copyright{} \yourName{}, \yourYear{}

\end{singlespacing}

\end{center}
\end{titlepage}

\currentpdfbookmark{Title Page}{titlePage}  



\newcommand{\committeeMemberOne}{Dr. Moinuddin K. Qureshi, Advisor}
\newcommand{\committeeMemberOneDepartment}{School of Electrical and Computer Engineering}
\newcommand{\committeeMemberOneAffiliation}{Georgia Institute of Technology}

\newcommand{\committeeMemberTwo}{Dr. Sudhakar Yalamanchili}
\newcommand{\committeeMemberTwoDepartment}{School of Electrical and Computer Engineering}
\newcommand{\committeeMemberTwoAffiliation}{Georgia Institute of Technology}

\newcommand{\committeeMemberThree}{Dr. Saibal Mukhopadhyay}
\newcommand{\committeeMemberThreeDepartment}{School of Electrical and Computer Engineering}
\newcommand{\committeeMemberThreeAffiliation}{Georgia Institute of Technology}

\newcommand{\committeeMemberFour}{Dr. Milos Prvulovic}
\newcommand{\committeeMemberFourDepartment}{School of Computer Science}
\newcommand{\committeeMemberFourAffiliation}{Georgia Institute of Technology}

\newcommand{\committeeMemberFive}{Dr. Onur Mutlu}
\newcommand{\committeeMemberFiveDepartment}{Department of Computer Science}
\newcommand{\committeeMemberFiveAffiliation}{ETH Z\"{u}rich}

\newcommand{\approvalDay}{6}
\newcommand{\approvalMonth}{April}
\newcommand{\approvalYear}{2017}


\begin{titlepage}
\begin{singlespacing}
\begin{center}

\textbf{\MakeUppercase{\thesisTitle}}\\
\vspace{10\baselineskip}

\end{center}
\vfill

\ifdefined\committeeMemberFour

Approved by:
\vspace{3\baselineskip}		

\vspace{1.5\baselineskip}		

\begin{tabular}{lr}
\begin{minipage}[t]{2.7in}

    \committeeMemberOne\\
    \committeeMemberOneDepartment\\
    \textit{\committeeMemberOneAffiliation}\\

\end{minipage}
&\hspace{0.2in}
\begin{minipage}[t]{3.5in}

	\committeeMemberTwo\\
	\committeeMemberTwoDepartment\\
	\textit{\committeeMemberTwoAffiliation}\\

\end{minipage}
\\
\\
\\
\begin{minipage}[t]{2.3in}

	\committeeMemberFour\\
	\committeeMemberFourDepartment\\
	\textit{\committeeMemberFourAffiliation}\\

\end{minipage}
&
\begin{minipage}[t]{3.5in}
	
	\committeeMemberThree\\
	\committeeMemberThreeDepartment\\
	\textit{\committeeMemberThreeAffiliation}\\

\end{minipage}
\\
\\
\\
\begin{minipage}[t]{2.3in}
	
	\committeeMemberFive\\
	\committeeMemberFiveDepartment\\
	\textit{\committeeMemberFiveAffiliation}\\

\end{minipage}
&
\begin{minipage}[t]{3.5in}

\end{minipage}
\end{tabular}

\vspace{0.2in}
\vfill
		
\hfill Date Approved: \approvalMonth{} \approvalDay, \approvalYear
\vspace{5\baselineskip}		

\else

\hspace{0.6\textwidth}
\begin{minipage}[b]{0.4\textwidth}
	
	Approved by:
	\vspace{2\baselineskip}		
	
	\committeeMemberOne\\
	\committeeMemberOneDepartment\\
	\textit{\committeeMemberOneAffiliation}\\
	
	\committeeMemberTwo\\
	\committeeMemberTwoDepartment\\
	\textit{\committeeMemberTwoAffiliation}\\
	
	\committeeMemberThree\\
	\committeeMemberThreeDepartment\\
	\textit{\committeeMemberThreeAffiliation}\\
	
	\vspace{2\baselineskip}		
	
	Date Approved: \approvalMonth{} \approvalDay, \approvalYear
	\vspace{\baselineskip}		
	
\end{minipage}

\fi

\end{singlespacing}
\end{titlepage}



\newcommand{\yourQuote}{``Imagination is more important than knowledge."}
\newcommand{\yourAuthor}{Albert Einstein}


\begin{titlepage}
\begin{center}

\vspace*{\fill}
\yourQuote\\
\textit{\yourAuthor}
\vspace*{\fill}

\end{center}
\end{titlepage}



\newcommand{\yourDedication}{Dedicated to my Amma, Acchan, and Unni.}


\begin{titlepage}
\begin{center}

\vspace*{\fill}
\yourDedication\\
\vspace*{\fill}

\end{center}
\end{titlepage}


\pagenumbering{roman}
\addcontentsline{toc}{chapter}{Acknowledgments}
\setcounter{page}{5} 
\clearpage
\begin{centering}
\textbf{ACKNOWLEDGEMENTS}\\
\vspace{\baselineskip}
\end{centering}
I would like to dedicate this thesis to my parents and brother, without whose support I would not be what I am today. I am especially indebted to my mother and father for their love and encouragement during the tough times in the course of my Ph.D. I grateful to my brother as he always believed in me and kept me motivated. My parents and my brother played a key role in shaping my future by enduring immense sacrifices; I cannot find words to express my gratitude towards them.

I am grateful to my advisor, Moinuddin K. Qureshi, for enabling me to become a successful researcher. I am grateful that he saw potential in me and motivated me to do useful research. He was key in shaping my personality and is instrumental in teaching me the art of effective communication. His words of wisdom and his philosophy towards life are something that will always drive my decisions. The relationship with my advisor has only strengthened during the six years of my Ph.D. Therefore; I am looking forward to a lifetime of good friendship with him.

I would like to thank my childhood friends, who are family to me, for making my Ph.D. pleasantly memorable. The time spent with my close friends Darshan and Amar, especially during my visits to India, were invaluable. I cannot forget the visits to meet my close friend Melroy; those visits helped us develop a strong relationship that will last a lifetime. I do not have words to completely express my gratitude towards my close friend Santosh. By supporting my brother and me during very difficult times, he stands out as someone whom I always look forward for advice. Even today, I continue to re-live these experiences whenever I meet my childhood friends. These experiences were instrumental in making my Ph.D. successful and joyful. Going forward, these experiences would be key in shaping my outlook towards life.

I am indebted to my Ph.D. committee members and my referees for providing me valuable inputs in several research directions. I am thankful to Sudhakar Yalamanchili and Saibal Mukhopadhyay for their directions in the areas of VLSI and Systems. I am grateful to Onur Mutlu for encouraging me to do good research while also supporting me for the academic job search. I am thankful to Milos Prvulovic for the insightful discussions and directions in the area of security. I am grateful to Murali Annavaram and Rajeev Balasubramonian for supporting me in my academic job search. Internships helped me understand the role the computing industry plays in implementing the ideas generated during my Ph.D. into products. To this end, I am grateful to Pradip Bose and Alper Buyuktosunoglu for providing me with great mentorship. I am also thankful to Vilas Sridharan and David Roberts for the technical discussions and the ideas that came out of them. I am thankful to Chris Wilkerson and Shih-Lien Lu for their insights and flexibility in exploring new ideas.

My high-school teachers played a key role in building my character and kindled my love for science. I would like to thank Babu E. P. and Elizabeth John for providing the initial motivation towards the path of scientific endeavors. Whenever I am stuck on a problem, I use them as inspiration and always try to maintain a positive outlook towards life. Thus far, the academic path I have chosen is heavily influenced by my high-school teachers.

During my Ph.D., I was fortunate to have met Daniel Wong, Elizabeth Wong, and Kevin Chang. Over the years, our friendship has only grown stronger. I am also thankful to my labmates for providing me with great technical inputs. My good friend Chia-Chen Chou has been a key part of my Ph.D. The discussions we had during our co-authored papers were one of the most memorable moments during my Ph.D. I am also grateful to Swamit Tannu, Vinson Young, Gururaj Saileshwar, and Dae-Hyun Kim for several insightful and detailed technical discussions. Without such insights from my labmates, I would have found it difficult to explore new research areas. I am also thankful to Srinivas Eswar and Prasun Gera for providing me with moral support during my time at Georgia Tech. By being such accommodating roommates and close friends, I will miss them when I graduate. I am grateful to Ankit Bansal, Amit Karande and Aniruddha Satoskar for their support during my initial years in the USA. My Ph.D. would not be as memorable if it were not for them.

I would also like to thank Priya, for supporting and standing by me during my Ph.D. Her support has helped me shape my decisions and overcome personal hurdles. Her encouraging words always helped me keep focus during my Ph.D. Priya has been my constant source of motivation as she always ensured that I see things clearly while believing in me.

As my Ph.D. ends, I look forward to a lifetime of shared happiness with her. My Ph.D. helped me develop goals that are focused on scientific contributions. However, there comes a point when one might have to make choices whether one can sustain making scientific contributions. As this chapter of my life concludes, I can only hope that the people I have met, the ones I am going to meet, and the decisions I will take, will enable me to sustain making scientific contributions.

\clearpage




\renewcommand{\cftchapdotsep}{\cftdotsep}  
\renewcommand{\cftchapfont}{\bfseries}  
\renewcommand{\cftchappagefont}{}  
\renewcommand{\cftchappresnum}{Chapter }
\renewcommand{\cftchapaftersnum}{:}
\renewcommand{\cftchapnumwidth}{5em}
\renewcommand{\cftchapafterpnum}{\vskip\baselineskip} 
\renewcommand{\cftsecafterpnum}{\vskip\baselineskip}  
\renewcommand{\cftsubsecafterpnum}{\vskip\baselineskip} 
\renewcommand{\cftsubsubsecafterpnum}{\vskip\baselineskip} 

\titleformat{\chapter}[display]
{\normalfont\bfseries\filcenter}{\chaptertitlename\ \thechapter}{0pt}{\MakeUppercase{#1}}

\renewcommand\contentsname{Table of Contents}

\begin{singlespace}
\tableofcontents
\end{singlespace}

\currentpdfbookmark{Table of Contents}{TOC}

\clearpage


\addcontentsline{toc}{chapter}{List of Tables}
\begin{singlespace}
	\setlength\cftbeforetabskip{\baselineskip}  
	\listoftables
\end{singlespace}

\clearpage

\addcontentsline{toc}{chapter}{List of Figures}
\begin{singlespace}
\setlength\cftbeforefigskip{\baselineskip}  
\listoffigures
\end{singlespace}

\clearpage


\clearpage
\begin{centering}
\textbf{SUMMARY}\\
\vspace{\baselineskip}
\end{centering}

High capacity and scalable memory systems play a vital role in enabling our desktops, smartphones, and pervasive technologies like Internet of Things (IoT). Unfortunately, memory systems are becoming increasingly prone to faults. This is because we rely on technology scaling to improve memory density, and at small feature sizes, memory cells tend to break easily. Today, memory reliability is seen as the key impediment towards using high-density devices, adopting new technologies, and even building the next Exascale supercomputer. To ensure even a bare-minimum level of reliability, present-day solutions tend to have high performance, power and area overheads. Ideally, we would like memory systems to remain robust, scalable, and implementable while keeping the overheads to a minimum. This dissertation describes how simple cross-layer architectural techniques can provide orders of magnitude higher reliability and enable seamless scalability for memory systems while incurring negligible overheads.

%
%


\clearpage
\pagenumbering{arabic}
\setcounter{page}{1} 

\titleformat{\chapter}[display]
{\normalfont\bfseries\filcenter}{\MakeUppercase\chaptertitlename\ \thechapter}{0pt}{\MakeUppercase{#1}}  
\titlespacing*{\chapter}
  {0pt}{0pt}{30pt}	
  
\titleformat{\section}{\normalfont\bfseries}{\thesection}{1em}{#1}

\titleformat{\subsection}{\normalfont}{\uline{\thesubsection}}{0em}{\uline{\hspace{1em}#1}}

\titleformat{\subsubsection}{\normalfont\itshape}{\thesubsection}{1em}{#1}


\chapter{Introduction}
\paragraph{}High capacity and scalable memory systems play a vital role in enabling our desktop machines, smartphones, and supercomputers. One of the key techniques to enable high-density memories is technology scaling. Technology scaling allows manufacturers to reduce the feature size of each memory cell. This enables manufacturers to fit a greater number of cells per unit area in each chip and increase their density. Apart from technology scaling, at the system level, computers are designed to accommodate a greater number of memory modules to increase their effective capacity. Furthermore, both industry and academia have also been investigating new memory technologies that offer very high densities and act as replacements to current technologies. However, akin to the scalability and reliability problems while maintaining Moore's Law in computing systems, memory systems are also facing challenges. One of the key challenges towards scalable memory systems is maintaining the reliability of its components.

To ensure even a bare-minimum level of reliability, current systems tend to incur high performance, power, and area overheads. Ideally, we would like to obtain strong memory reliability and seamless scalability with negligible overheads. Based on the choice of technologies, this dissertation broadly classifies these concerns into two problems.

\section{Problem 1: Scalability concerns for Current Memory Systems}
\subsection{Low-Cost Reliability for Sub-20nm DRAM Scaling}
\paragraph{}Dynamic Random Access Memory (DRAM) has been the basic building block for main memory systems since the 1980s. Each DRAM cell uses a capacitor to store binary data in as an electric charge. As we scale DRAM, the width of its cell-capacitors reduces and their height increases, leading to high aspect ratios. At sub-20nm nodes, the DRAM cell-capacitor aspect ratio becomes impractically high and tends to cause cells to turn faulty at manufacture time. Thus, these broken cells are unable to store charge and therefore any data that is present in these cells become erroneous. As DRAM scales, its chips are expected to have bit-error rates as high as 10$^{-4}$. At these high error rates, traditional techniques tend to have high overheads and therefore become ineffective. To this end, this dissertation explores low-cost architecture-level solutions for tackling scaling-related faults in current memory systems.

\subsection{Strong Runtime Reliability Using Commodity DRAM-Based Systems}
\paragraph{}Several field studies have shown a high incidence of multi-granularity faults within DRAM modules during their operation. For instance, a recent study showed that single-bit failures tend to be as common as chip failures at runtime. Due to this, one technique would be to protect DRAM modules against chip failures and improve reliability. Currently, protecting against chip-failures involve employing costly error correction techniques like Chipkill that uses a larger number of chips. While most DRAM modules that use error correction codes (ECC) use 9 chips, Chipkill requires activating 18 chips and therefore incurs high performance and power overheads. To tackle this problem, this dissertation describes a simple architectural solution that can tolerate chip-failures by using only 9-chips, while making no changes in the memory interface and incurring negligible overheads. 

\section{Problem 2: Challenges in adopting New-Memory Technologies}
\subsection{Enabling Reliable Stacked Memories}
\paragraph{}Stacked memories are a new-memory technology that enables manufacturers to place memory dies over one another. This technology enables manufacturers to increase the effective density and bandwidth of the memory system. To enable stacking, manufactures use through silicon vias (TSVs) as conduits to send data and addresses within stacked memories. Therefore, one can improve the effective bandwidth within each stacked memory by increasing the density of TSVs. Unfortunately, the TSV technology is relatively new and is therefore prone to failures. Furthermore, each DRAM die within the stacked memory is also susceptible to large-granularity failures. Simply employing Chipkill within the stacked memory is costly as it would require activating multiple dies in the stack to fetch a single cacheline. This would result in lowering the effective bandwidth, thereby reducing performance. Furthermore, as multiple dies are being activated, naively employing Chipkill also increases the total power consumption of the stacked memory. This dissertation proposes techniques that enable runtime reliability for TSVs and robust stacked memories that have minimal performance and power costs.

\subsection{Scalable Memories That Can Tolerate High-Rates of Transient Faults}
\paragraph{}Memory system can also incur intermittent faults as they scale. At high rates, the intermittent or transient failures will require new and efficient error correction strategies. For instance, Spin-transfer torque magnetic random-access memory (STTRAM) is a promising new-memory technology that is widely viewed as a replacement for SRAM. The benefits of STTRAM include 4x-6x higher density as compared to SRAM and low static power consumption. Unfortunately, the data retention time of STTRAM cells decreases exponentially as they scale. Even after scrubbing every 100ms, STTRAM based memory systems are projected to show bit-error rates (BER) as high as 10$^{-5}$. Furthermore, akin to alpha particle strikes, scaling-related errors in STTRAM are transient in nature and any cell can turn faulty over time. Due to this, one cannot simply disable faulty STTRAM cells, as that would render the entire memory to be disabled within a few hours. To enable scalable STTRAM, this dissertation describes simple ECC based solutions that minimize area, performance, and complexity overheads while offering very high reliability.

\section{Thesis Statement}
Cross-layer architectural techniques act as enablers for scalable and reliable memory systems. By scripting cross-layer error correction strategies at the architecture level, the system can obtain 100x-1000x higher memory reliability while incurring negligible overheads.

\section{Contributions}
This dissertation makes the following contributions.
\begin{enumerate}
    \item This dissertation proposes architectural techniques to handle high rates of permanent faults. To this end, it advocates exposing these faults from within the memory to the architecture-level.
    \item This dissertation proposes architectural techniques to handle high rates of transient faults. It advocates designing systems that use simple and efficient ECC to fix common cases of faults and use strong ECC only in the uncommon cases of faults.
    \item In systems with multiple levels of error codes, this dissertation describes how these error codes can be designed to interact and increase the overall robustness of the entire memory system.
    \item This dissertation highlights techniques to efficiently encode RAID-based schemes within stacked memories. This dissertation describes how runtime TSV repairing and ECC can be tuned to cater to the granularity of faults that occur at runtime.
\end{enumerate}

\section{Thesis Organization}
This dissertation is organized into seven chapters. Chapter 2 describes the related work on memory-reliability. Chapter 3 tackles the issue of technology scaling in DRAM. Chapter 4 addresses the issue of large-granularity runtime faults in memory systems. Chapter 5 investigates how to implement reliable stacked memories. Chapter 6 describes how to implement reliable and scalable memory systems with high rates of transient faults. Chapter 7 concludes this dissertation and describes some future work.


\chapter{Related Work}
\paragraph{}Several prior work have looked at important reliability concerns that plague memory systems. Most prior work rely on error correction codes (ECC) and creative data organizations to fix faulty memories. This chapter describes relevant prior work that have tried to tackle scaling-related and run-time faults for current and future memory systems. This dissertation also provides qualitative and quantitative comparisons of key prior work with respect to the proposed techniques in the upcoming chapters.

\section{Studies For Identifying and Characterizing Failures}
\subsection{Studies on DRAM} 
Field studies on supercomputing and server clusters help obtain real world data. Some field studies on DRAM based main memory systems have investigated data errors in commercial clusters~\cite{Schroeder:2009,Schroeder:2010}. Contrary to reporting fault rates, these studies report data error rates which depend on the application that the system executes and its memory mapping. For instance, a memory system with a single bit with permanent fault can result in billions of errors if the bit remains uncorrected and if the application frequently accesses the faulty memory bit. Similarly, systems can also report billions of errors if the OS naively maps pages into such faulty locations without decommissioning the region. However, to evaluate reliability, fault statistics provide an clear metric when compared to error statistics. 

To address this, Sridharan et. al.~\cite{failures:sridharan12,failures:sridharan13} present a clearer distinction between errors and faults and report memory faults and their positional affects by studying supercomputer clusters. Although these studies present detailed failure data, they do not use this data to suggest quick reliability exploration techniques. Commercial solutions like Chipkill present specific results for certain FIT rates; however they do not estimate memory reliability as these systems scale~\cite{chipkill}. In an attempt to estimate reliability, recent studies have investigated integrating field data into analytical models~\cite{ChipKill:Jian,ChipKill:DeBardeleben}.

Instead of charactering the faults within the memory system of an entire datacenter, some prior work have also looked at characterization of an individual memory modules. For instance, DRAM based memory modules tend to have DRAM cells with variable rentention times. A prior work bu Liu et.al.~\cite{process3} charcterized the memory cells based on their retention times. This work was instrumental in pointing out that only a few DRAM cells exhibit rentention time that is lower than 256ms. Thereafter, a prior work by Khan et.al.~\cite{process-sig} pointed out the variable retention time (VRT) phenomenon in DRAM. VRT cells vary their retention times and therefore cannot easily be statically profiled. 

Thereafter, several prior work such as AVATAR~\cite{avatar} and PARBOR~\cite{Khan:2016} have looked at efficient ways to profile and fix VRT cells in DRAM. Furthermore, a recent work, by Khan et.al.~\cite{Khan:CAL} has also looked at how data patterns affect the retention times for DRAM. Additionally, technology scaling in DRAM also exposes security vulnerabilities in the memory system. For instance, at lower technology nodes, DRAM cells are sensitive and therefore they tend to be suseptible to bit-flips based on its activity. This issue is called rowhammer and it tends to not only be a reliability concern, but also a security concern. To this end, prior work have profiled memory devices and described low-cost techniques to fix these reliability and security problems~\cite{rowhammer:yoongu,rowhammer:kim}. While these studies are instrumental in highlighting the challenges in reliability, they tend to not provide a strong solution towards enabling a scalable DRAM.

As DRAM systems scale, their effective characterization can also be used to implement schemes that can have an interplay between reliability, latency, performance, and power of the memory system~\cite{Chang:2016-sig,Chang:2017-sig,lee:2017-sig}. For instance, Lee et. al.~\cite{adaptivelat} investigated the dependence of retention time of DRAM with temperatures and thereby modulated its latency.

\subsection{Studies on Flash}
Similar to DRAM, several prior work have also looked at faults in Flash. For instance, Cai et. al.~\cite{Cai:Date2012} showed that the error rates for Flash devices tend to depend on their data patterns and based on these insights, we can write optimal patterns to reduce data errors. Cai et. al.~\cite{nand:cai2013} also highlighed the retention time problems in Flash memories and their resulting errors. To this end, Cai et. al.~\cite{nand:cai2013} investigated how to improve the error correction capability of Flash based devices using their neighboring cells. In similar spirit, Cai et. al.~\cite{cai2015hpca} characterized Flash and investigated an optimized read design that can overcome high rates errors. Cai et. al~\cite{Cai:Date2013} also investigated the distrubution of threshold voltage to help reduce errors in Flash memories.

Akin to row-hammer in DRAM, Flash suffers from the read-disturb problem. Read-disturb occurs when the the cells lose their contents during reads and become errorneous. To mitigate the read-disturb problem in Flash, Cai et. al.~\cite{cai2015read} characterized Flash devices and highlighted the effects of read-disturb. Furthermore, Cai et. al.~\cite{cai2017hpca} exploited the read-disturb problem to highlight reliability and security vulnerabilities in Flash. To mitigate these concerns, Cai et.al.~\cite{cai2017hpca} propose using circuit and architectural techniques like buffering, adaptive read voltages for LSB cells and multiple pass though voltages. Even programming Flash cells can reduce the reliability of Flash, Cai et. al.~\cite{cai2013program} investigated this program interference and characterized Flash devices. These prior work are key in exposing the reliability and security implications of having error prone memories like Flash. To enable researches to gain more insights, Meza et.al~\cite{facebook} performed field studies on the Facebook datacenter on their flash devices. While these studies have provides insights into retention errors and their types, there is still potential for cross-layer solutions to provide higher reliability.

\section{Handling Scaling-Related Faults}
\paragraph{}A DRAM-based memory system with scaling-related BER of 10$^{-4}$ would have nearly 0.1\% of the cachelines exhibiting multi-bit faults. Furthermore, even new-memory technologies like STTRAM are projected to have transient BER of 10$^{-5}$ and would likely encounter instances of 6-bit errors during their operational lifetime. Therefore, for scalability, the memory system must be capable of handling multi-bit faults. 

\subsection{Related Work on Multi-bit ECC Schemes} 
Several multi-bit ECC schemes have been proposed to mitigate high rates of faulty cells. For instance, Alamelden et. al. and Wilkerson et. al.~\cite{Alameldeen:2011,wilkerson:isca10} investigated using Multi-Bit ECC to fix multi-bit failures that result from reducing cache voltages. This enabled reusing reliability mechanisms like ECC to save power.
 
In similar spirit, one can tolerate a high error-rate by employing multi-bit error correction in DRAM memories. For instance, to tolerate an error-rate in the regime of 100ppm, we need three bit error correction, i.e. ECC-3 for each word (ECC-4 if we want soft error protection). Employing such high levels of error correction would require storage overhead of 37\% of memory space.  This would need the DIMM to have three extra ECC chips, resulting in prohibitive cost.  It will also result in lower performance due to higher decode latency of ECC-4 ~\cite{chris:isca08}.

\subsection{Related Work on Parity-Based Schemes}
Rather than using hamming codes and BCH codes, one can use simple RAID-type correction by using parity~\cite{bch_code:ibm,RAID5:1995}. Correctable Parity Protected Cache  (CPPC)~\cite{manoochehri2011cppc} uses a parity-based detection of a single bit error on a per-line (or per-word basis), and tracks a global parity of the data using a separate buffer. When the parity associated with the line detects an error, the global parity is used to restore the data of the faulty line (much like a RAID-4 scheme).  However, CPPC was designed for a fairly low bit-error rate (evaluated with a per-cell mean time to failure of 1 million hours) and cannot tolerate BER as high as 10$^{-4}$. CPPC also does not scale well as the size of its buffer will become a tens of Megabytes in size for a DRAM-based system that is a few Gigabytes in size.

Two-Dimensional Error Coding (2DP)~\cite{kim2007multi} is another parity-based scheme that keeps both horizontal parity and vertical parity to perform correction of single bit errors by using only a parity-bit per line (or word). This scheme is low-cost and is highly effective at low-error rates and when the tracked regions have correlated errors. Unfortunately, both DRAM and New-Memory technologies are projected to have high error rates and 2DP is ineffective at tolerating high rates of bit-failures. 

\subsection{Related Work on Error Correction for New-Memory Technologies} 
Several recent studies have looked at error correction in Phase Change Memories (PCM). These solutions range from replicating pages with faulty cells~\cite{drm:asplos10}, to correcting hard errors with pointers or data inversion~\cite{ecp:isca10,safer:micro10}, to efficiently using non-uniform levels of error correcting pointers~\cite{payg:micro11}, to sparing lines with faulty cells with embedded pointer~\cite{freep:hpca11}. FREE-p decommissions a line with faulty cells (more than what can be handled by the per-line ECC) and stores a pointer in the line to point to the spare location.  It relies on the read-before-write characteristics of PCM memory to read the pointer before writing to the line.

Prior studies~\cite{sachin-stt} have looked at using multibit ECC to mitigate errors in STTRAM to improve the overall density of the STTRAM technology. However, they incur the significant cost of multi-bit ECC for each line.  To tolerate transient failures in scaled-down STTRAM, prior studies~\cite{STTRAM:MICRO2011} have proposed DRAM-style refresh. As the failure mode of STTRAM is like transient error due to particle strike, DRAM-style refresh is ineffective for STTRAM. Smullen et al.~\cite{sudhanva-stt}  proposed a refresh policy that reads every line of the cache iteratively and writes it back again within the retention time. They also used a single-bit error correction mechanism, so that in the worst-case scenario, they can writeback data after detecting an error and correcting it. Unfortunately, for BER as high as 10$^{-5}$, having only  ECC-1 with each line is insufficient.  Naeimi et al.~\cite{ITJ:2013}  suggested using 5EC6ED for a 64MB STTRAM-based cache to guarantee fixing 5-bit errors. Unfortunately, a 5EC6ED code involves large transcoding latencies, complex circuitry, and a 10\% area overhead.

\section{Handling Large-Granularity Runtime Faults}
Several prior work have proposed techniques to handle runtime faults in memory systems.

\subsection{Related Work on Strong ECC Schemes for Runtime Faults}
A recent work, Virtual and Flexible ECC (VFECC)\cite{vfecc:asplos10}, allows systems to implement high levels of ECC without relying on ECC based DRAM Modules. It incorporates the ECC storage within the main memory. Unfortunately, VFECC does not reduce the storage overhead associated with high levels of error correction, as the ECC level is not dependent on the number of faults in the word. To implement ECC-3, VFECC would still need to dedicate about 37\% of memory capacity making it unappealing for practical implementations. Similarly, Memguard~\cite{Memguard} tries to use ordinary Non-ECC memory modules to provide strong reliability by storing hashes of data and check-pointing data. Memguard stores hashes of data values to detect errors. Memguard incurs checkpointing overheads for tolerating chip-failures. In a similar vein, COP~\cite{COP} and Frugal-ECC~\cite{frugal} can use ordinary memory modules to provide ECC protection by storing ECC alongside compressed lines. However, COP and Frugal-ECC are vulnerable to cachelines are incompressible. 

Bamboo-ECC~\cite{bambooecc} and ARCC~\cite{arcc} tries to tradeoff reliability with the storage and performance overheads of maintaining ECC. Unfortunately, these schemes do not provide complete robustness for the memory system. Another prior work proposes a low overhead ChipKill code that can be used with current commodity ECC-based memory modules without using additional chips \cite{Jian:2013}. This work uses a combination of error detection and correction codes, but does not talk about efficient memory sparing and low-latency correction. Furthermore, RAID type ECC schemes have been proposed to mitigate large-granularity faults, however if designed improperly, they incur high bandwidth overheads ~\cite{RAID5:1995}.

\subsection{Related Work on OS-Based Reliability Techniques} 
Memory errors can be tolerated in software as well. For example, with memory page retirement~\cite{MPR,cosmicray:asplos2012}, the OS can retire a faulty page from the memory pool, once the fault is
detected.  Unfortunately, these schemes operate at a coarse granularity of page size. Given that the typical page size is 4KB, these schemes are unable to tolerate error-rates higher than one error for every several tens of thousand of bits. To operate at high error-rate, a fine grained approach such as at word-granularity or line-granularity is needed.

\subsection{Related Work on Reliable Stacked Memories}
Several techniques have been proposed for ``swapping in'' such redundant TSVs to replace faulty TSVs in a 3D die stack \cite{TSV:DATE12}. Similarly, two prior works try to address stacked memory reliability without considering TSV faults. The first prior work proposes techniques to reliably architect stacked DRAM caches~\cite{GABEISCA:2013}.  It uses
CRC-32 to detect errors in caches. However, correction is performed simply by disabling clean lines and replicating dirty lines.  While such correction can be useful for caches, disabling random locations
of lines is an impractical option for main memory.  Furthermore, replicating all the data for main memory leads to a capacity loss of 50\% and doubles the memory activity.


\chapter{Enabling Robust Technology Scaling of DRAM}
\paragraph{}Dynamic Random Access Memory (DRAM) scaling has been the prime driver for increasing the memory capacity over the past three decades. Unfortunately, scaling DRAM to smaller technology nodes has become challenging due to the inherent problem in designing smaller geometries, coupled with device variation and leakage. Future DRAM devices are likely to experience significantly high error-rates. Techniques that can tolerate errors efficiently can enable DRAM to scale to smaller technology nodes. However, existing techniques such as row/column sparing and ECC become prohibitive at high error-rates. To develop cost-effective solutions for tolerating high error-rates, this chapter suggests a cross-layer approach in which the faulty cell information within the DRAM chip is exposed to the architectural level.

\section{Introduction}

DRAM has been the basic building block for main memory systems for the past three decades.  Scaling of DRAM to smaller technology nodes allowed more bits in the same chip area, and this has been a prime driver for increasing the main memory capacity. Data is stored in a DRAM cell as charge on a capacitor. As we scale down the feature size, the amount of charge that must be stored on the capacitor must still remain constant in order to meet the retention time requirements of DRAM.  DRAM technology has already reached sub-30nm regime, and it is becoming increasingly difficult to further scale the cells to smaller geometries.  The challenge lies not only in inherent problems of fabricating small cylindrical cells for
the capacitor but also from the increased variability and leakage
across cells.  Recently, DRAM scaling challenges have caused the
community to look at alternatives technologies for main memory~\cite{lee:isca09, qureshi:isca09}. Unfortunately, a viable DRAM
replacement that is competitive in terms of cost and performance is
still not commercially available.  Therefore, scaling DRAM to smaller feature sizes remains critical for future systems.

The smaller geometry and increased variability for future technologies are likely to result in higher error-rates.  To maintain system integrity, faulty DRAM cells must either be decommissioned or corrected.  If the cost of tolerating faulty cells is significantly higher than the capacity gains from moving from a given technology node to a smaller technology node, future technology nodes may be deemed not viable, thus halting DRAM scaling. Thus, techniques that can tolerate high error-rates at low cost can allow DRAM technologies to scale to smaller technology nodes than otherwise possible.

\begin{figure}[ht]
  \centering
  \centerline{\psfig{file=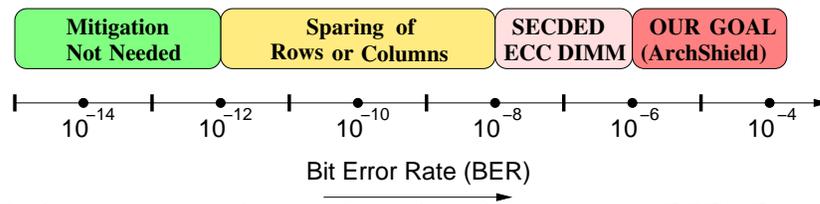,width= 4.3 in}}
  \vspace{-0.15in}
  \caption{Fault mitigation technique depends on bit error-rate (BER).
    Row sparing works well at low error-rates and SECDED-based DIMMs
    can tolerate BER of approximately $10^{-6}$. This dissertation targets a BER that is about 100x higher.}
      \vspace{-0.15in}
  \label{fig:intro1}
\end{figure}

Figure~\ref{fig:intro1} shows different schemes to mitigate errors in
DRAM (without loss of generality, this chapter considers an 8GB Dual Inline Memory Module (DIMM) for its design and evaluation studies). If the bit error-rate (BER) of DRAM cells is less than $10^{-12}$ then the memory system may not need any error correction for faulty cells.  Current DRAM systems rely on sparing of rows/columns to tolerate faulty cells.  For example, with row sparing, the DRAM row containing the faulty DRAM cell is replaced by one of the spare rows. This method incurs an overhead of about 10K-100K bits (and several laser fuses) for tolerating one faulty bit.  While seemingly expensive, this method works quite well at low bit error-rates that are typical in current DRAM chips. Unfortunately, the high cost makes this technique impractical for high error-rates.

Another alternative to tolerate errors in DRAM is to use Error
Correcting Code (ECC).  Commodity DIMMs are also available with ECC,
which can correct one bit out of the 8-byte word.  While these DIMMs
are aimed at tolerating soft errors, we can also use it to tolerate
faulty DRAM cells. However, using such DIMMs to tolerate random bit
errors, is still ineffective for high bit error-rates. Our analysis
shows that ECC DIMMs can tolerate an error-rate of only in the regime of about 1 faulty cell per million.  To tolerate higher error-rates, we would need higher levels of ECC. For example, for tolerating an error-rate of $10^{-4}$ we need 3-bit error correction per 64-bit word. Such high level of ECC is expensive in terms of both storage and latency. Furthermore, this approach sacrificed soft error resilience for tolerating faulty cells, and would need additional ECC to tolerate soft errors.  Ideally, one would want to use ECC DIMMs to tolerate both faulty cells due to manufacturing and soft errors due to alpha particles.

This dissertation advocates exposing the information about the faulty DRAM cells to the hardware, so that the amount of error tolerance can be tailored to the vulnerability level of each word.  This chapter describes such an architecture-level framework called {\em ArchShield}.  ArchShield is built on top of commodity ECC DIMMs, and is geared towards tolerating 100x higher error-rates than can be handled by ECC DIMMs alone, while retaining the soft error tolerance. When a new DIMM is configured in the system, ArchShield performs a runtime testing of the DIMM to identify its faulty cells. In particular, it tracks if the given 64-bit word has no error, one error, or more than
one error.

ArchShield contains a {\em Fault Map} that stores information about
faulty words on a per line basis.  All faulty words (including the
ones with one-bit error) are replicated in a spare region.  Such {\em
  Selective Word Level Replication (SWLR)} allows decommissioning for
words with multi-bit error, while providing soft error protection for
words with one-bit error. On a memory access, the fault map entry is
consulted.  If the line is deemed to have a word with more than 1
error, the replication area is accessed to obtain the replicated words
for the corresponding line. Whereas, if the line is deemed to have a
word with 1-bit error, the replicated copy is accessed only when an
uncorrectable fault is encountered at the original location, which
allows fast access in common case.  Thus, ArchShield can tolerate
multi-bit errors, while retaining soft error protection of 1-bit error
correction per word.

The Fault Map and word-level repair of ArchShield is inspired, in
part, by similar approach to dealing with high error-rate in current
Solid State Disk (SSD). Similar to SSD, we propose to embed the Fault
Map and Replication Area in reserved portion of the DRAM memory.  This
reduces the effective main memory visible to the operating system.
Fortunately, the visible address space provided by ArchShield is
contiguous, so ArchShield can be employed without any software changes
(except that the memory is deemed to have smaller
capacity). Similarly, ArchShield does not require any changes to the
existing ECC DIMMs, and only minor changes to the memory controller to
do runtime testing, orchestrate Fault Map access, and update and
access replicas.

This chapter showcases evaluations for ArchShield with 8GB DIMM.  To tolerate a high error-rate of $10^{-4}$, ArchShield requires 4\% memory space, and causes a performance degradation of less than 2\% due to the extra memory traffic of Fault Map and SWLR.  ArchShield provides this while maintaining a soft error protection of 1-bit error per word.

\section{Background and Motivation}

The ITRS road-map for the next decade projects DRAM technology node of 10nm in 2022, in essence a new technology node every three years.  If DRAM technology could be kept on this scaling curve, we can expect a doubling of memory capacity of DRAM modules every three years.  Unfortunately, scaling DRAM to smaller technology nodes has become quite challenging. In addition to the typical problems of scaling to smaller geometries, DRAM devices face several additional barriers.

\subsection{Why DRAM Scaling is Challenging}

\begin{figure}[htp]
  \centering
  \centerline{\psfig{file=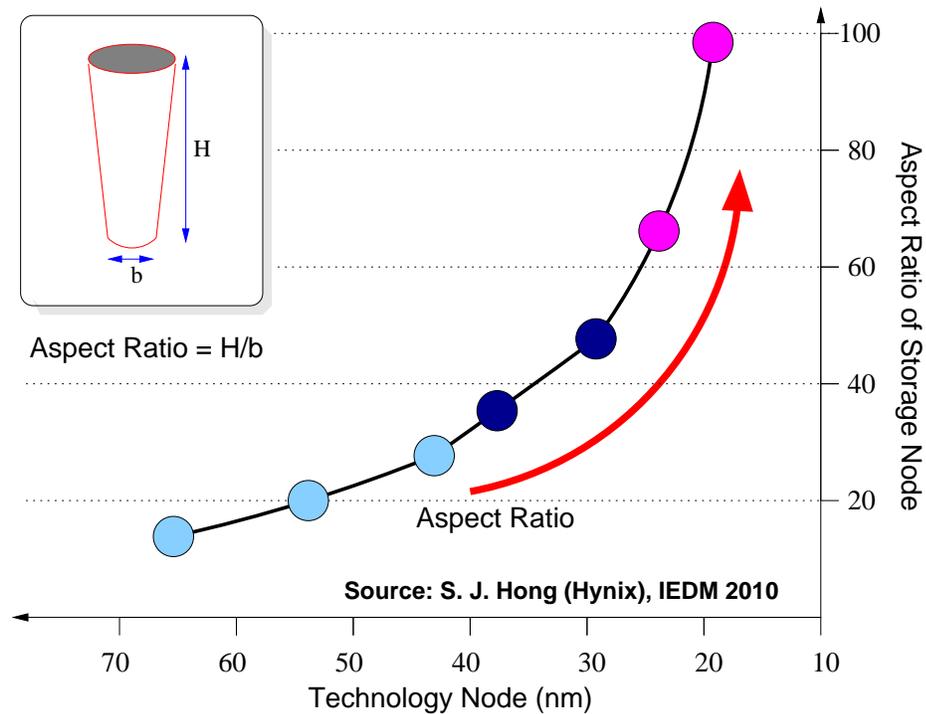,width=4.8in }}
  \caption{Exponential increase in aspect ratio of DRAM cells with
    scaling to smaller technology nodes (redrawn
    from~\cite{hong:iedm2010})}
  \label{fig:aspect} 
\end{figure}

The capacitive element used to store charge in DRAM is typically made
as a vertical structure to save chip area (as shown in the inset in
Figure~\ref{fig:aspect}). To meet the DRAM retention time, the
capacitance stored on the DRAM device needs to be approximately
25fF. When DRAM technology is scaled to smaller node, the linear
dimensions scale by approximately 0.71x, the surface area of the cell
reduces to approximately 0.5x, which means the depth of the vertical
structure must be doubled to obtain the same capacitance. Let {\em
  Aspect Ratio} be the ratio of the height of the cell to the
diameter. As shown in Figure~\ref{fig:aspect}, the aspect ratio has
been increasing exponentially and is expected to reach more than 100x
at sub-20nm~\cite{hong:iedm2010}. Such narrow cylindrical cells are
inherently unstable due to mechanical reasons, hence difficult to
fabricate reliably~\cite{future}.

The second problem is reduction in the thickness of the dielectric
material of the DRAM cell. This makes it challenging to ensure the
same capacitance value, given the unreliability of the ultra-thin
dielectric material. The third problem is the increase in gate induced drain leakage and increased variability, which means that to obtain the same retention time we may be forced to increase the capacitance of the DRAM cell, exacerbating the problem of cell geometry and reliability of the dielectric material.

Due to the challenges from shrinking dimensions and variability,
future DRAM cells will be expected to have much higher rate of faulty
cells than current designs. To assist DRAM scaling, cost effective
solutions must be developed to tolerate such high rate of faulty
cells, otherwise it may become prohibitive to scale DRAM to smaller
nodes. Unfortunately, the exact data about error-rates in DRAM memories tend to be proprietary information and is guarded closely by DRAM
manufactures. So, in this chapter, we assume that error-rates exceed
significantly than what are handled by traditional techniques. This chapter also assumes that these errors are persistent, and that they are distributed randomly across the chips. This chapter targets a bit error-rate in the regime of 100 parts per million (ppm), or equivalently $10^{-4}$.

\subsection{Drawbacks of Existing DRAM Repair Schemes}

\begin{figure}[htb]
  \centering
  \centerline{\psfig{file=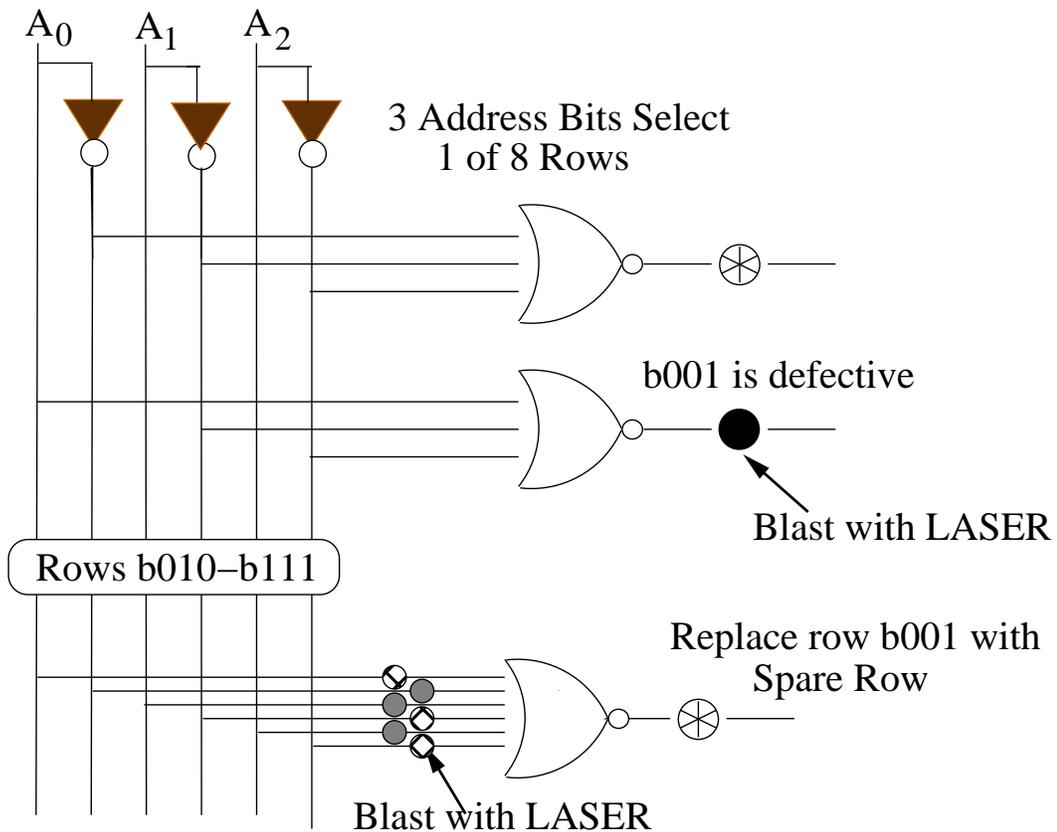,width= 5.5in}}
  \caption{Typical row sparing design relies on laser fuses and
    sacrifices an entire row for masking a faulty cell.}
  \label{fig:rowspare} 
\end{figure}

Current DRAM chips tolerate faulty cells by employing row sparing and
column sparing. These mechanisms tend to mask the faulty cell at a
large granularity. For example, with row sparing, the entire DRAM row
containing the faulty cell gets decommissioned and replaced by a spare
row. Given that DRAM rows contain in the regime of 10K-100K bits,
masking each faulty cell incurs a significant overhead. Further-more
disabling the faulty row and enabling the spare row must be done at
design time, hence it must rely on non-volatile memory.  Typically
laser fuses are used to disable the row with faulty cell, and enable
the spare row for the given row address, as shown in
Figure~\ref{fig:rowspare} (derived from~\cite{Jacob2008}).  To handle
a memory array containing few thousand rows, each spare row requires
fuse memory of few tens of bits.  Unfortunately, each bit of laser
fuse incurs an area equivalent few tens of thousands of DRAM
cells~\cite{FUSEAREA}.  Thus, sparing incurs an overhead of
approximately several hundred thousand DRAM cells to fix one faulty
cell.  While this overhead may be acceptable at very small error-rate,
it is prohibitive to tolerate error-rates in the regime of several
parts per million.

\subsection{Limitations of Tolerating Faulty Cells with ECC DIMM}

Instead of masking faulty cells, one can correct them using ECC.  Commodity memory modules are typically also
available in ECC enabled versions, in a (72,64) configuration.  Such
modules contain an extra ECC chip in addition to the eight data chips,
and can correct up-to one error (and detect up-to two errors) in the
64-bit word.  While the typical applications for ECC DIMM tend to be
to tolerate soft errors, we can potentially use it to tolerate faulty
DRAM cells as well.  However, even with an ECC DIMM the error-rates
that can be tolerated is low.

The studies in this chapter consider an 8GB DIMM, containing one billion 8-byte words.  The expected number of random errors that would result in a word with two errors can be computed using the Birthday Paradox analysis~\cite{bday}.  For example, if balls are randomly thrown into
N buckets, on an average after 1.2$\times\sqrt{N}$ throws, we can
expect at-least one bucket to have more than one ball.  Similarly, on
average, a memory with 1 billion words would tolerate approximately
40K errors before getting a word with two errors.  Thus, the error
rate tolerated with ECC DIMM is 40K divided by the number of bits in
memory (77 billion), or equivalently 0.5 ppm, approximately 200x lower
than the error-rate we want to handle.  Furthermore, such usage of ECC
DIMM to tolerate faulty cells increases the vulnerability of the
system to soft errors. Ideally, one should tolerate faulty cells while
retaining soft error protection of ECC DIMMs.

\subsection{Need for Handling Multiple Faults/Word}

A higher rate of faulty cells can be tolerated with the ECC approach
if we correct multiple errors per word. To estimate the amount of
multi-bit error protection required, one can compute the expected number of
words for a given number of faults. Let {\em p} be the probability of
bit failure.  Let there are {\em b} bits in the word. The expected
number of faulty bits per word is $p \cdot b$. If $p \cdot b
<< 1$, then the probability ($P_k$) that the word has $k$ errors $(k
\ge 1)$ can be approximated by Equation 1.
\begin{equation}
P_k = \frac{(p \cdot b)^{k}}{k!}
\end{equation}
The studies in this chapter consider a traditional (72,64) ECC DIMM. So, the
number ECC word has 72 bits. Table~\ref{table:multibit} shows
the expected number of words in an 8GB memory that have 0, 1, 2, 3,
and 4 or more errors for a probability of bit failure of 100 ppm.  The
episodes of 4 or more errors are rare, but we need to tolerate three
faulty cells per word.

\begin {table}[ht]
\begin{center} 
\caption{Percentage of words with multiple faulty cells (and expected number of words in 8GB memory, i.e. $2^{30}$ words).}
\vspace{0.3in}
\resizebox{0.8\columnwidth}{!}{
\begin{tabular}{|c|c|c|c|c|c|} \hline
Num Faulty bits  &   0         &  1       & 2         & 3       & 4+   \\ \hline \hline
Probability      &     0.993   & 0.007    &  $26 \cdot 10^{-6}$  &  $62 \cdot 10^{-9}$    & $10^{-10}$        \\ \hline 
Num words        &    0.99 Bln &  7.7 Mln & 28K       &    67      &     0.1      \\ \hline 
\end{tabular}
}
\label{table:multibit}
\end{center}
\end{table}

\subsection{Low Cost Fault Handling by Exposing Faults}

To handle 3-bits per word, the ECC overhead would be approximately 24
bits per word, or approximately 37\%.  Thus, the storage overhead of
uniform fault tolerance is prohibitive at high error-rates.  The
problem with both row sparing and ECC schemes is that they try to hide
the faulty cell information from the architecture, hence they incur
significant storage overhead.  To develop a cost-effective solution,
we take inspiration from the fault tolerant architecture typically
used in Solid State Drives (SSD)~\cite{ssd}.  SSD are made of Flash
technology, that tends to have high error-rates. The management layer
in SSD keeps track of bad blocks and redirects access to good
location. A similar approach can also allow DRAM systems to tolerate
high error-rates.

From Table~\ref{table:multibit} we see that only a small fraction of
words have more than 1 faulty cell.  If we can expose the information
about faulty cells to the architecture layer, then we can tolerate
faulty words by decommissioning and redirecting at a word granularity
and thus significantly reduce the storage overhead of tolerating
faulty cells. Note that we cannot arbitrarily disable words in memory,
as the operating system relies on having a contiguous address space.
We propose the {\em ArchShield} framework that can efficiently
tolerate high rate of faulty cells, provides contiguous address space
to the Operating System (OS), does not require changes to the existing ECC DIMMs, while still retaining soft error tolerance.

\section{ArchShield Framework}
\begin{figure}[htb]
\vspace{-0.15in}
  \centering \centerline{\psfig{file=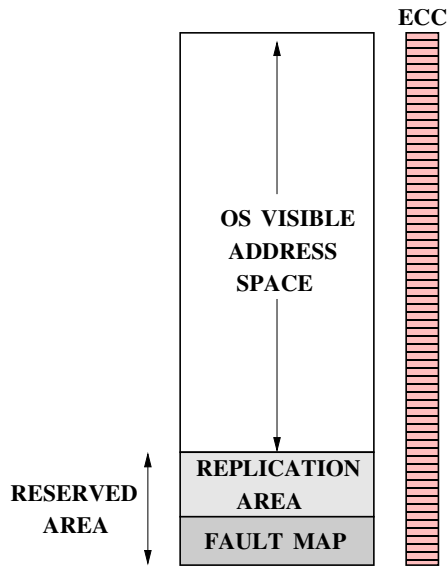,width=2.3in}} \caption{Overview of ArchShield (Figure not to scale)} \label{fig:highlevelarch}
\end{figure}

ArchShield leverages existing ECC DIMMs and enables them to tolerate
high-rate of faulty DRAM cells.  Figure~\ref{fig:highlevelarch} shows
an overview of ArchShield.  ArchShield divides the memory into two
regions: one that is visible to the OS, and the other reserved for
handling faulty cells.  Thus, the OS is provided with a contiguous
address space, even though this space may have faulty cells.
ArchShield contains two data structures: Fault Map (FM) and
Replication Area (RA).  The Fault Map contains information about the
number of faulty cells in the word.  ArchShield employs {\em Selective
Word Level Replication (SWLR)}, whereby only faulty words are
replicated in the Replication Area.  On a memory access, ArchShield
obtains the Fault Map information associated with the line.  If the
line contains word with faulty cells, it is repaired with the replicas
from the Replication Area.

For implementing ArchShield several challenges must be addressed. For
example, having Fault Map entry for every word incurs high
overhead. Similarly, accessing Fault Map from memory on every access
incurs high latency.  Also, the replication area must be architected
to reduce the storage and latency overhead associated with obtaining
replicas. Ideally, one would want almost all of the memory address available for demand usage (visible to OS), and keep the performance
penalties associated with Fault Map access and Replication Area to be
small-level, while retaining soft error protection.

\subsection{Testing for Identifying Faulty Cells}

ArchShield relies on having the location of faulty cells available.
If the error-rate was small, then this information can be supplied by
the manufacturer using some non-volatile memory on the DRAM
module. Unfortunately, this method does not scale well to high error
rates, as it incurs high storage overhead and cost (especially if the
non volatile memory is employed with laser fuses as done with row
sparing). So, for tolerating high error-rates, this chapter suggests runtime runtime testing.  This chapter assumes that there is a Built-In Self Test (BIST) controller present in the system that performs testing on the memory module when the module is first configured in the system.  Testing can be done by writing a small number of patterns (such as ``all ones'' and ``all zeros'') as done in~\cite{rapid:hpca06,raidr:isca} or by using well-known testing
algorithms such as MARCH-B, MARCH-SS, and pseudo random
algorithms for testing Active Neighborhood Pattern Sensitive Faults
(ANPSFs)~\cite{Test1,Test2}.

As ECC protection exists at the word granularity, testing is also
performed at word granularity. During the testing phase, the words are
classified into three categories: Words with no faulty cells (NFC),
Words with single faulty cell (SFC), words with multiple faulty cells
(MFC). This chapter assumes that testing is able to identify all faulty
cells,\footnote{Given that ArchShield provides a protection of 1-bit
soft error per word, it can tolerate a small probability of faults
escaping the testing procedure. In particular, the system can tolerate
one untested fault per word. A persistent soft error in the word can
be notified to the Fault Map.} and the Fault Map and Reserved Area are
populated with the results of testing.

\subsection{Architecting Efficient Fault-Map}

ArchShield makes a separation between words with single faulty cell
(SFC) and multiple faulty cells (MFC) as words with SFC can be handled
with ECC in the absence of soft error. Thus, the Fault Map entry for
each word must provide a tertiary value: NFC, SFC, or MFC.  If one
keeps 2-bits per 64-bit word, this would result in a storage overhead
of 1/32 of the entire memory. Furthermore, there may be faulty cells
in the Fault Map as well, so additional redundancy would make the
storage overhead of Fault Map prohibitive.

\subsubsection{Line Level Fault Map}
This chapter suggests reducing the storage overhead of Fault Map by exploiting the observation that memory is typically accessed at a cache line
granularity (64 bytes).  So, one can keep the information about faulty
words at the cache line granularity as well.  To ensure correctness,
the fault level of all the words in the line is determined by the word
with the most number of errors.  If the line contains no faulty cell,
it will be classified to be an NFC line.  If the line contains
at-least one SFC word, but no MFC word, the line is classified as an
SFC line.  Whereas, if the line has a MFC word, the line is classified
as an MFC line.

As the line contains eight words, the probability of SFC line is
approximately 8x higher than SFC word, increasing from 0.7\% of words
to 5.6\% of the lines.  Similarly, the probability that the line is
classified as MFC line is increased by approximately 8x as well,
increasing from 26ppm to 200ppm.  The increase in SFC line does not
impact performance significantly, as the replicated information is not
accessed on a read (unless there is soft error).  The dual read
because of increase in MFC line is negligible to have any meaningful
impact system performance, as it affects one out of 5000 accesses.

\subsubsection{Fault Tolerance and Overhead of Fault Map}
ArchShield assumes that the entire memory can contain faulty cells,
including the area used to store the Fault Map. Therefore, this chapter proposes using redundancy in storing the Fault Map entry.  Each Fault Map entry
consists of 4-bits.  If it is 0000, the line is deemed to have no
faulty cells.  If it is 1111, the line is deemed to have at-least one
(or more) word with at-most one faulty cell.  For any other
combination, the line is conservatively deemed to be a MFC line.  The MFC line is stored as 1100 in the Fault Map.

An error in Fault Map results in reading the replicated version of the
word. The Fault Map area is also protected by ECC, so on any detected
(or corrected) fault, the design conservatively tries to read from the
replicated region.  With 4-bits per 64-byte line, the storage overhead
of Fault Map would be 1/128 of the entire memory, or equivalently 64MB
for a 8GB DIMM. The address of the Fault Map entry can be obtained by
simply adding the line address to the Fault Map Start Address (which
is kept in a register of ArchShield).

\subsubsection{Caching Fault Map Entries for Low Latency}

The Fault Map must be consulted on each memory access. A naive
implementation of probing Fault Map in main memory on every memory
access would result in high performance overhead.  So, this chapter recommends
caching the Fault Map entries in the on-chip cache, on a demand basis.
Each Fault Map access can bring in a cache line worth of Fault Map
information and cache it in the Last Level Cache (LLC). Given each Fault Map
entry is only 4-bits, each cache line of Fault Map contains Fault Map
information for 128 lines, resulting in high spatial and temporal
locality. The analysis in this chapter shows that the Fault Map hit rate in the
on-chip LLC to be in the regime of 95\% on average,
thus significantly reducing the memory accesses for Fault Map and
associated performance penalties.

\subsection{Architecting Replication Area}

The Replication Area stores a replica for all the words with a faulty
cell. The Fault Map only identifies if the line has a word with faulty
cell, it does not identify the location of the replicated copy of this
word. Therefore, the Replication Area must also contain a tag entry
associated with each word. The tag size depends on the ratio of Replication Area to Memory size. To tolerate a BER of $10^{-4}$,
the Replication Area needs to store 7.74 million faulty words for an
8GB DIMM.  If one could configure the Replication Area as a fully
associative structure, then one would need only 7.74 million entries,
incurring about 1\% of memory capacity. Unfortunately, this
configuration would incur unacceptably high latency overheads. Replication Area is provisioned to be $\frac{1}{64}$th of main memory for BER of 10$^{-4}$. So we have 6 bits for line address, 3 bits for word in line, 1 valid bit and 2 overflow bits (replicated) for every entry, hence we get 1.5 bytes for tag. Thus, each entry in the replication region would be 9.5 bytes (1.5
bytes for tag and 8 bytes for data). This section identifies the appropriate structure for Replication Area to reduce latency while keeping the storage overhead
manageable.

\subsubsection{A Set Associative Structure}

\begin{figure}[htpb]
  \centering
  \centerline{\psfig{file=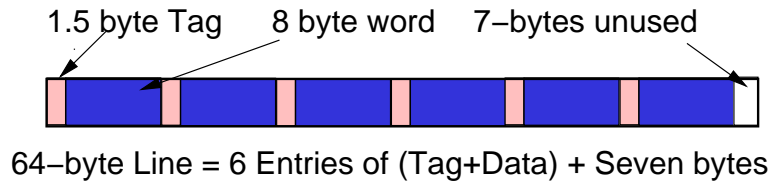,width=4in}}
  \caption{A 64-byte line configured as one set in the replication region. It
can hold six entries and have seven bytes unused.}
  \label{fig:spareset} 
\end{figure}

This sections aims to keep the interaction between the memory and the memory controller
to be at a cache line granularity. Therefore, even the memory of the
Replication Area can be accessed at a cache line granularity. Given
that the cache line is 64 bytes, and each Replication Area entry is
9.5 bytes (1.5 bytes tag + 8 bytes data), one can store six entries in
each line of 64 bytes, and have two bytes of unused storage, as shown
in Figure~\ref{fig:spareset}.

Since each line can hold six entries, one can configure the
Replication Area as a 6-way set associative structure. If the access
across sets was uniform, then only 1.3 million sets (7.74
Million divided by six) sets would be required. Unfortunately, as errors are spread randomly
throughout the memory space, the allocation of this structure is
non-uniform. We want to avoid the overflow of any of the set, as it
would mean that we are unable to accommodate all faulty cells, and that
module may be deemed unusable.  

One can reduce the probability of overflow by increasing the number of
sets. For the described configuration, to avoid the overflow of any set, we need 12x more sets. This incurs a storage overhead of approximately 15\%, and is unappealing.

\subsubsection{Efficiently Handling Overflow of Sets}

\begin{figure}[htb]
  \centering \centerline{\psfig{file=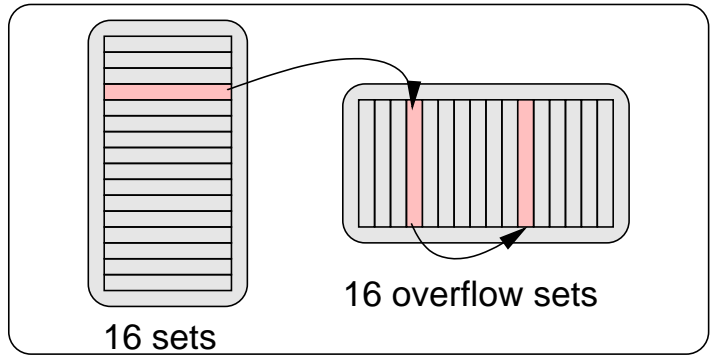,width=3.7 in}}
  \caption{An RAgroup with 16-sets and 16 overflow sets.
  An overflow set can overflow into another set of same RAgroup.}
  \label{fig:ragroup}
\end{figure}

Given that the overflow of the set associative structure are
infrequent, we can tolerate these with a flexible organization that
handles overflows in the set associative structure.  We provide the
set associative structure with a victim-cache like structure.  Each
group of 16-sets is provisioned with a 16 additional overflow
sets. The 7-bytes unused in each set is used to link to one of the
entries in the overflow region.  The location of the overflow set can
be identified with 4-bits and coupled with a valid bit, the pointer to
overflow sets would take 5-bits.  This chapter proposes using triple modulo redundancy on
the pointer for fault tolerance.  Furthermore, this chapter calls such a structure of 16 sets
+ 16 overflow sets as a {\em Replication Area group}, or simply
RAgroup. Figure~\ref{fig:ragroup} shows the overview of RAgroup.

Note that even though there is linkage between the normal sets and
overflow sets, this does not impact the deterministic latency of
existing memory interfaces. We first access the normal sets in the
group. If no words for the given line is present, and there is a link
to the overflow sets, then we send another memory request for
obtaining the overflow set.  Thus, our proposed structure can be
easily incorporated in existing memory
controllers. 

\begin{figure}[htpb]
  \centering
  \centerline{\psfig{file=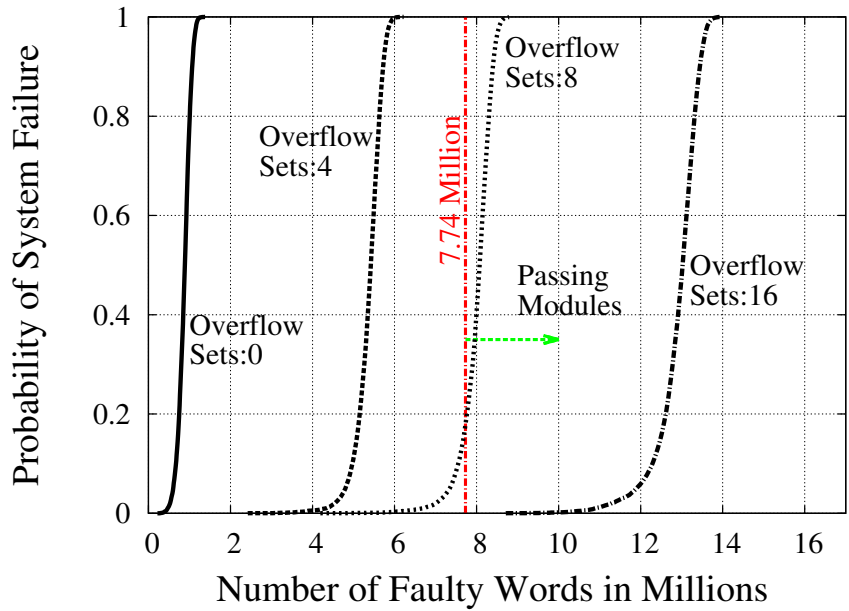,width=4.7in}}
  \caption{Probability that a Replication Area group structure is unable to handle given
  number of errors (in million). We recommend the structure with 16
  overflow sets to tolerate 7.74 million errors in DIMM.}  
\label{fig:overflowprob}
\end{figure}

Given that the normal sets occupy a storage of 1KB and the overflow
sets also occupy a storage of 1KB, the entire RAgroup can reside within the same 2KB row-buffer.
Thus, the access to overflow set is guaranteed to get a row buffer
hit, reducing the access latency. To handle 7.75 million faulty words,
we use 128K RAgroups (each with 16-set + 16 overflow sets).  As each
RAgroup incurs a storage overhead of 2KB, the proposed structure for  Replication Area incurs an overhead of 256MB.

Figure~\ref{fig:overflowprob} shows the probability that this
structure will not be able to handle a given number of random errors,
for different value of overflow sets in the group. Monte-Carlo simulation is used to perform this analysis, by using 100K
runs. Even in 100K simulations, the structure with 16 overflow sets
was unable to handle 8 Million errors only once. Thus, the structure
has low variance which means the probability of deeming the DIMM
unusable is negligible (10ppm).

\subsection{ArchShield Operation: Reads and Writes}

ArchShield extends the memory controller to do read and write
operations appropriately.  On a read request that misses in the LLC,
the request is sent to memory. In parallel, the address for the Fault
Map entry is computed and the LLC is probed with the Fault Map
address.  In case there is a LLC hit for the Fault Map address (common
case), the Fault Map entry is retrieved.  Otherwise, another request
is sent to memory to obtain the line containing the Fault Map (an
uncommon case) and is installed in the LLC.  If the Fault Map entry
shows that the line does not have any faulty cell, one can use the data
supplied from the main memory. If the line is deemed to have
single faulty cell words, and ECC operation on the line does not
result in uncorrectable error, one does not require reading the replicated
copy. However, if there is one bit soft error and the ECC operation
results in uncorrectable error, the replicated copy is read, thus
providing soft error protection.  If the line is deemed to have a word
with multiple faulty cells, then the replicated copy is read and the
matching words are incorporated in the line. Thus, accessing a line
with multiple faults causes extra latency, however this is a rare
event. For an error-rate of $10^{-4}$, extra read is performed for
less than one in few thousand read operations.

We add a bit called {\em Replication bit (R-bit)}  to the tag-store
entry in each line of the LLC to mark if the line requires replication
on writeback.  If, on the demand read, the line was determined to have
a single faulty cell or multiple faulty cells the R-bit is set. A
write to two locations (a good location and the replicated location)
in case of word with single fault ensures that soft errors can be
corrected by reading the copy from the Replication Area.

When a dirty line is evicted from the cache, and the R-bit is not set,
writeback is done in normal manner. However, if the R-bit is set, we
also need to update the replicated region.  After the normal write is
performed, the memory controller probes the replicated area for
obtaining the set containing the replicated words for the given line.
It then updates the data value for the corresponding words of the
line, and updates the replicated region. Thus, while the Fault Map is
cached in LLC, the replicated region is updated by the memory
controller on a demand basis, and is not cached. Also note that the
latency for doing the multiple writes is not in the critical path,
however the extra operations can cause contention and thus impact
performance indirectly.  For an error-rate of $10^{-4}$, 5.6\% of the
memory lines will require extra write operations.

\begin{figure}[H]
  \centering
  \centerline{\psfig{file=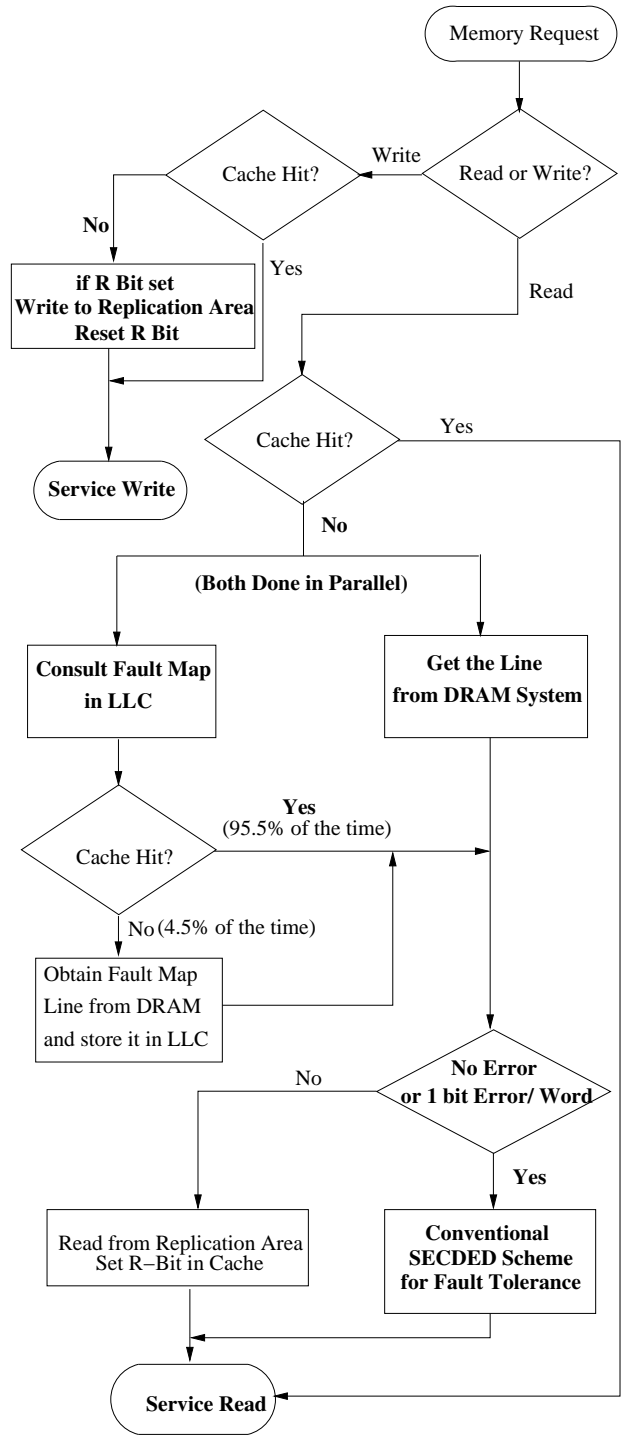,width=3.2in}}
  \caption{A Flowchart of the read and write operations in ArchShield. The decisions in `Bold' words indicate the most frequent path for requests in case of a LLC miss}  
\label{fig:flowchart}
\end{figure}

Figure~\ref{fig:flowchart} shows the flowchart depicting the events involved when a memory request arrives. The performance is impacted by the hit rate of the fault map for high MPKI benchmarks. As the Fault Map is organized with high locality, for a read request, 95.5\% of the time, we need only one main memory transaction.

\newpage
The proposed implementation assumes an R-bit in each cache line. If the cache does not support this, we can still implement ArchShield by making dirty lines that are evicted from the cache probe the Fault Map in order to determine if dual writes must be performed. Currently, Fault Map requires 4-bit per line (64MB for 8GB chip). This structure is designed to handle high BER.  When the BER is low, an altenative implementation (such as Bloom filters and lookup-tables) can be used to reduce the storage overhead.

\subsection{ArchShield: Tying it All Together}

\begin{figure}[htb]
  \centering
  \centerline{\psfig{file=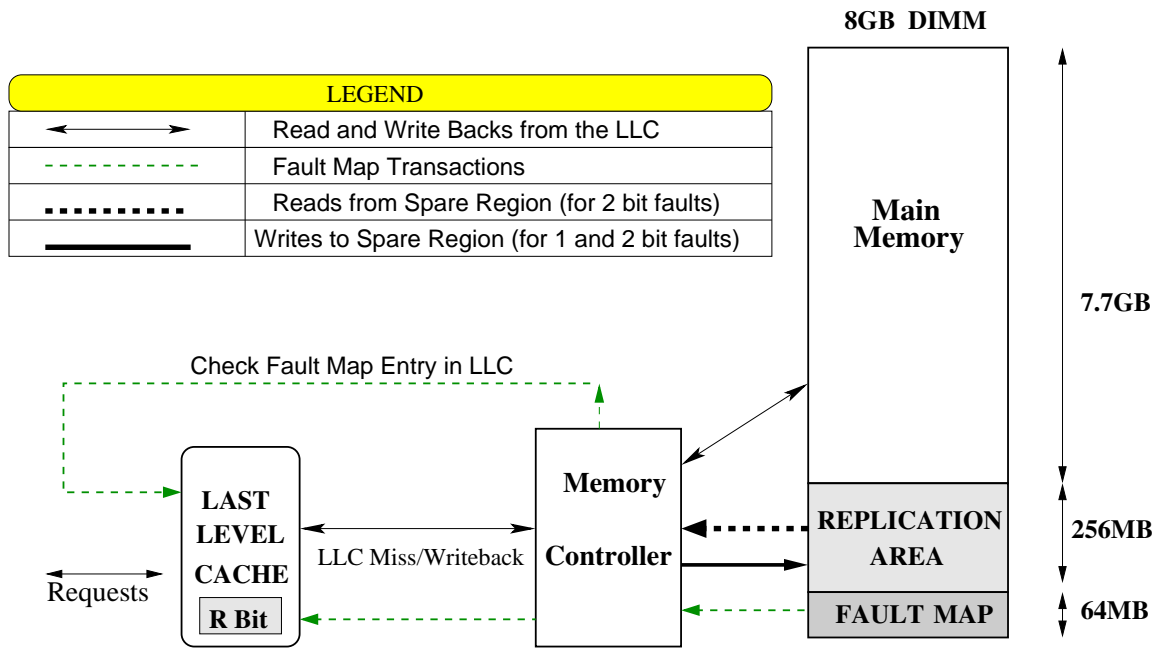,width=\columnwidth}}
\caption{Memory System with ArchShield} 
\label{fig:tyingtogether}
\end{figure}

Figure~\ref{fig:tyingtogether} shows a memory system with ArchShield.
The main memory consists of traditional ECC DIMMs and does not require
any changes.  The memory space is divided into addressable space,
Replicated Area and Fault Map. 

The memory controller is extended to
compute the address of the Fault Map entry, check that entry in the
LLC, and in cases of an LLC miss for the Fault Map, read the required
line with Fault Map information and cache it in the LLC.  On an LLC
read miss, the memory controller obtains the Fault Map entry, and
determines if a second read from the replicated region is required.
If so, it reads the replicated region and repairs the line with
replicated words. In case of an LLC writeback, the memory controller
determines if the replicated region must be updated. If so, the extra
write operations are performed. This check for replicated writeback is
assisted by the R-bit in the LLC. Thus, ArchShield requires changes to
the memory controller and minor changes to the cache structure (to add
the R-bit to the tag store entry).

The data-structures for ArchShield are kept in main memory. For 8GB
memory, the Fault Map requires 64MB storage, and the Replication Area
requires 256MB storage, for a total storage overhead of 320MB. Thus,
ArchShield provides remaining 7.7GB (or 96\% of the 8GB memory)
available as visible address space.

\section{Experimental Methodology}
\subsection{Configuration}

For evaluating ArchShield, this chapter uses an in-house memory system simulator for our studies. The
baseline configuration is described in
Table~\ref{table:system_config1}.  There are 8 cores sharing an 8MB
LLC. The memory system contains two channels, each with one 8GB DIMM.
The virtual to physical translation is performed using a first touch policy,
with 4KB page size. The Fault Map entries are cached on a demand
basis and evicted using LRU replacement of LLC. The scaling-related error-rate is assumed to be 10$^{-4}$, and that faulty cells are spread randomly across the
memory space. For accessing replicated region, the simulation requires extra 3 DRAM
cycles for parsing the tag-store, and one additional DRAM cycle for
access to overflow set.

\begin {table}[ht]
\begin{center} 
\caption{Baseline System Configuration for ArchShield}
\vspace{0.1in}
\resizebox{4in}{!}{
\begin{tabular}{|c|c|}
\hline
				 \multicolumn{2}{|c|}{\textbf{Processors}} \\ \hline
                 Number of cores               & 8         \\ 
                 Processor clock speed         & 3.2 GHz    \\ \hline
                 \multicolumn{2}{|c|}{\textbf{Last Level Cache}} \\ \hline
                 L3 (shared)     & 8MB   \\
                 Associativity   & 8 way \\
                 Latency		 & 24 cycles \\ 
                 Cache line size & 64Bytes    \\ \hline
                 \multicolumn{2}{|c|}{\textbf{DRAM 2x8GB/channel-DDR3}} \\ \hline
                 Memory bus speed             & 800MHz (DDR3 1.6GHz)\\
                 Memory channels              & 2         \\
                 DIMM capacity per channel    & 8GB         \\
                 Ranks per channel            & 2         \\
                 Banks per rank               & 8         \\
                 Row Buffer Size			  & 8KB (DIMM)\\
                 Bus width 					  & 64 bits per channel\\
				 t$_{CAS}$-t$_{RCD}$-t$_{RP}$-t$_{RAS}$ 		  & 9-9-9-36\\ \hline

\ignore{
				 \multicolumn{2}{|c|}{\textbf{ArchShield Parameters}} \\ \hline
				 Spare Set Associativity & 6 \\
				 Overflow Set Associativity & 6\\
				 Entries Per Spare Set Group & 16\\
				 Entries Per Overflow Set Group & 16\\
				 Fault Map Redundancy Scheme & Mod-4\\ 
				 Spare/Overflow Set Search Latency & 3 cycles (@ 800 MHz)\\ 
				 Spare Set Overflow Latency & 1 cycle (@ 800 MHz)\\ \hline

}
\end{tabular}
}
\label{table:system_config1}
\end{center}
\end{table}

\subsection{Workloads}

A representative slice~\cite{simpoint} of 1 billion
instructions for each benchmark from the SPEC2006 suite is used. Evaluations are performed by executing the benchmark in rate mode, where all the
eight cores execute the same benchmark. The Read and Write MPKI of these workloads indicate their memory
activity. Workload footprint is computed by the number of unique (4KB)
pages touched by the workload.  Since there are 8 copies of the benchmark,
the total footprint is increased by 8x. Timing simulation is performed
till all the benchmarks in the workload finish execution. Thereafter,
the average execution time over 8 cores is computed.

\section{Results}
\subsection{Impact on Execution Time}

ArchShield has two sources of performance overhead. One is caching of
the Fault Map. A read operation for a line from main memory will not
complete until the Fault Map entry is available.  So, Fault Map miss
in the the LLC causes increase in the read latency.  The other is the extra traffic due to updates to the Replication Area.  To, better understand the performance implications from these two factors, we conducted experiments with three ArchShield configurations. First, an ideal Fault Map (which does not consume LLC area or memory traffic). Second, a configuration in which the extra traffic for the
Replication Area is ignored. Third, ArchShield with realistic Fault Map and Replication Area.

\begin{figure*}[htpb]
  \centering
  \centerline{\psfig{file=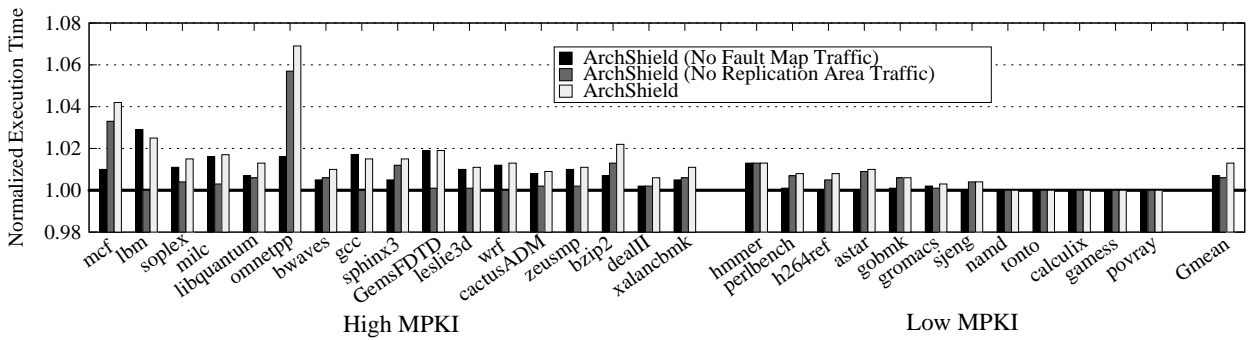,width=6.5in}}
  \caption{Impact on Execution Time for three ArchShield configurations: 1. Ideal Fault Map, 2. No extra writes, 3. Realistic }
  \label{fig:speedup}
\end{figure*}

Figure~\ref{fig:speedup} shows the execution time of the three
ArchShield configurations.  The execution time is normalized to the
baseline with fault-free memory.  The bar labeled Gmean shows the
geometric mean over all the workloads.  On average, ArchShield causes
an execution time increase of 1\%.\footnote{In our analysis we have
assumed that the performance loss due to the unavailable memory
capacity (4\%) is negligible, which is accurate given the footprint
  of our workload.  However, for workloads with larger footprints
  there may be a minor (negligible) performance loss due to reduced
  capacity.}  The Fault Map and Replication Area are each responsible
for approximately half of the performance loss.  However, the impact
depends on the workloads. For several workloads the performance loss
is primarily because of extra traffic to the Replication Area. For
\textit{omnetpp}, the performance loss is due to non-ideal Fault Map.

\subsection{Fault Map Hit Rate Analysis}

\begin{figure*}[htpb]
  \centering
  \centerline{\psfig{file=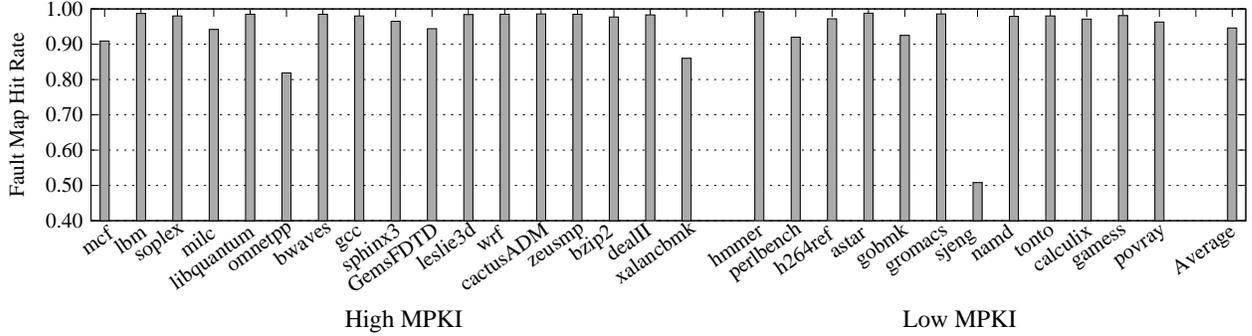,width=6.5in}}
  \caption{Fault Map Hit Rate in Last Level Cache}
  \label{fig:hitrate1}
\end{figure*}

The locality of the Fault Map is central to efficient operation of
ArchShield.  Given that each line of Fault Map contains information
about 128 contiguous lines, we expect high spatial and temporal
locality for the Fault Map line in the LLC. Figure~\ref{fig:hitrate1}
shows the hit rate of the LLC for Fault Map accesses.  On average, the
Fault Map hit rate for LLC is 94\%.

For benchmarks that have high MPKI, the Fault Map hit rate is
reduced. This happens because the cache is contended for both the
demand lines as well as the lines from the Fault Map. For example,
\textit{omnetpp} has a Read MPKI of 20.8, and FM hit rate of 82\%,
hence it has the highest performance degradation with
ArchShield. Other high MPKI workloads such as \textit{mcf} and
\textit{xalancbmk} show similar behavior. For \textit{sjeng}, the low
hit rate of the Fault Map does not impact performance because it has
very low MPKI, hence the system performance is not sensitive to memory
performance.  Overall, the Fault Map caching for ArchShield is
quite effective as only three benchmarks out of 29 show a FM hit rate
of less than 90\%,

This chapter also analyses the occupancy of Fault Map entries in the LLC. On average, 6\% of the LLC contains lines from the Fault
Map. Thus, the spatial locality of Fault Map entries helps the Fault
Map to get high hit rate without occupying significant area in the
LLC.  Note that, while performing cache replacement in the LLC, we do
not differentiate between lines from the main memory and lines from
the Fault Map. So, even a simple demand-based caching policy for the
Fault Map works quite well.

\subsection{Analysis of Memory Traffic}

In addition to the normal memory traffic from LLC misses and
writebacks, ArchShield increases the memory traffic due to extra
activity.  In particular, the memory traffic is increased because of
Fault Map misses in the LLC and the extra writes to the Replication
Area for the faulty lines.  Furthermore, caching the Fault Map entries
in the LLC may increase the LLC miss rate and writebacks for the
demand accesses.

\begin{figure*}[htp]
  \centering
  \centerline{\psfig{file=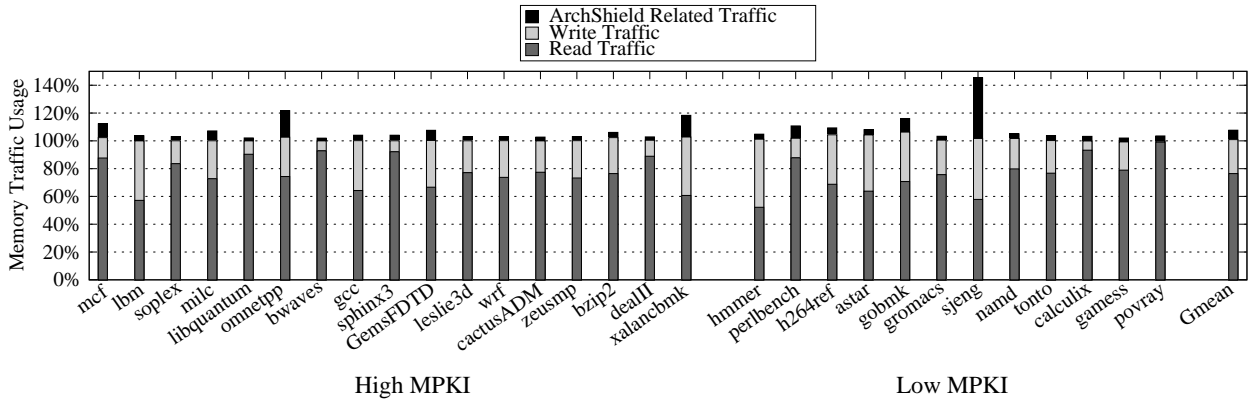,width=6.5in}}
  \caption{Memory Traffic Breakdown with ArchShield}
  \label{fig:traffic1} 
\end{figure*}

To capture the impact of ArchShield on memory traffic we divide the
total memory traffic into three components.  The read traffic
emanating from LLC misses, the writebacks from LLC, and the traffic
related to ArchShield (Fault Map and extra writes). Figure~\ref{fig:traffic1} shows the breakdown of these three
components. The total memory traffic is normalized to the memory
traffic with the fault-free memory.

The traffic due to ArchShield shows a negative correlation with Fault
Map hit rate. The benchmark \textit{sjeng} has the highest traffic
overhead due to ArchShield of around 35\%. This happens because of low
hit rate of the Fault Map.  However, as this benchmark has low MPKI,
the impact on performance is insignificant.  For \textit{astar}, the
traffic due to demand accesses is higher compared to the baseline
because of extra LLC misses and writebacks due to caching of Fault Map
entries.  

Due to the replication of lines with fault cells, we can expect the
writeback traffic to increase by 5.6\%, as 5.6\% of the lines are
expected to have a faulty cell. On average, ArchShield increases the
total memory traffic by 6\%.

\subsection{Analysis of Memory Operations}

For lines with multiple faults, ArchShield requires that multiple
accesses be done on a read: one to the normal location and the
other to the Replication Area. The access to the Replication Area can
itself result in multiple accesses, if the set in the Replication Area
overflows to another set.  However, this happens rarely.
Table~\ref{table:mem_analysis} shows the breakdown of memory
operations in terms of number of accesses to memory. This subsection analyzes three operations: a read operation due to LLC miss, a writeback from LLC and a Fault Map miss in the LLC.  All numbers are
relative to the total memory operations.

\begin {table}[htpb]
\begin{center} 
\caption{Analysis of Memory Operations for ArchShield}
\begin{tabular}{|c|c|c|c|}\hline
				 \textbf{Transaction}&\textbf{1 Access(\%)}&\textbf{2 Access(\%)}&\textbf{3 Access(\%)}\\ \hline \hline
                 Reads& 72.13& 0.02& \textasciitilde 0 \\
                 Writes& 22.07& 1.18 & 0.05\\
                 Fault Map& 4.55 & N/A & N/A\\\hline
                 Overall & 98.75 &1.2 & 0.05\\\hline
\end{tabular}
\label{table:mem_analysis}
\end{center}
\end{table}

On average, 72.15\% of all memory accesses are read operations, out of
which only 0.02\% accesses require two memory accesses.  Thus, almost
all read operations get satisfied with single access.  Writebacks
account for 23.3\% of all memory operations on average.  As we can
expect 5.6\% of lines to cause extra writes (due to replication), the
number of writes that require two accesses are
5.6\%*23.3\%=1.18\%. Only a negligible number of write operations
require three accesses.  On average, 4.55\% of the memory operations
are due to Fault Map miss, each of which get satisfied in one memory
operation.  Thus, ArchShield satisfies 98.75\% of all memory
operations with single memory access.

This section also analyzes the read latency for the baseline and ArchShield. ArchShield obtains an average read latency of 200 cycles with a baseline of 197 cycles. This 1.5\% increase in the read latency causes only a 1\% reduction in performance.

\subsection{Sensitivity of ArchShield to Bit Error-Rate}

We have selected parameters for ArchShield to tolerate a bit error-rate of 10$^{-4}$.  ArchShield can be tuned to handle a different
error-rate.  For example, to handle a bit error-rate of 10$^{-5}$, we
can reduce the size of Replication Area by 8x, as we expect 10x fewer
faulty cells. This reduces the storage overhead of ArchShield to
96MB, making 98.8\% of memory capacity available for normal usage.
Also, fewer faulty cells also reduces the traffic due to extra writes.
The overall increase in execution time is 0.5\%, instead of 1\% at
error-rate of 10$^{-4}$.

Conversely, to handle 2x higher error-rate ($2\times10^{-4}$), the
storage overhead would get doubled to 7\%, making only 93\% of memory
capacity available for use. It will also cause higher performance
degradation due to increased write traffic from replication, as 11\%
of the lines would require an extra write.

\subsection{Quantitative Comparison with Prior-Work: FREE-P} 
The work that is most closely related to ArchShield is FREE-p
(Fine Grained Remapping with ECC and Embedded
Pointers)~\cite{freep:hpca11}. FREE-p decommissions a line with faulty cells (more than what can be
handled by the per-line ECC) and stores a pointer in the line to point
to the spare location.  It relies on the read-before-write
characteristics of PCM memory to read the pointer before writing to
the line (to avoid destroying the pointer).  While this may be a
reasonable assumption for PCM because of high write latency, such
read-before-write operations cause significant performance degradation
in DRAM memories. 

\begin{figure*}[htpb]
  \centering
  \centerline{\psfig{file=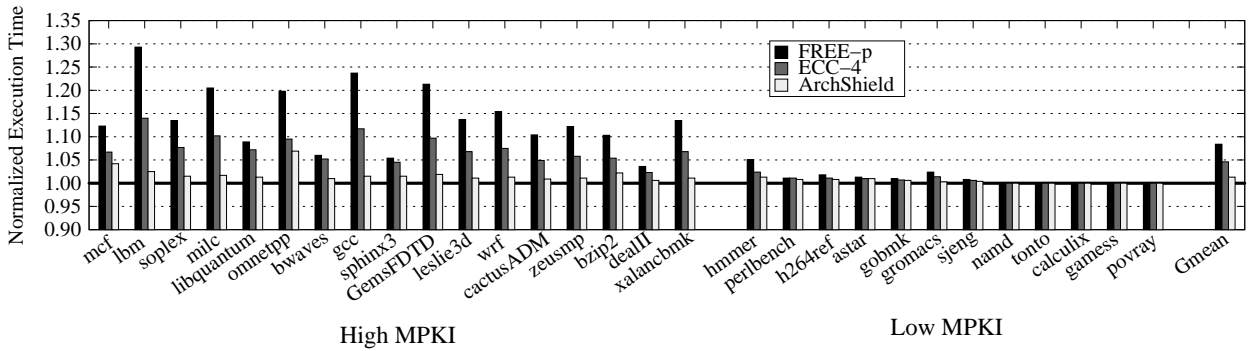,width=6.5in}}
  \caption{Execution time impact of different schemes. Providing ECC-4 per word incurs prohibitive storage overhead (37\% memory capacity),
    whereas the read-before-write requirement of FREE-p causes
    significant performance degradation. }
  \label{fig:related1} 
\end{figure*}

Figure~\ref{fig:related1} compares the performance
of FREE-p with ArchShield. This dissertation implements the \textit{Baseline} FREE-p system. FREE-p causes 8\% performance degradation on average (and sometimes as high as 29\%, such as for lbm), whereas ArchShield causes negligible performance impact.  Furthermore, FREE-p assumes a fault indicator bit with each line, which is not present in traditional DIMMs. Even if one chooses other implementations of FREE-p (\textit{pCache}, \textit{pIndexCache}), they would incur the high latency of their multi-bit ECC decoder. Since multi-bit ECC decoder delay is not present in ArchShield, it gives a better performance when compared with FREE-p.

\section{Summary}

Scaling of DRAM memories has been the prime enabler for higher
capacity main memory system for the past several decades.  However, we
are at a point where scaling DRAM to smaller nodes has become quite
challenging.  If scaling is to continue, future memory systems may be
subjected to much higher rate of errors than current DRAM systems. Unfortunately, tolerating high error
rates while concealing the information about faulty cells within the
DRAM chips results in high overhead.  To sustain DRAM scaling,
efficient hardware solutions for tolerating high error-rates must be
developed. To that end, this chapter makes the following contributions:

\begin{enumerate}

\item This chapter proposes {\em ArchShield}, an architectural framework that exposes the information about faulty cells to the hardware.  It uses a {\em Fault Map} to track lines with faulty cells, and employs {\em Selective Word Level Replication (SWLR)}, whereby only faulty words are replicated for fault tolerance.

\item This chapter shows that embedding the data structure of ArchShield in memory still renders (96\%) of the memory capacity useful, even at high error-rate.

\item This chapter shows that the performance degradation of ArchShield from
  extra traffic due to Fault Map and SWLR is only 1\%.  This is
  achieved by demand-based caching of Fault Map entries on processor
  chip, and by architecting the replication structure to reduce access
  latency.

\end{enumerate}


\chapter{Low-Cost ECC for Strong Runtime Reliability}
Large-granularity memory failures continue to be a critical impediment to system reliability. To make matters worse, as DRAM scales to smaller nodes, the frequency of unreliable bits in DRAM chips continues to increase. To mitigate such scaling-related failures, memory vendors are planning to equip existing DRAM chips with On-Die ECC. For maintaining compatibility with  memory standards, On-Die ECC is kept invisible from the memory controller.  This chapter explores how to design memory systems in presence of On-Die ECC to improve runtime reliability.

\section{Introduction}
Technology scaling has been the prime driver of increasing the capacity of the DRAM modules. Unfortunately, as technology scales to smaller nodes, DRAM cells tend to become unreliable and exhibit errors~\cite{Prashant_ISCA:2013,cidra}. The industry plans to continue DRAM scaling by placing Error Correcting Codes (ECC) inside DRAM dies, calling it {\em On-Die ECC} (also known as {\em In-DRAM ECC})~\cite{uksong}. On-Die ECC enables DRAM manufacturers to correct errors from broken cells~\cite{HAMMING:1950}. Consequently, DRAM chips with On-Die ECC are already proposed for systems with DDR3, DDR4 and LPDDR4 standards~\cite{uksong,synopsys_ondie,lpddr4_ondie}. For maintaining compatibility with DDR standards and to reduce the bandwidth overheads for transmitting On-Die ECC information, manufacturers plan to conceal the On-Die error information to remain within the DRAM chips~\cite{uksong,lpddr4_ondie}. Thus, On-Die ECC is invisible to the system and cannot be leveraged to improve resilience against runtime faults. This chapter looks at how to design systems with stronger memory resilience in the presence of On-Die ECC.

Recent field studies from super-computing clusters show that DRAM reliability continues to be a critical bottleneck for the overall system reliability~\cite{failures:sridharan12,failures:sridharan13,failures:sridharan15}. Furthermore, these studies also highlight that large-granularity failures that happen at runtime, such as row-failures, column-failures and bank-failures, are almost as common as bit failures. DRAM modules can be protected from single bit failures using an ECC-DIMM that provisions an extra chip for error correction. However, tolerating large-granularity failures in the memory system is expensive and high-reliability systems often need to implement Chipkill to tolerate a chip failure at runtime. Unfortunately, implementing Chipkill requires activating 18 chips, which necessitates either using a non-commodity DIMM (x4 devices), and or accessing two memory ranks (x8 devices) simultaneously, which increases power and reduces parallelism. Ideally, we want to implement Chipkill using commodity memory modules and without the storage, performance, and power overheads.

\begin{figure}[htb]
  \centering \centerline{\psfig{file=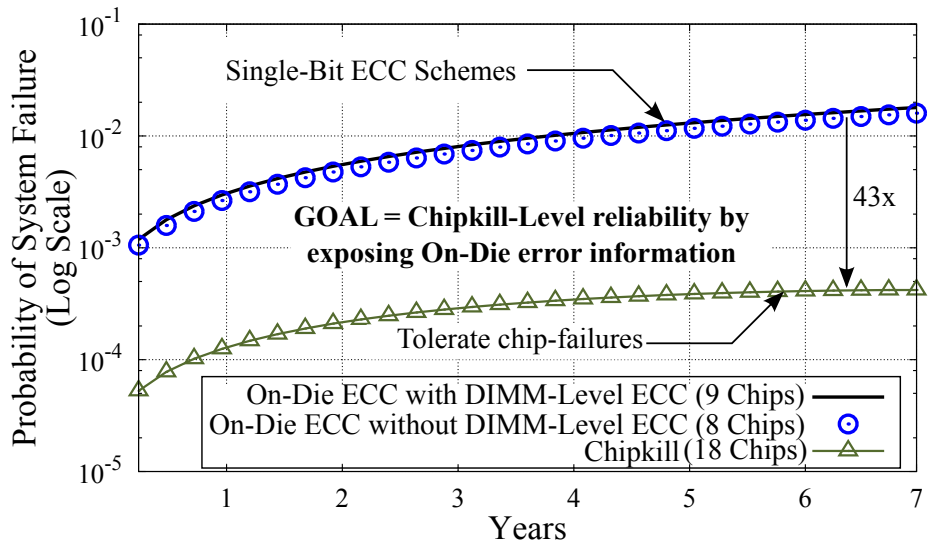,width=4.8in}}
  \caption{Effectiveness of reliability solutions in presence of On-Die ECC.}
  \label{fig:intro2}
\end{figure}

This chapter analyzes how On-Die ECC affects the reliability of DIMM-based ECC and Chipkill. Figure~\ref{fig:intro2} shows the probability of system failure, by considering real world failure-rates, for the memory system over a period of 7 years. This chapter compares three systems: (a) Non-ECC DIMM with 8 chips, (b) ECC-DIMM with 9 chips, and (c) Chipkill-based system with 18 chips. It is observed that if the system is provisioned with On-Die ECC there is almost no benefit of having the DIMM-level ECC. Furthermore, Chipkill-based systems provide 43x more reliability than ECC-DIMM. From this analysis, one may conclude that the 9-chip ECC-DIMM solution is superfluous in the presence of On-Die ECC. This dissertation argues that this is an effect of concealing the On-Die ECC information from the external system. This dissertation shows that revealing the On-Die ECC error detection to the memory controller can enable Chipkill-level reliability while avoiding the associated overheads.

Unfortunately, exposing On-Die ECC to the memory system requires that more bits be transferred from the DRAM chips to the memory controller~\cite{uksong,synopsys_ondie,lpddr4_ondie}. This can accomplished by either providing more lanes or using additional bursts, both of which are incompatible with existing DDR standards~\cite{DDR3:2015,DDR4:2015}. Ideally, one would like to expose the On-Die error information without any overheads and without changing the existing standards. This dissertation leverages the observation that the memory controller does not need to have visibility of the On-Die ECC bits; it simply needs to know if the On-Die ECC has detected an error. The memory system can then use the On-Die error detection information in conjunction with the DIMM-Level parity and correct errors in a manner similar to RAID-3. To this end, this dissertation proposes \emph{XED} (pronounced as ``zed'', the British pronunciation of the letter ``z''), a technique that e\underline{X}poses On-Die \underline{E}rror \underline{D}etection information while avoiding the bandwidth overheads and changes to the memory standards.

To efficiently communicate that On-Die ECC has detected an error to the memory controller, XED relies on {\em Catch-Word}. A Catch-Word is predefined randomly selected data-value that is transmitted from the chip to memory controller to convey that a fault has occurred in a given DRAM chip. Both the memory controller and the DRAM chip a priori agrees on the given Catch-Word. XED uses the 9th chip in the ECC-DIMM to store parity information of all the other chips.  When the On-Die ECC identifies an error, the DRAM chip transmits the Catch-Word instead of the requested data-value to the memory controller. When the memory controller recognizes the Catch-Word, it ignores the value from the associated chip and uses the parity from the 9th chip to reconstruct the data of the faulty chip. XED exploits the observation that typically a chip (x8 devices) provides a 64-bit data-value on each memory access. However, the chip cannot store all possible $2^{64}$ data-values. In fact, even a chip as large as 8Gb stores only $2^{27}$ 64-bit words. Even if all these words had unique values, the likelihood that the chip stores a data-value that matches with the Catch-Word is negligibly small ($2^{-37}$, or 1 in 140 billion).\footnote{While the likelihood of data-value matching the Catch-Word is negligibly small (once every million years for an x8 DIMM), XED can continue to operate reliably even when this occurs. In fact, XED can reliably detect the episode of a data-value matching the Catch-Word, and use this information to change the Catch-Word. This chapter discusses detecting collisions and updating Catch-Words in Section~\ref{sec:collisionx8}} 

This chapter discusses how XED can mitigate a chip failure by using an ECC-DIMM. It also discuss how XED can mitigate scaling faults in multiple chips. Thereafter, this chapter shows how XED can perform correction when runtime chip failure occurs concurrently with scaling faults. This chapter also presents evaluations which show that XED provides 172x higher reliability than ECC-DIMM alone. Furthermore, XED incurs negligible performance overheads ($<$ 0.01\%) and provides a 21\% lower execution time compared to traditional Chipkill. This chapter also analyzes XED for a system that implements Chipkill and show that XED enables this system to achieve Double-Chipkill level reliability without the overheads of Double-Chipkill. 

Overall, this chapter makes the following contributions to the dissertation:
\begin{enumerate}
\item It shows that DIMM-Level ECC provides no added reliability benefit to a memory system with On-Die ECC. This is because large-granularity runtime-faults are the main cause of memory  failures~\cite{failures:sridharan12,failures:sridharan13,vfecc:asplos10,LOTECC:2012,arcc,citadel:2014,Citadel}. Therefore, implementing the conventional DIMM-level SECDED with the 9th chip incurs area and power overheads without providing any reliability benefits.

\item It proposes XED, a technique that uses Catch-Words to reveal On-Die ECC error detection information to the memory controller without relying on extra bandwidth and changes to the memory interface.

\item It proposes a simple correction scheme for XED that uses the ECC-DIMM to store parity information and relies on RAID-3 based correction to tolerate a chip failure. It also shows that XED is effective at tolerating chip failure in the presence of scaling failures. This chapter also presents evaluations which show that XED enables Chipkill-level reliability without the power and performance overheads of traditional Chipkill implementations.

\item It shows that XED can enable conventional Chipkill systems to provide Double-Chipkill level reliability while obviating the storage, performance, and power overheads of Double-Chipkill.
\end{enumerate}

\section{Background}
This section provides a brief background on the DRAM organization, memory modules and On-Die ECC. This section also discusses the sources of errors and discuss the typical techniques for error mitigation.

\subsection{DRAM Module Organization}

DRAM memory is typically implemented as Dual Inline Memory Modules (DIMM), consisting of eight chips (x8 devices) providing a 64-bit wide databus. Each chip is further divided into banks, and each bank is further divided into rows and columns~\cite{Jacob2008,SALP}. An access to a DRAM DIMM activates a given bank in all of the chips. The access may activate a row of cells in the DRAM and then only a small portion from this row (corresponding to a cache line size, typically 64 bytes) is streamed out over the data bus~\cite{half_dram,minirank}. Thus, each chip is responsible for providing 64-bit per access, which is sent using 8 bursts of 8 bits each.

If the DIMM is equipped with ECC, it will have a 9th chip and will support 72 data lines (64 for data and 8 for ECC).  Each chip is still responsible for providing 64 bits for each memory access.

\subsection{On-Die ECC: The Why and the How}

As technology scales to smaller nodes, the number of faulty cells in a DRAM chip is expected to increase significantly. Mitigating these design-time faults with row-sparing or column-sparing will be prohibitively expensive. To increase yield, DRAM companies would like to use DRAM chips with scaling faults while still ensuring reliable operation and without significant overheads.  To achieve this, DRAM companies are planning to equip each chip with {\em On-Die ECC} (also called as {\em in-DRAM ECC}), whereby each 64-bit data within the chip is protected by an 8-bit SECDED code. DRAM errors are handled internally within the DRAM chip and this information is not made visible to the memory controller. As such, the On-Die ECC works transparently without requiring any changes to the existing memory interfaces and without making the memory controller aware that the chip is equipped with On-Die ECC.

\subsection{Fault Modes: Birthtime versus Runtime}

This dissertation classifies the faults into two categories: birthtime faults and runtime faults. Birthtime faults are those that occur at manufacturing time and can be detected by the memory vendors.  To ensure reliable operation of the chips, it is important that the memory vendors mitigate the birthtime faults or simply discard the faulty chips. Scaling faults~\cite{Prashant_ISCA:2013,cidra,hong:iedm2010,uksong,gu2003challenges} are birthtime faults and the On-Die ECC is designed such that these faults do not become visible to the external system. To ensure that a chip with On-Die ECC is not faulty, the manufactures will need to ensure that no 64-bit word has more than 1 faulty bit (if a word had multi-bit scaling-faults then use row sparing or column sparing to fix those uncommon cases).  In this chapter, we assume that scaling faults are limited to at most 1 bit per 64-bit word.

Runtime faults are those that occur during the operation of the DRAM chip. Runtime failures can be either transient or permanent, and can occur at different granularities, such as bit-failure, word-failure, column-failure, row-failure, bank-failure or rank-failure~\cite{failures:sridharan12,failures:sridharan13,failures:sridharan15}.  Recent field studies show that large-granularity runtime failures are almost as common as bit-failures. Therefore, we need solutions to efficiently  handle not only bit-failures  but also large-granularity failures. The next subsection describes the typical error mitigation techniques that are used in current systems.

\subsection{Typical Error Mitigation Techniques}

\subsubsection{SECDED} Memory systems may develop single-bit faults due to alpha-particle strikes and weak cells~\cite{alpha_dram,avatar}. To protect against single-bit faults, memory systems can use a variant of ECC codes that corrects single-bit errors and detect two-bit errors (SECDED)~\cite{bch_code:ibm,bch_decoder,BCCode}. DIMMs equipped with SECDED typically provide 8 bits of ECC for every 64 bits of data, and while activating  a single rank.

\subsubsection{Chipkill} Large-granularity failures, such as chip failures, can be tolerated by Chipkill, which employs symbol-based error correction code. Each data chip provides one symbol and there are extra chips provisioned for storing "check" symbols that are used locate and correct the faulty symbol (chip). With two check symbols, Chipkill can correct one faulty symbol (chip) and detect up to two faulty symbols (chips)~\cite{RSCode}.  As Chipkill needs two extra chips for storing these symbols, commercial implementations of Chipkill require that 18 chips be activated for each memory access (16 for data and two for check symbols). Unfortunately, this would mean that memory systems either use non-commodity chips (x4 devices) or obtain two cachelines for each access (x8 devices), causing a 100\% overfetch which increases power consumption and reduces parallelism.

\subsubsection{Erasures} The Chipkill design tries to do both, locate the faulty chip as well as correct the faulty chip. If we have an alternative means of knowing which chip is faulty, then we can tolerate a chip failure by simply relying on one chip (in general, for tolerating N chip failures we  need only N extra chips). This is called as {\em Erasure Coding}~\cite{RSCode,Erasure1,Erasure2}.

\subsection{The Goal of this Chapter}

The goal of this chapter is to propose a technique that can obtain Chipkill-level reliability without the associated overheads of area, power, and performance. This chapter leverages on the key observation  that if the DRAM chips already have On-Die ECC, then having the information about which chip encountered a fault can help us design an Erasure-based scheme to tolerate chip failures. However, one would want to expose the On-Die error detection information from inside the DRAM chip to the memory controller without incurring extra bandwidth and changing the memory interfaces. To that end, this dissertation proposes e\underline{X}posed On-Die \underline{E}rror \underline{D}etection (XED). Before describing XED, this section describes the reliability evaluation infrastructure.


\section{Reliability Evaluation}

To evaluate reliability of our proposed schemes we use \textsc{FaultSim}, an industry-grade fault and repair simulator~\cite{faultsim}. This study extends \textsc{FaultSim} to  accommodate scaling-faults faults. Based on prior studies, this study also assumes a scaling-fault rate of 10$^{-4}$~\cite{Prashant_ISCA:2013,uksong}. To model runtime-faults, this analysis uses real-world field data from Sridharan et al.~\cite{failures:sridharan12} as shown in Table~\ref{table:FIT2}.

\begin {table}[htb]
\begin{center}
\caption{DRAM failures per billion hours (FIT)~\cite{failures:sridharan12}}{
\vspace{0.15in}
\resizebox{0.60\columnwidth}{!}{
\begin{tabular}{|c|c|c|} \hline &\multicolumn{2}{c|}{Fault Rate (FIT)} \\
       \cline{2-3} DRAM Chip Failure Mode & Transient & Permanent \\ \hline 
        Single bit & 14.2 & 18.6\\
        Single word & 1.4 & 0.3 \\ 
        Single column & 1.4 & 5.6 \\ 
        Single row & 0.2 & 8.2\\ 
        Single bank & 0.8 & 10\\ 
        Multi-bank & 0.3 & 1.4\\ 
        Multi-rank & 0.9 & 2.8\\ \hline 
\end{tabular}
}
}
\label{table:FIT2}
\end{center}
\end{table}

The memory system has 4 channels, each containing dual-ranked DIMM of 4GB capacity (x8 devices of 2Gb each). FaultSim performs Monte-Carlo simulations over a period of 7 years and check if the system encounters an uncorrectable, mis-corrected, or silent error at any-time during the 7-year period. If so, the system is deemed as a ``failed'' system. The {\em Probability of System Failure} is computed as the fraction of systems that failed at any-time during the 7-year period. FaultSim simulates a total of 1 billion systems and report the average Probability of System-Failure as the figure of merit.

\section{XED: An Overview} 

This chapter investigates a memory system in which all DRAM chips are equipped with On-Die ECC. The key observation is that exposing the information about On-Die error detection to the memory controller can enable high-reliability memory systems at low cost. XED exposes the information that the On-Die ECC detected (or corrected) an error to the memory controller without requiring any changes to the bus interface or requiring extra bandwidth. This chapter describes how to implement XED using a conventional ECC-DIMM consisting of 9 chips.

\begin{figure}[htb]
  \centering \centerline{\psfig{file=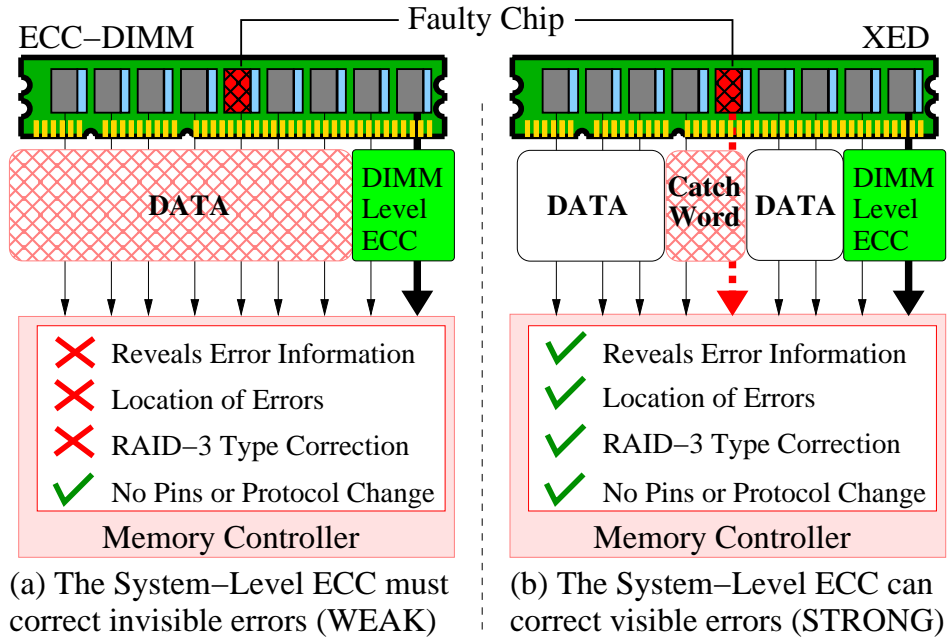,width=0.85\columnwidth}}
  \caption{(a) Conventional ECC-DIMM is not useful in presence of On-Die ECC (b) XED exposes detection information of On-Die ECC to provide stronger reliability (using RAID-3) without any interface changes.}
  \label{fig:insight}
\end{figure}

Figure~\ref{fig:insight} shows an overview of XED. Unlike conventional ECC-DIMM, which uses the 9th chip to store the ECC code, XED uses the 9th chip to store the parity information computed across the remaining eight chips. XED transfers error information by replacing data with Catch-Words. Thereafter, XED can correct data errors with the help of the error location and the parity information stored in the 9th chip (similar to RAID-3~\cite{raid3}). This enables XED to identify and reconstruct the data of a faulty chip.

This dissertation provides an interface that enables exposing the On-Die error detection information using {\em Catch-Words} that act as error indicators. XED relies on the observation that a typical memory chip (with x8 devices) provides a 64-bit data-value on each transfer. However, the chip does not store all possible $2^{64}$ values. In-fact, even a relatively large 8Gb chip stores only $2^{27}$ words of 64-bit each. Even if all the stored 64-bit words were unique, the likelihood of the chip storing the data-value that matches the randomly selected Catch-Word is negligibly small ($2^{-37}$, or 1 in 140 billion). So the appearance of Catch-Word at the memory controller signals that an episode of error detection or correction by On-Die ECC occurred within the DRAM chip. This dissertation also analyzes the effectiveness of XED in the presence of chip failures and scaling faults.

\section{Efficient Chipkill With XED}
This section describes the implementation of XED. This section also discusses how correction is performed using the position of faulty chip and the DIMM-level parity stored in the 9th chip of XED. In this section, we assume that at most one chip is faulty. The case of multiple scaling faults (Section ~\ref{sec:scalenew}) and a chip failure in the presence of scaling faults (Section ~\ref{sec:chipscalenew}) are discussed in later sections.

\subsection{Implementing XED using an ECC-DIMM}

To implement XED, each chip is equipped with two registers:  {\em XED-Enable}  and {\em Catch-Word-Register (CWR)}. To enable XED on the DIMM, the XED-Enable register is set to 1. Furthermore, the CWR is also set to a randomly selected 64-bit value by the memory controller. Fortunately, DRAM DIMMs use a separate interface to update internal parameters using {\em Mode Set Registers (MRS).} XED-Enable and CWR registers can also be configured using the MRS. As the Catch-Word is 64-bits long and XED-Enable is 1-bit long, the total storage overhead for enabling XED is only 65 bits per chip. 

\begin{figure}[htb]
  \centering \centerline{\psfig{file=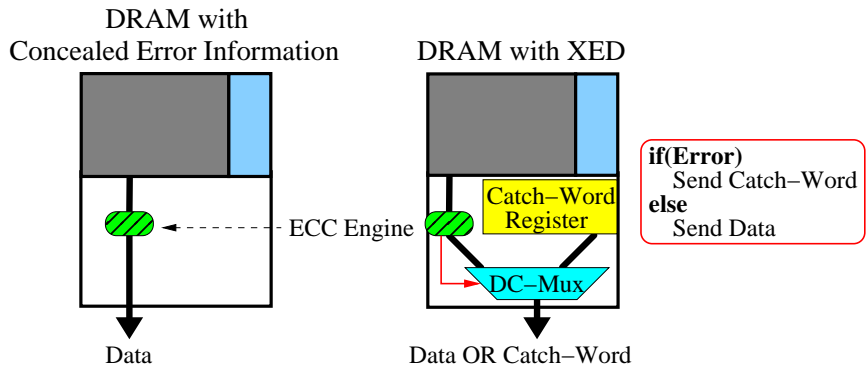,width=4.5in}}
  \caption{XED uses a multiplexer to provide the Catch-Word or the data value, depending on if the error is detected or corrected by On-Die ECC.}
  \label{fig:detcor}
\end{figure}

XED-Enable register is set at boot time and the memory controller generates a unique random Catch-Word and stores it in each chip. The memory controller also retains a copy of CWR.  This helps the memory controller in deciding if the data provided by the chip matches with the Catch-Word. To implement XED, DRAM chips are also equipped with a {\em Data-Catch-Word Multiplexer (DC-Mux)} that dynamically selects between the requested data value and Catch-Words based on the correction or detection of errors. Figure~\ref{fig:detcor} shows the internals of a DRAM chip equipped with a DC-Mux. 

If no error detected or corrected by the On-Die ECC then DC-Mux selects the data. However, if the On-Die ECC detects or corrects an error, the DC-Mux to selects the Catch-Word.  Note that this selection happens on if the XED-Enable bit is set.  If XED-Enable is not set, then the DRAM Chip supplies the data value and acts as the baseline ECC-DIMM.

\subsection{Detection: A By-Product Of On-Die ECC}
The SECDED code corrects one faulty bit and detects of two faulty bits. As SECDED code always detect the error before correcting it, we can also reuse the SECDED code to find out if an error was detected. The distance between valid code-words is called as the hamming distance and any valid data would always land on valid code-words~\cite{HAMMING:1950}. However, if the data is erroneous then it tends to land on an invalid code-word. An ECC scheme mitigates errors by selecting a unique nearest valid code-word for the detected invalid code-word. Therefore, On-Die ECC can implicitly serve as a strong detection code if it informs the memory system whenever an invalid code-word is encountered. 

\begin{figure}[htb]
  \centering \centerline{\psfig{file=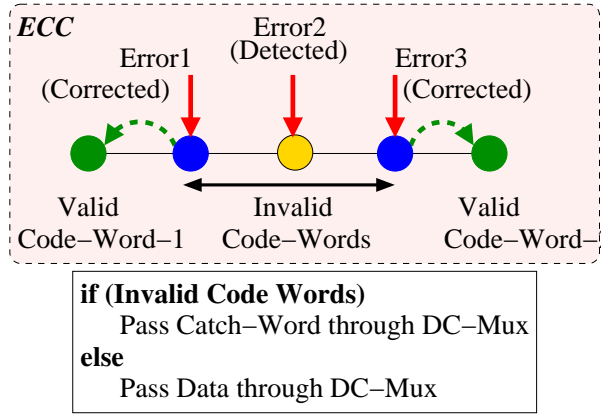,width=3.1in}}
  \caption{Leveraging ECC-based correction for stronger detection. For example, a three-bit error may get mis-corrected with conventional SECDED DIMM, but XED will be able to correct it.  }
 \label{fig:hamdet}
\end{figure}

For example, Figure~\ref{fig:hamdet} depicts a scenario where an invalid code-word is encountered by the ECC engine, so XED would use the DC-Mux to transmit a Catch-Words instead of the requested data.  Thus, the DC-Mux transmits the requested data only when the On-Die ECC engine detects a valid code-word, or when XED-Enable is set to 0.

\subsection{Mitigate a Chip Failure Using XED}
XED uses of Catch-Words to identify the faulty chip and the DIMM-level ECC to correct erroneous data of the faulty chip. This subsection describes how error correction is performed by XED.

\subsubsection{Using Catch-Words and Parity To Locate Errors}
The ninth chip in a XED is provisioned to store ``Parity'' of the data words in a burst.
A parity code enables the memory controller to identify any single erroneous data word.
For example, if data words D0 to D7 form a data burst, then Parity is computed as an XOR ($\oplus$) of all words between D0 to D7, as shown in Equation~\eqref{eqn_pa}.
\begin{multline} \label{eqn_pa}
Parity =D0 \oplus D1 \oplus D2 \oplus D3 \oplus D4 \cdots \oplus D7
\end{multline}

Therefore, in case no data is errorneous, the XOR ($\oplus$) of all words between D0 to D7 and Parity will yield ``0'' as shown in Equation~\eqref{eqn_pa2}.
\begin{multline} \label{eqn_pa2}
Parity \oplus D0 \oplus D1 \oplus D2 \oplus D3 \oplus D4 \cdots \oplus D7 = 0
\end{multline}

\begin{figure}[htb]
  \centering \centerline{\psfig{file=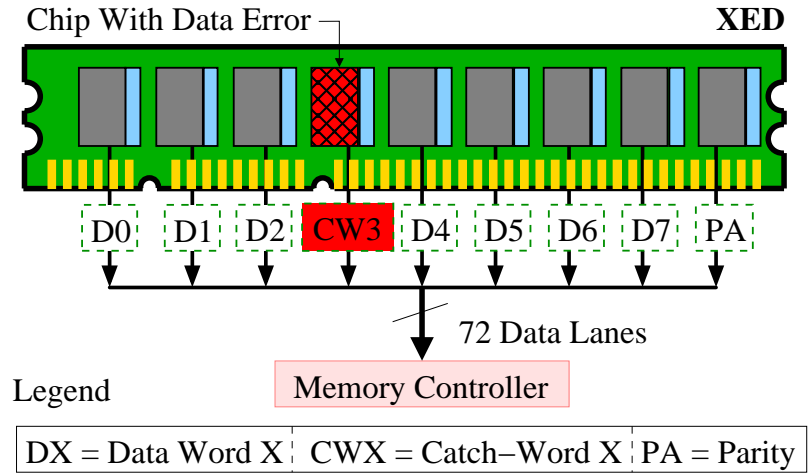,width=4.2in}}
  \caption{Catch-Words are transmitted instead of Data if On-Die error code detects errors. 
  When used with Parity, Catch-Words enable the memory system to identify the faulty chip.}
  \label{fig:catchwordtx}
\end{figure}

During a write, the parity is stored in the 9th DRAM-chip. On a subsequent read, if any data word or the Parity gets corrupted, then Equation~\eqref{eqn_pa} will not be satisfied. Consequently, memory system detects a data error. The key drawback of this technique is that, using Parity alone, a memory system cannot identify which data was erroneous. To identify the faulty chips, we use the On-Die error code that is provisioned to act as a strong error detection code within each chip.
On detecting an error, the chip relays the Catch-Word rather than transmitting the erroneous data.
As the memory controller can identify the Catch-Word, it can detect the faulty chip.  For example, Figure~\ref{fig:catchwordtx} shows a faulty chip that sends a Catch-Word (CW3) instead of Data (D3).

In this case, using Equation~\eqref{eqn_pa2}, we get Equation~\eqref{eqn_pacw} which represents the case of a single erroneous chip.
\begin{multline} \label{eqn_pacw}
Parity \oplus D0 \oplus D1 \oplus D2 \oplus \textbf{CW3} \oplus D4 \cdots \oplus D7 \neq 0
\end{multline}
Using the Catch-Words as an identifier and Equation~\eqref{eqn_pacw} to detect errors, the memory system can identify that Chip-3 is the faulty chip. As catch-words are transmitted instead of valid data, there is no change in the memory protocol. Therefore, XED is  compatible with existing memory interfaces and can relay the error information without any changes in the memory protocols.

\subsubsection{Using Parity to Correct Errors}
On detecting only a single Catch-Word, the memory controller can correct the erroneous data by using Parity. 
For example, a corrupted data-word D3 can be recovered using Parity as shown in Equation~\eqref{eqn_rec}.
Using Parity and other valid data-words, the memory controller reconstructs the corrupted data-word that is pointed by the Catch-Word.

\begin{multline} \label{eqn_rec}
Parity \neq D0 \oplus D1 \oplus D2 \oplus CW3 \oplus D4 \cdots \oplus D7 \cdots [\text{from \eqref{eqn_pacw}}]
\end{multline}

Solving for D3 instead of CW3, we get Equation~\eqref{eqn_rec2}
\begin{multline} \label{eqn_rec2}
D3 = D0 \oplus D1 \oplus D2 \oplus Parity \oplus D4 \cdots \oplus D7
\end{multline}

Therefore,  XED-based systems can achieve Chipkill-level reliability by activating only one rank of 9 chips. Thus, XED enables computer systems to obtain Chipkill-level reliability by using commodity x8 DRAM-chips.

\subsection{Collisions of Catch-Words with Data}
It is possible that a legitimate data-word matches a Catch-Word. Such incidents are referred to as collisions of Catch-Words with data-words. Note that occurrence of a collision does not indicate loss of reliability with XED. If collision happens, XED will ignore the data value from the given chip assuming it as a Catch-Word, and recreate the same value using the parity information stored in the ninth chip. So, even in the rare case of a collision, XED still provides the correct value, albeit with unnecessary correction.

\subsubsection{Identifying a Collision}
A collision can easily be identified if a Catch-Word is encountered, and the value corrected from XED (using the parity stored in the 9th chip)  matches with the Catch-Word.

\subsubsection{Chances of Collision}
\begin{figure}[htb]
  \centering \centerline{\psfig{file=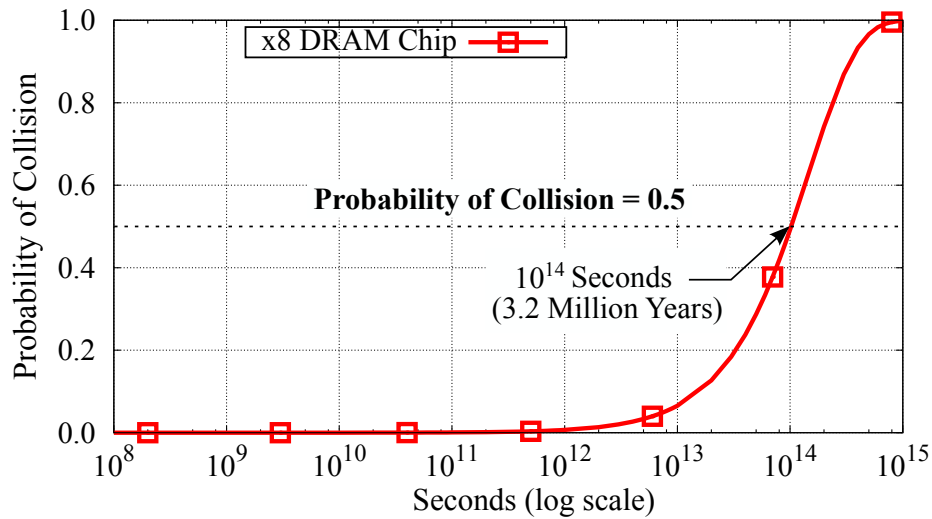,width=4.8in}}
  \caption{XED with x8 chips is likely to encounter collisions once every 3.2 million years, on average.}
  \label{fig:collision}
\end{figure}

This section quantitatively identifies the chance of a collision for a XED-based DRAM chip. If one conservatively assumes that a different data-word is written in every transaction, then one can measure the probability of collision of Catch-Words for each DRAM chip. Figure~\ref{fig:collision} depicts the probability of collision over time. As the system uses x8 DRAM-chips and a randomly selected 64-bit Catch-Word, the probability that a given data value being written to the DRAM chip matches with the Catch-Word is 1 out of 2$^{64}$, an extremely unlikely event. On average, an x8 DRAM-chip will have a collision once every 3.2 million years, assuming a memory write every 4ns.

\subsubsection{Updating Catch-Words on Detecting a Collision}\label{sec:collisionx8}
When a collision with Catch-Word is detected, this dissertation recommends that the memory controller regenerate a new Catch-Word and update all the DRAM chips with new Catch-Words. Doing so, would increase the average time between collisions. For updating the Catch-Word, the memory controller does not have to read the entire data from the chip or update all ECC values within each chip. This is because, randomly generating a Catch-Word will reduce the average chances of collision to be every 3.2 million years irrespective of the data value within each chip.

\subsection{The Need for Strong On-Die Error Detection}
This chapter assumes 8-bits of On-Die ECC for every 64-bits of data, with an aim of implementing SECDED on 64-bit granularity~\cite{uksong}. If there is freedom in choosing the code for On-Die ECC, this chapter explores codes that not only guarantee single-bit correction but also are highly effective at multi-bit detection.  While Hamming-Code~\cite{HAMMING:1950} is popular for implementing SECDED in memory systems, this dissertation recommends that the On-Die ECC use CRC8-ATM code~\cite{crc8atm,crc8itu} for implementing SECDED. CRC8-ATM has previously been used in computer networks~\cite{crc8atm,crc8itu}. Both Hamming-Code and CRC8-ATM provide the functionality of SECDED, however CRC8-ATM code has stronger error detection capabilities.  Table~\ref{table:detcode} shows the invalid-code detection capability of the Hamming Code and the CRC8-ATM code under both random errors as well as burst errors.

\begin {table}[htb]
\begin{center}
\caption{Detection-Rate of Random and Burst Errors with Single-Bit ECC}{
\vspace{0.1in}
\resizebox{0.70\columnwidth}{!}{
\begin{tabular}{|c||c|c||c|c|} \hline &\multicolumn{2}{c||}{(72,64) Hamming Code} & \multicolumn{2}{c|}{(72,64) CRC8-ATM Code}\\
       \cline{2-5}Errors & Random & Burst & Random & Burst\\ \hline \hline
        1 & 100\% & 100\% & 100\% & 100\%\\
        2 & 100\% & 100\% & 100\% & 100\% \\ 
        3 & 100\% & 100\% & 100\% & 100\% \\ 
        4 & 98.3\% & \cellcolor{red}50.73\% & 99.2\% & 100\%\\ 
        5 & 100\% & 100\% & 100\% & 100\% \\ 
        6 & 99.1\% & 100\% & 99.22\% & 100\%\\
		7 & 100\% & 100\% & 100\% & 100\%\\ 
        8 & 99.16\% & \cellcolor{red}50.75\%	 & 99.22\% & 100\%\\ \hline 
\end{tabular}
}
}
\label{table:detcode}
\end{center}
\vspace{-0.15in}
\end{table}

Hamming Code has as low as 50.7\% detection-rate for invalid code-words in the presence of burst errors. On the other hand, a CRC8-ATM code has 100\% detection-rate of invalid code-words in the presence of burst errors. Therefore, the CRC8-ATM code is more effective than Hamming Code for detecting burst errors. Therefore, this dissertation recommends using CRC8-ATM code as a design choice for the On-Die ECC.

The SECDED code should incur low latency for encoding and decoding. Fortunately, CRC8-ATMs implementations can be performed within one cycle by using only 256 entry lookup tables~\cite{crc8atm,crc8itu}. The CRC8-ATM computation consumes only a single cycle latency as it only uses a tree of XOR-gates for encoding-decoding. The current the On-Die ECC specifications do not provision any additional latency for encoding or decoding and leverage the timing slacks within DRAM chips for ECC computation (1 to 2 cycles).

\section{Mitigating Chip Failures When On-Die ECC Fails to Detect an Error}\label{sec:chipscale}
XED relies on On-Die ECC to detect faults within the DRAM chips.  Unfortunately, this detection is imperfect, and there is a small (0.8\%) likelihood that a multi-bit error within the chip remains undetected.  If the system encounters a multi-bit failure in a chip, and the On-Die ECC fails to detect this fault, XED will still be able to detect this fault at the system level because of the parity mismatch at the DIMM-level ECC. Such a scenario is deemed be an uncorrectable error, and the system is informed that an uncorrectable error has occurred. Unfortunately, the resilience of such a design would be much worse than Chipkill, as we are unable to correct the faulty chip. However, if one could identify the faulty chip, then one can use system-level parity to reconstruct the data of the faulty chip. This section describes two schemes to identify the faulty chip when the On-Die ECC fails to detect an error.

\subsection{Inter-Line Fault Diagnosis}

A multi-bit failure can occur at runtime due to large-granularity faults, such a row-failure, column-failure or bank-failure. Such error modes cause not only the requested line to fail, but also the spatially close lines to fail. This dissertation uses the insight that even if the error in a single cacheline goes undetected by On-Die ECC, it is highly unlikely that errors in the neighboring faulty lines will also go undetected by On-Die ECC. Therefore, if one reads multiple neighboring lines, then we are likely to notice errors in the  neighboring lines for the faulty chip. The chip with the highest number of faults in the neighboring lines is deemed as the faulty chip.  This dissertation proposes to stream out the entire row buffer (128 lines), and use a threshold of 10\% faulty lines to identify the faulty chip.  The analysis in Section VIII shows that using this threshold is sufficient to avoid identifying chips as faulty simply due to scaling faults. This scheme is termed as {\em Inter-Line Fault Diagnosis}.

Performing Inter-Line Fault Diagnosis incurs high latency (128 reads), so one would want to avoid performing this diagnosis frequently. This dissertation proposes to store the result of this diagnosis in a hardware structure called the {\em Faulty-Row Chip Tracker (FCT)} that tracks the location of the faulty row and the corresponding faulty chip identified using Inter-Line Fault Diagnosis. An FCT-entry is a tuple of the row-address (32-bits) and the faulty chip (4 bits). The design uses a small FCT with few entries (4-8) as the system is either likely to encounter 1 or 2 faulty rows (due to a row failure) or thousands of faulty rows (due to column failure or bank failure).  If only a single row-failure occurs, only one FCT entry is updated and the chip is not marked as faulty. However, for column or bank failure, all FCT entries would get used and point to the same chip.  This chip is permanently marked as faulty, and for all subsequent accesses to this chip, XED would reconstruct the data for this chip using parity information.  

\subsection{Intra-Line Fault Diagnosis}

While Inter-Line Fault Diagnosis is effective at detecting errors that span across multiple lines, it is ineffective when the multi-bit error is constrained to be within the given line. In such scenarios, the neighboring lines will be error free and the Inter-Line Fault Diagnosis will be unable to identify the faulty chips. When this occurs, we perform an {\em Intra-Line Fault Diagnosis} that tries to detect permanent errors in the requested line. To accomplish this, XED first copies the data of the requested line in a buffer.  A diagnosis is then performed by writing sequences of `\texttt{all-zeros}' and `\texttt{all-ones}' into the requested memory line and reading the value. The chip with the permanent word faults or bit faults will get detected by this diagnosis. If the fault occurred in only one chip, then the data for the chip can be recovered using parity information.


Note that Intra-Line Fault Diagnosis will be unable to detect word failures that are transient. Fortunately, the rate of a transient word fault is relatively small (7.7$\times$10$^{-4}$ over a period of 7 years) and the likelihood that the On-Die ECC will be unable to detect it is also quite small (0.8\%), so these cases happen with a negligibly low rate (6.1$\times$10$^{-6}$, two orders of magnitude smaller than a multi-chip failure).\footnote{There is a small probability that two words within a line will each have 1-bit scaling fault. If a single-bit runtime fault occurs in either of these two words, it would result in an detectable uncorrectable error (DUE). Fortunately, the rate of this event is negligibly small (10$^{-15}$ over 7 years).}

\subsection{Results: Effectiveness of XED}
\begin{figure}[htb]
  \centering \centerline{\psfig{file=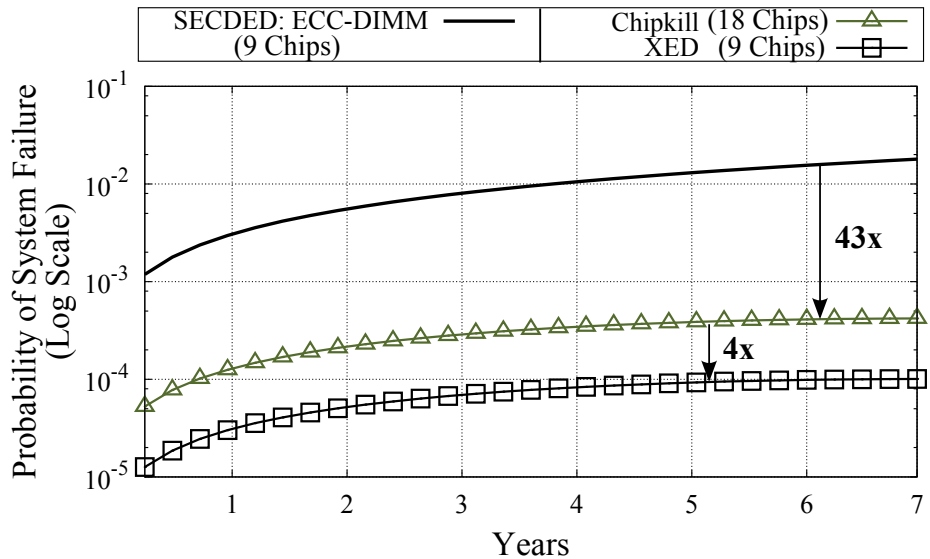,width=4.8in}}
  \caption{Reliability of ECC-DIMM, XED, and Chipkill. XED is 172x more reliable than ECC-DIMM and 4x more reliable than Chipkill.} 
  \label{fig:xk_ns}
\end{figure}

Reliability evaluations employ a system that employs DRAM chips with On-Die ECC. Figure~\ref{fig:xk_ns} shows, that XEDs provide 172x more reliability than Ordinary DIMMs. XEDs are also more 4x more resilient than any ECC-DIMM based Chipkill. This is because, Chipkill operates over 18-DRAM chips, whereas  XEDs operate over only 9-DRAM chips. A larger number of chips reduces the mean time to failure (MTTF) for a system.

Note (in Figure 4.1) that if error detection information of On-Die ECC is not exposed to the external system, the having the 9th chip in the  ECC-DIMM does not provide any added reliability benefits. This is because, once the chips can tolerate single bit failures, the dominant source of failure is due to large-granularity failures such a row or column or bank failures. Simply using a 9th chip to store SECDED is ineffective at mitigating such large-granularity faults.

\section{XED for Mitigating Scaling Errors}

The On-Die ECC is meant to protect the DRAM chip against scaling faults. While the DRAM manufactures will ensure that there are no two faulty bits are placed within the same 64-bit word of the given chip, it is possible that two separate chips can each encounter 1 faulty bit while providing data for a single 64 byte access. Ideally, XED should correct all of these scaling faults when there are no runtime errors.  This section analyzes the effectiveness of XED at  mitigating scaling faults for both when they occur without runtime faults and in the presence of runtime faults.

\subsection{Chance of Receiving Multiple Catch-Words}
\label{sec:scale}

It is possible that two or more DRAM chips can detect scaling errors simultaneously and relay Catch-Words. As scaling errors are single-bit failures, they will always be detected by the On-Die Error Code. Fortunately, the chances of two Catch-Words for any memory transaction are extremely low. Table~\ref{table:DODC} shows that even at an error rate of 10$^{-4}$, there is only 2$\times$10$^{-5}$ chance of getting multiple Catch-Words in a given access. On receiving  Catch-Words from multiple chips, XED is able to correct the data for all these chips, as longs as the errors are only due to scaling faults.

\begin {table}[htp]
\begin{center} 
\caption{Likelihood of Directed On-Die Correction with XED}{
\vspace{0.1in}
\resizebox{0.75\columnwidth}{!}{
\begin{tabular}{| c | c | c|}
\hline
Scaling-Fault Rate		&		 Chance of Receiving Multiple Catch-Words\\ \hline 
10$^{-4}$         			&		 	2$\times$10$^{-5}$\\
10$^{-5}$					&			2$\times$10$^{-7}$\\
10$^{-6}$					&			2$\times$10$^{-9}$\\ \hline
\end{tabular}}
}
\label{table:DODC}
\end{center}
\end{table}

\subsection{Correcting Scaling Errors in Multiple Chips}\label{sec:scalenew}
To correct scaling-faults, XED relies on the error correction capability of On-Die ECC, which is guaranteed to correct the single bit error.  On receiving a line with multiple Catch-Words, the memory controller enters a serial mode, where it allows only one request to go through the DIMM.  The memory controller resets the XED-Enable bit, reads the data from the given location (as XED-Enable is not set, the DIMM will send the corrected values), and then set the XED-Enable bit. It will then use the parity information in the 9th chip to ensure that the data read from this operation matches with the parity. Note that correcting scaling errors requires multiple read and write operations. Fortunately, this overhead is incurred infrequently -- once every 200K accesses even for a high error rate of 10$^{-4}$.

\subsection{Correcting Runtime Failures along-with Scaling Errors}\label{sec:chipscalenew}
A runtime failure in one chip can occur concurrently with scaling-related faults in other chips to generate multiple Catch-Words.  This can be detected as the system-level, as the parity of the 9th chip will cause a mismatch. In this case, the memory controller needs to identify the chip with the large granularity fault and use the parity to recover correct data for the chip failure. To achieve this, the memory controller instructs the On-Die ECC to correct these errors and performs Inter-Line and Intra-Line diagnosis on the faulty chip. A scaling fault is corrected by On-Die ECC and the chip failure is identified by using Inter-Line Fault Diagnosis  and Intra-Line Fault Diagnosis. If the diagnosis is successful at identifying a faulty chip, the memory controller can recover the data of the faulty chip using parity information.  However, if the diagnosis cannot determine a faulty, then XED signals an episode of Detected Uncorrectable Error (DUE) so that the system can restart or to restore an earlier checkpoint.

\subsection{Results: XED for Runtime Errors and Scaling Errors}

\begin{figure}[htb]
  \vspace{-0.3in} 
  \centering \centerline{\psfig{file=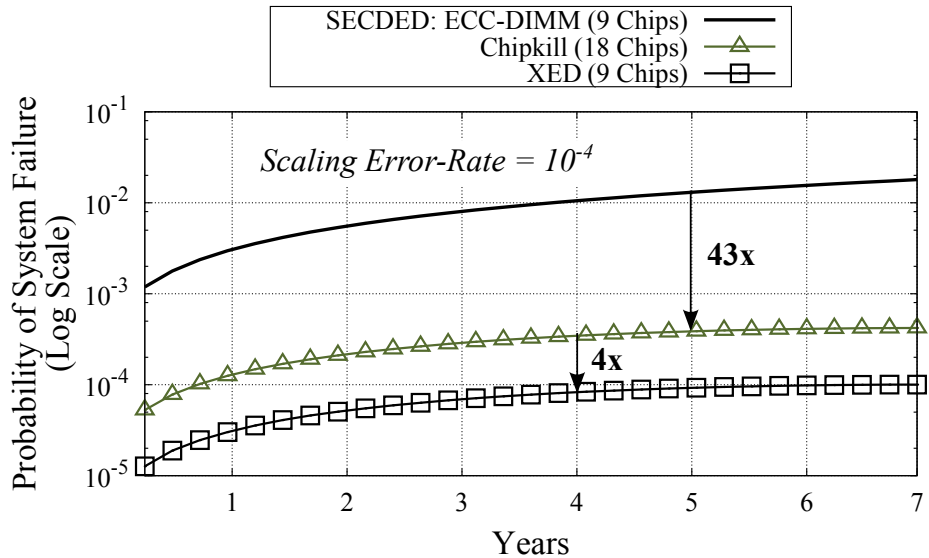,width=4.8in}}
  \caption{Reliability of ECC-DIMM, XED and Chipkill for runtime faults occurring in the presence of scaling-faults (10$^{-4}$). }
  \vspace{-0.2in} 
  \label{fig:xk_s4}
\end{figure}

Figure~\ref{fig:xk_s4} shows the effectiveness of XED, ECC-DIMM, and Chipkill in the presence of scaling errors.  This study assumes the rate of scaling errors to be 10$^{-4}$. This dissertation observes that, even in the presence of scaling errors, XED continues to provide stronger reliability than even Chipkill. Chipkill provides 43x stronger reliability than ECC-DIMM, whereas XED provides 172x stronger reliability than ECC-DIMM. This is because, On-Die ECC enables the memory system to correct scaling-faults in addition to runtime-faults.

\section{SDC and DUE Rate of XED}

XED is guaranteed to correct scaling errors in any  number of chips. However, for a chip failure, there is a small likelihood that the error may go unnoticed, resulting in a mis-correction, or cause a detectable error which cannot be corrected.  This section quantifies the vulnerability of XED using two metrics: {\em Detected Uncorrectable Error (DUE)} and {\em Silent Data Corruption (SDC)}. DUE indicates the scenario when the system encounters an uncorrectable error, whereas SDC captures the scenarios where the error remains undetected or gets mis-corrected.

\textbf{DUE}: The dominant cause of DUE are transient word-faults. When this occurs, XED first performs Inter-Line Fault Diagnosis followed by Intra-Line Fault Diagnosis, both of which fail to identify the faulty chip. In this case, even though XED detects the error due to parity mismatch of the DIMM-level parity, XED is unable to perform correction and reports an uncorrectable error. Fortunately, the rate of encountering a transient word-fault during a 7 year period is only 7.7$\times$10$^{-4}$. Furthermore, the likelihood that this fault is undetected by On-Die ECC is only 0.8\%. Therefore, the rate that XED reports an uncorrectable error due to transient word-fault, over a period of 7 years, is 6.1$\times$10$^{-6}$.

\textbf{SDC}: The dominant cause of SDC is an incorrect identification of a faulty chip by Inter-Line Fault Diagnosis. This diagnosis relies on a faulty chip encountering a large number of errors and the other chips not encountering as many errors. We use a threshold of 10\% faulty-lines within a row to identify the faulty chip. Under high rate of scaling-related faults, there is a small probability that 10\% of the lines in the row will have scaling errors. This may cause the diagnosis to deem the incorrect chip as faulty. Fortunately, even at a high error rate of scaling related fault, the chance that 10\% of the lines in a row will have errors is negligibly small (10$^{-12}$ under scaling-related fault rate of $10^{-4}$). 

Table~\ref{tab:sdcdue} shows the DUE and SDC rate for XED, assuming runtime failures are constrained to be within one chip. The SDC rate is 1.4$\times$10$^{-13}$ and the DUE rate is 6.1$\times$10$^{-6}$.  Note that the DUE rate is two orders of magnitude smaller than the likelihood of data loss due to multi-chip failure. Given that our solution is not designed to tolerate multi-chip failures, such failures will determine the overall reliability of the system, rather than the SDC and DUE rates of XED.

\begin {table}[htp]
\begin{center} 
\caption{SDC and DUE Rate of XED}{
\vspace{0.1in}
\resizebox{0.7\columnwidth}{!}{
\begin{tabular}{| c | c |}
\hline
Source of Vulnerability		& Rate over 7 years\\ \hline
XED: Scaling-Related Faults & \textbf{No SDC or DUE}\\ \hline
XED: Row/ Column/ Bank Failure & 1.4$\times$10$^{-13}$ (\textbf{SDC})\\ \hline
XED: Word Failure& 6.1$\times$10$^{-6}$  (\textbf{DUE})\\ \hline \hline
Data Loss from Multi-Chip Failures& 5.8$\times$10$^{-4}$\\ \hline
\end{tabular}
}
}
\label{tab:sdcdue}
\end{center}
\vspace{-0.3in}
\end{table} 

\section{Double-Chipkill With XED}

Memory systems that seek stronger reliability than Chipkill implement {\em Double-Chipkill} to correct up-to two faulty chips. Double-Chipkill requires four extra symbols, two each for identifying the faulty chips and for correcting the data of these faulty chips. Therefore, it is typically implemented with 36 chips, whereby 32 chips store the data and 4 chips store the check symbols. Unfortunately, accessing 36 chips requires activation of upto two ranks over non-commodity DIMMs consisting of x4 DRAM-chips. Thus, even with x4 devices, Double-Chipkill requires overfetch of 100\%. It would be desirable to obtain Double-Chipkill level reliability on a single cache line, without activating multiple ranks or channels. This section shows how XED can be applied to conventional Chipkill designs (with x4 devices) to obtain the reliability similar to Double-Chipkill. For this section only, it is assumed all systems are designed with x4 devices.

\subsection{Use Erasure Coding For Error Correction}

When XED is implemented on the top of conventional Chipkill design, one would require to have two extra chips (16 data chips plus two extra symbol chips). Given that XED can provide the location of the faulty chips, one can perform \textit{erasure} based error correction using the two symbol chips to correct upto two chip failures.  As this implementation uses 18 chips of x4 devices, each access obtains only a single cacheline, and avoids the power and performance overheads of Double-Chipkill. We note that, with x4 devices, the Catch-Word is only 32-bits, so the expected time to collision is approximately 6.6 hours (fortunately, the latency to update the Catch-Word is only a few hundred nanoseconds).  

\subsection{Results: Double-Chipkill with XED}
\begin{figure}[htb]
  \centering \centerline{\psfig{file=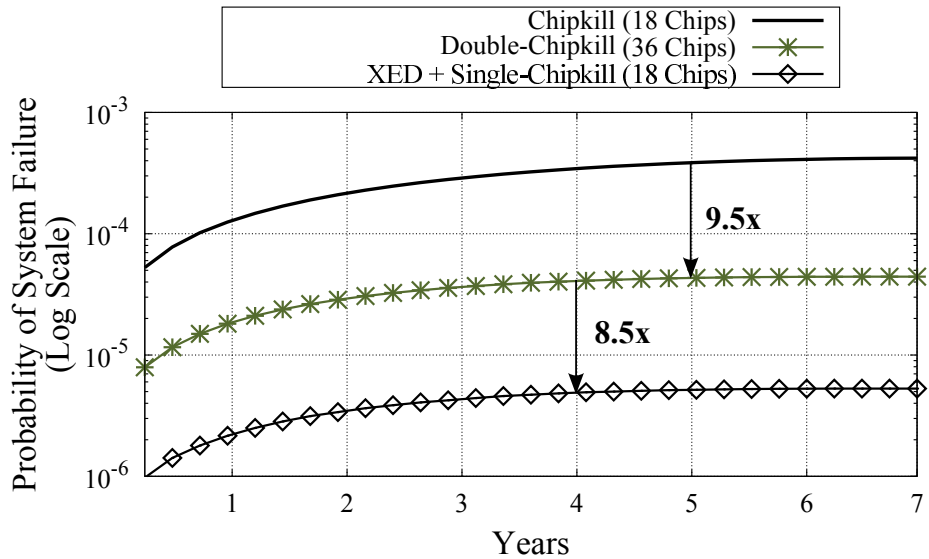,width=4.8in}}
  \caption{Reliability of Single-Chipkill, Double-Chipkill, and XED-based Single-Chipkill. Even with hardware similar to Single-Chipkill, XED provides 8.5x more reliability than Double-Chipkill.} 
  \label{fig:xdk_ns}
\end{figure}

Figure~\ref{fig:xdk_ns} compares the reliability of Double-Chipkill, Single-Chipkill, and XED implemented with Single-Chipkill systems, all evaluated in the absence of scaling errors.  Overall, Double-Chipkill provides almost an order of magnitude improvement over Single-Chipkill.  Unfortunately, it incurs significant power and performance overheads compared with Single-Chipkill. XED allows the memory system  to get Double-Chipkill level reliability while retaining the hardware of Single-Chipkill. In fact, given that XED on the top of Chipkill has only 18 chips instead of the 36 chips for Double-Chipkill, it is observed that XED provides almost 8.5x higher reliability than Double-Chipkill while obviating the performance and power overheads of Double-Chipkill.

\begin{figure}[htb]
  \centering \centerline{\psfig{file=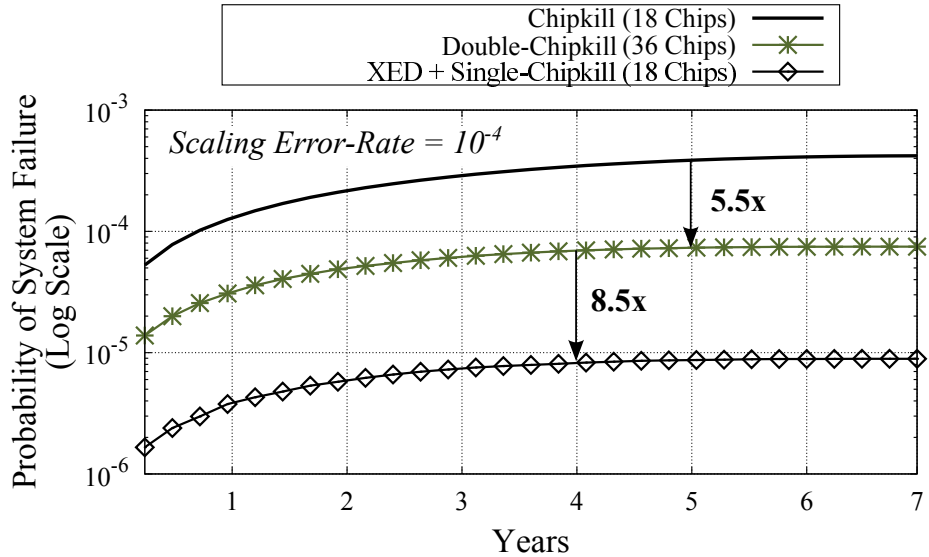,width=4.8in}}
  \caption{Reliability of Single-Chipkill, Double-Chipkill, and XED-based Single-Chipkill in the \textbf{presence of scaling faults}. XED on Single-Chipkill  provides 8.5x more reliability than Double-Chipkill.} 
  \label{fig:xdk_s4}
\end{figure}

Figure~\ref{fig:xdk_s4} compares the reliability of Double-Chipkill, Single-Chipkill, and XED on top of Single-Chipkill in the presence of scaling errors. This section assumse the rate of scaling errors to be  10$^{-4}$. Note that, in the presence of scaling errors, Double-Chipkill is 5.5x more effective than Single-Chipkill.  XED implemented with Single-Chipkill continues to provide 8.5x better reliability than Double-Chipkill, primarily due to fewer chips.

\section{Experimental Methodology}
\label{sec:methodology2}
To evaluate memory power and performance impact, this chapter uses USIMM, a cycle accurate memory system simulator~\cite{usimm,msc}. USIMM enforces strict timing and also models all JEDEC DDR3 protocol specifications. USIMM is configured with the power parameters from industrial 2Gb x8-DRAM chips and x4-DRAM chips~\cite{ddr3_power}. As On-Die ECC needs 12.5\% more DRAM cells per die, the background current and the current for refreshes, activation and precharge are increased by 12.5\%. Since error detections require only a syndrome check, it is assumed to consume 1 core cycle. The error correction at the memory controller is assumed to consume 4 core cycles. For erasure codes, the error correction is conservatively assumed to incur 60 core cycles. Table~\ref{table:system_config2} shows the parameters for the baseline system.

\begin {table}[ht]
\begin{center} 
\caption{Baseline System Configuration for XED}{
\vspace{0.1in}
\resizebox{0.6\columnwidth}{!}{
\begin{tabular}{|c|c|}
\hline
                 Number of cores              & 8        \\ \hline
                 Processor clock speed        & 3.2GHz    \\
                 Processor ROB size           & 160       \\
                 Processor retire width       & 4         \\
                 Processor fetch width        & 4         \\ \hline
                 Last Level Cache (Shared)   & 8MB, 16-Way, 64B lines \\ \hline
                 Memory bus speed             & 800MHz    \\
                 DDR3 Memory channels         & 4         \\
                 Ranks per channel            & 2         \\
                 Banks per rank               & 8         \\
                 Rows per bank                & 32K \\ 
                 Columns (cache lines) per row & 128       \\ \hline

\end{tabular}}
}
\label{table:system_config2}
\end{center}
\vspace{-0.2in}
\end{table}

The evaluations use benchmarks which have greater than ``1 Miss Per 1000 Instructions'' from Last Level Cache, from the SPECCPU 2006~\cite{spec}, PARSEC \cite{PARSEC} and BioBench~\cite{BIOBENCH} suites. We also include five commercial applications~\cite{msc}. For simulations, a representative slice of 1 billion instructions using Pinpoints is generated. The evaluations execute the benchmark in rate mode and all cores execute the same benchmark. This study performs timing simulation until all the benchmarks in the workload finish execution, and measures the average execution time of all cores.

\section{Results}
\subsection{Impact on Performance}
\begin{figure*}[htp]
  \centering \centerline{\psfig{file=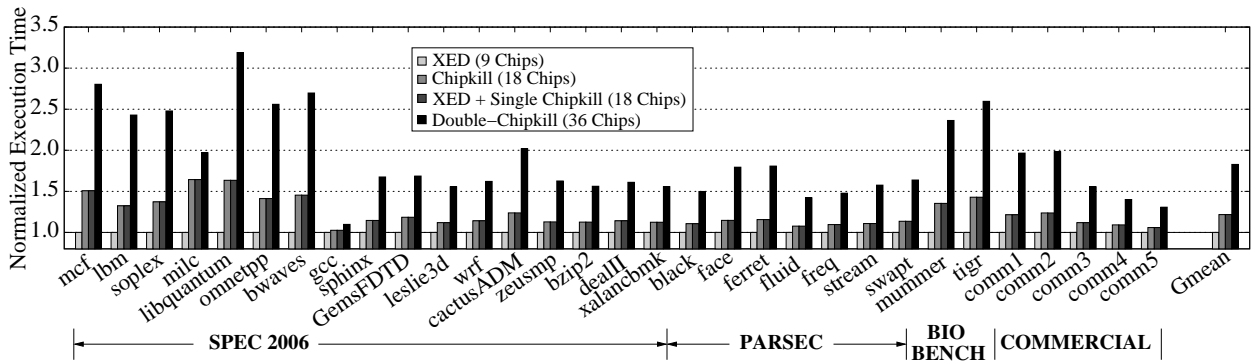,width=6.5in}}
  \caption{Normalized Execution Time (with respect to ECC-DIMM) for XED, Chipkill, XED on the top of Chipkill and Double-Chipkill. XED activates 2x fewer ranks and has 21\%  (61\%) lower execution time than Chipkill (Double-Chipkill).} 
  \label{fig:perf2}
\end{figure*}

Figure~\ref{fig:perf2} shows the impact on execution time for Chipkill and Double-Chipkill-level protection using ECC-DIMMs and compares them to their XED implementations. On a baseline that is normalized to a ECC-DIMM based SECDED, a conventional Chipkill reduces the rank-level parallelism by 2x (by activating two ranks) and increases execution time by 21\% on an average. Furthermore, applications that are bandwidth bound ( eg. \texttt{libquantum}) shows upto 63.5\% increase in execution time. Furthermore, even latency sensitive applications like \texttt{mcf} shows upto 50.7\% increase in execution time. XED activates only a single rank and consumes no performance overheads. The overheads of XED happen only on receiving multiple Catch-Words, something that happens rarely (once every 200K  accesses).

For Double-Chipkill, XED on the top of Chipkill activates 18 DRAM-chips (by activating two ranks) instead of to 36 DRAM-chips (by activating four ranks) in traditional Double-Chipkill. Consequently, by activating 18 DRAM-chips, XED based Double-Chipkill has the same overheads as traditional ECC-DIMM based Chipkill. Due to this, XED based Double-Chipkill increases the execution time by 21\% which is similar to conventional Chipkill. Unfortunately, traditional Double-Chipkill systems increase the execution time by 82\%. Furthermore, bandwidth sensitive applications such as \texttt{libquantum} increase the execution time by 220\%. Even in latency sensitive benchmarks like \texttt{mcf}, a Double-Chipkill increases the execution time by 180\%. 

\subsection{Impact on Power}
\begin{figure*}[htb]
  \centering \centerline{\psfig{file=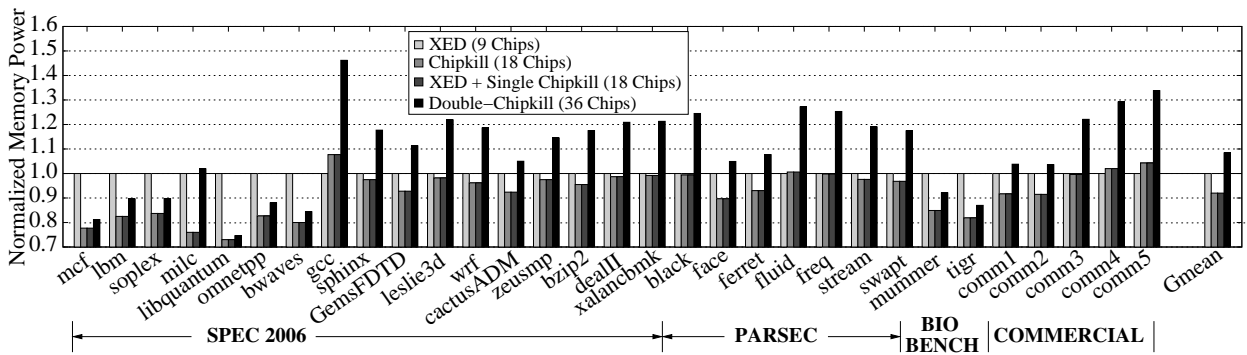,width=6.5in}}
  \caption{Normalized Memory Power (with respect to ECC-DIMM) for XED, Chipkill, XED on the top of Chipkill and Double-Chipkill.. The reduction in memory power in Chipkill is  due to the increased execution time. Double-Chipkill activates two channels and consumes significantly more power. }  
\label{fig:pow}
\end{figure*}

Figure~\ref{fig:pow} shows the impact of memory power while providing Chipkill and Double-Chipkill using ECC-DIMMs when compared to XED based systems. On a baseline that is normalized to an ECC-DIMM based SECDED, a conventional Chipkill not only activates two ranks but also increases execution time. Since power is ``energy spent over the total execution'' of the application, ECC-DIMM based Chipkill reduces the memory power consumption by 8\%. On the contrary, XED consumes the same amount of power as ECC-DIMM based SECDED implementation as it activates only a single rank. Furthermore, because it activates only a single rank, XED also takes almost the same amount of execution time as SECDED systems.

Conventional Double-Chipkill systems consume 8.4\% more memory power than ECC-DIMM based SECDED implementation. This is because, even though ECC-DIMM based Double-Chipkill systems increase execution time by 63.5\%, they also activate 36-DRAM chips (by activating four ranks). This higher execution time does not compensate for the activation overheads and increases the memory power consumption by 8.4\%. XED based Double-Chipkill reduces the memory power consumption by 8\% by activating only 18 DRAM-chips instead of 36 DRAM-chips for traditional Double-Chipkill. Furthermore, the likelihood of receiving multiple Catch-Words are rare (1 in every 200K accesses) and therefore they consume negligible power overheads.

\subsection{Impact of adding a Burst or Transaction}
\begin{figure}[htb]
  \centering \centerline{\psfig{file=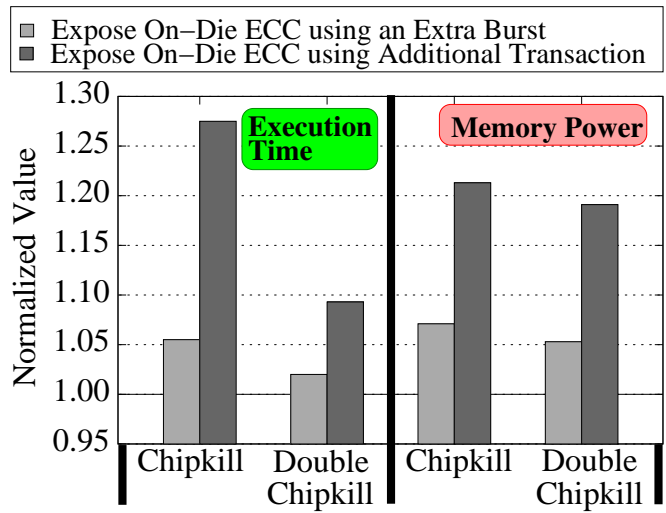,width=3.5in}}
  \caption{The performance and power overheads of exposing On-Die ECC using adding an additional two bursts or a transaction, instead of XED.}
  \label{fig:mod}
\end{figure}

XED relies on Catch-Word to convey error detection information. There are alternative ways to convey this information such as using additional bursts or transactions. The memory vendors can change the DDR protocol to expose On-Die ECC information by adding a burst. Adding another burst incurs a 25\% overhead in current memory systems as it increases the burst size from 8 to 10. Furthermore, DRAM vendors are reducing the burst-size to one or two~\cite{HBM:2013,WIDEIO:2013} which would increase this overhead to about 50\%-100\%. Alternatively, the memory controller can issue another transaction to fetch the On-Die ECC. Figure~\ref{fig:mod} shows the normalized execution time and power for these two alternatives (additional burst or additional transaction) compared to XED for both Chipkill and Double-Chipkill. Both these alternative implementations increase power consumption and execution time significantly compared to XED implementations for both Chipkill and Double-Chipkill.

The recently introduced DDR4 standards provide an ALERT\_n pin~\cite{DDR4:2015,lpddr4_ondie} to indicate errors in address, command, or write operations.  As there is only one ALERT\_n pin provisioned for the entire DIMM, the ALERT\_n signal can only convey that one of the chip is faulty, however it cannot identify the chip that encountered the fault. If future standards~\cite{alert} could extend the ALERT\_n pin to also convey the location of the faulty chip, then XED can be implemented using ALERT\_n instead of using Catch-Words.

\subsection{Comparison to Prior Proposals: LOT-ECC}
\begin{figure}[htb]
  \centering \centerline{\psfig{file=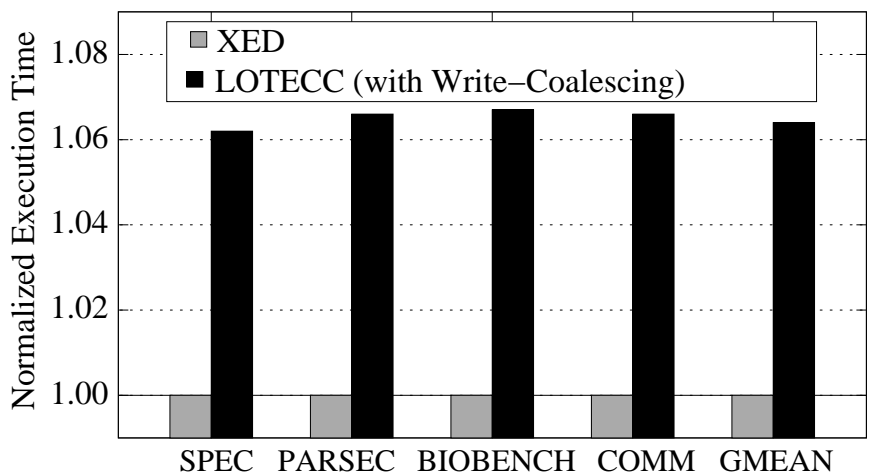,width=4.5in}}
  \caption{Execution time of LOT-ECC~\cite{LOTECC:2012} with respect to XED. LOT-ECC causes a slowdown of 6.6\%.}
  \label{fig:lotecc}
\end{figure}

A related work, LOT-ECC~\cite{LOTECC:2012}, explores a design that uses x8 chips to provide Chipkill by having tiers of error detection and correction code. This chapter compares LOT-ECC with XED. Figure~\ref{fig:lotecc} shows the execution time of LOT-ECC and XED when compared to a baseline ECC-DIMM. LOT-ECC has 6.6\% higher execution time compared to XED, as it increases the number of writes to the memory system.

\section{Summary}
As DRAM technology scales to smaller nodes, the rate of unreliable bits within the DRAM chips is increasing~\cite{hong:iedm2010,uksong}. Memory vendors are planning to provision {\em On-Die ECC} to handle the scaling-induced faulty bits~\cite{uksong,synopsys_ondie,lpddr4_ondie}. To maintain compatibility with  DDR standards, and to avoid the bandwidth overheads of transmitting the ECC code, the On-Die ECC information is not currently exposed to the memory controller and therefore, this information cannot be used to improve memory reliability. To enable low-cost higher-reliability memory systems in presence of On-Die ECC, this dissertation proposes proposes \emph{XED} (pronounced as ``zed'', the British pronunciation of the letter ``z''), a technique that e\underline{X}poses On-Die \underline{E}rror \underline{D}etection information to the memory controller while avoiding the bandwidth overheads and changes to the memory standards. The proposed implementation of {\em XED} has the following features:

\begin{enumerate}
\item XED exposes On-Die error detection information using {\em Catch-Words}, thereby avoiding any changes to the DDR protocol or incurring bandwidth overheads.  
\item XED uses the 9-th chip in the ECC-DIMM to store parity information of all the chips, and uses the error detection information from the On-Die ECC to correct the data from the faulty chip using a RAID-3 scheme.
\item XED not only tolerates chip-failure, but also mitigate scaling faults even at very high error rates (10$^{-4}$).
\end{enumerate}

XED provides Chipkill-level reliability using only a single 9-chip ECC-DIMM, and Double-Chipkill on a conventional implementation of Single-Chipkill. The reliability evaluations show that XED provides 172x higher reliability than an ECC-DIMM and reduces execution time by 21\% compared to traditional Chipkill implementations. As DRAM technology ventures into sub 20nm regime, solutions such as XED that spans across multiple sub-systems will become necessary to provide high reliability at low-cost.


\chapter{Enabling Robust and Efficient Stacked Memories}
Stacked memory modules are likely to be tightly integrated with the
processor. It is vital that these memory modules operate reliably, as
memory failure can require the replacement of the entire socket.  To
make matters worse, stacked memory designs are susceptible to newer
failure modes (for example, due to faulty through-silicon vias, or
TSVs) that can cause large portions of memory, such as a bank, to
become faulty.  To avoid data loss from large-granularity failures,
the memory system may use symbol-based codes that stripe the data for
a cache line across several banks (or channels).  Unfortunately, such
data-striping reduces memory level parallelism causing slowdown and higher power consumption.

This dissertation describes {\em Citadel}, a robust memory architecture that
allows the memory system to retain each cache line within one bank.
By retaining cache lines within banks, Citadel enables a high-performance and low-power memory system 
and also efficiently protects the stacked memory system from large-granularity failures.

\section{Introduction}
The emerging 3D stacked DRAM technology can help with the challenges
of power consumption, bandwidth demands and reduced footprint. One of
the key enablers of stacked memory is the \textit{through-silicon via}
(TSV) technology, which makes it possible to cost-effectively stack
multiple memory dies on top of each other~\cite{TSV_ISSCC:2009}. The
shorter internal data paths afforded by TSVs reduce capacitance and
active power.  By exploiting wide buses
\cite{HMC:2013} or high-frequency SerDes interfaces~\cite{HBM:2013}
and higher levels of internal parallelism, both bandwidth and
random-access latency are improved.  It is anticipated that
high-performance stacked memories often will be permanently attached
to host processors via direct stacking, silicon interposers or other
hard-wired interconnects.  In such a system, memories that develop
permanent faults must continue to work, in order to avoid
replacement of multiple chips which tends to be expensive. These factors motivate the adoption of a
{\em fail-in-place} philosophy for designing stacked memory systems.

Recent work on DRAM reliability~\cite{failures:sridharan12} showed
that large-granularity DRAM chip failures, such as bank failures,
occur nearly as frequently as single-bit failures in commodity
DIMMs. Stacked memory designs would not only be subject to these
failures but also to newer fault models, such as arising from faulty
TSVs. TSV faults can cause failures of several dies, often manifested as column failures or
bank failures. Thus, stacked memory systems will be more
vulnerable to large-granularity failures. Unfortunately, conventional
error correction schemes such as ECC DIMMs~\cite{siliconpower:eccdimm}
are targeted towards correcting random bit errors and are ineffective
at tolerating large-granularity faults. Memory systems can tolerate
large granularity failures using symbol-based coding schemes like
ChipKill~\cite{chipkill}. However, this increases the number of
activated chips and total power consumption.

To optimize performance and power for stacked memory, one would want to
retain the data for a cache line within a single bank.  However, a
bank failure would then cause loss of data for the whole cache line.
One can adopt a philosophy similar to ChipKill for tolerating
large-granularity failures for stacked DRAM. In such a design, the
data for a cache line would be striped across several banks (or
channels), and a symbol-based coding can be applied, in which the size
of each symbol would be equal to the amount of data stored in each
bank. Unfortunately, such a data mapping would require the memory
system to activate several banks to service a single request.  This
causes performance degradation (10\% to 25\%) due to loss of
bank(channel) level parallelism, and power consumption (as
high as 6x in the evaluations conducted by this dissertation) due to activation of several banks to
service one request.

\begin{figure}[htb]
  \centering \centerline{\psfig{file=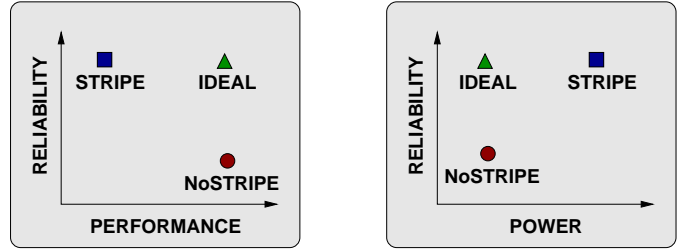,width=3.5in}}
  \caption{ Striping
  enhances reliability but sacrifices performance and power
  efficiency.  Ideally, we want to tolerate large-granularity
  failures at high performance and low power.}
  \label{fig:introfig}
\end{figure}

As shown in Figure~\ref{fig:introfig}, ideally one would want a system that
has the performance and power efficiency of storing the entire cache
line in one bank (NoStripe), and yet maintains robustness to large
granularity faults (Stripe). To that end, this dissertation proposes {\em
Citadel}, a robust memory architecture that allows the memory system
to retain each cache line within one bank (delivering high
performance and low power) and yet efficiently protects the stacked
memory from large-granularity failures.  

Like ECC DIMMs which have one additional chip per 8 chips, in our
study, Citadel has one extra die (ECC die) with smaller rows along with eight data dies. Similar
to an ECC-DIMM that provides 64 bits of ECC for every 512-bit cache line, Citadel
uses the 64 bits of metadata associated with each 512-bit cache
line. Based on key insights, Citadel employs a three-pronged approach for fault tolerance.

\subsubsection*{Insight 1- Protect Against Runtime TSV Faults}

As faulty TSVs tend to be a major cause of multi-bank failures in stacked
memories, our first idea, {\em TSV-Swap}, specifically targets TSV
faults that happen at runtime. DRAM vendors can use manufacture-level spare
TSVs~\cite{tsv_red}, to repair faulty TSVs at design
time. Unfortunately, manufacture-level sparing does not protect against runtime failures. 
Citadel proposes TSV-SWAP, a technique that does not rely on any manufacturer-provided spare
TSVs. Instead, TSV-Swap dynamically exchanges faulty TSVs with
non-faulty TSVs with a remapping circuit. This study found that while a data
TSV typically affects only one bit in a data line (albeit across many
lines), a failure of one of the address TSVs can make half of the
memory unreachable. Thus, address TSVs are much more critical than
data TSVs for system reliability. The proposal, TSV-Swap, can repair upto 8 faulty TSVs which can be
data, address or command TSVs.

\subsubsection*{Insight 2- Detect and Correct Large Granularity Failures}

Even after mitigation of TSV related faults, the stacked memory is
still vulnerable to internal DRAM die faults.  One would like to protect
stacked memory not only from small granularity failures (such as bit-fault or word-fault) 
but also from large granularity faults such as
column-fault, row-faults or even complete bank failures. The second
idea, {\em Tri Dimensional Parity (3DP)}, provides highly effective
and storage efficient correction for both small and large granularity
failures.  The 3DP proposal maintains parity in three dimensions: 1)
Across all banks and dies for individual rows. 2) Across all rows in
all banks within a die. 3) Across all rows in single bank across all
dies. Each line is equipped with CRC-32~\cite{CRC:1961} to detect data
errors. If any error is detected, it is corrected using the parity
information of 3DP. 3DP provides 130x higher resilience than just applying 2D-ECC. 3DP achieves this with only 1.6\% storage overhead, compared to the 25\% storage required for prior 2D schemes.

\subsubsection*{Insight 3- Isolate Faulty Memories with Efficient Sparing}

When a fault is detected, data is restored using the correction
capability of 3DP.  However, modules with permanent faults would incur
the correction overheads frequently.  To avoid such frequent
correction, one would like to redirect a faulty memory unit to a spare
area. Unfortunately, if the sparing granularity is too fine, then it
incurs significant tracking overheads (for example, if a bank fails
then thousands of rows get spared to the spare area).  If the sparing
granularity is too coarse then it results in significant wasted space
(for example, sparing at a bank granularity would be wasteful if only
one row is faulty).  This dissertation makes a key observation that a bank typically
has either one or two row failures, or has thousands of row failures
(due to a sub-array or bank failure). The third idea, {\em Dynamic
Dual-Grained Sparing (DDS)}, exploits the bimodal behavior of faulty
units and efficiently spares either at a row or bank granularity. The
proposed design of DDS can spare two faulty banks along with several
row failures.

This dissertation performs reliability studies using real field data and perform
sensitivity studies when field data is unavailable (e.g. for TSVs). The
evaluations, with an industry-grade fault simulator~\cite{faultsim}, show that
Citadel provides 100x-1000x higher reliability while still retaining
power and performance similar to a system that maps the entire cache
line in the same bank. Citadel achieves this using a storage
overhead similar to ECC DIMMs (14\% vs. 12.5\%).

\section{Background and Motivation}

Stacked memory systems have lower energy per bit and higher bandwidth
when compared to their 2D counterparts. However, to obtain the
power-efficiency and high bandwidth of stacked memory, the system must
first address reliability challenges. As shown in
Figure~\ref{fig:faults}, failures can occur in a memory system at
different granularities~\cite{failures:sridharan12,Schroeder:2009,Schroeder:2010,failures:sridharan13}.

\subsection{Memory Faults for Traditional Systems}

\begin{figure*}[htb]
  \centering
  \centerline{\psfig{file=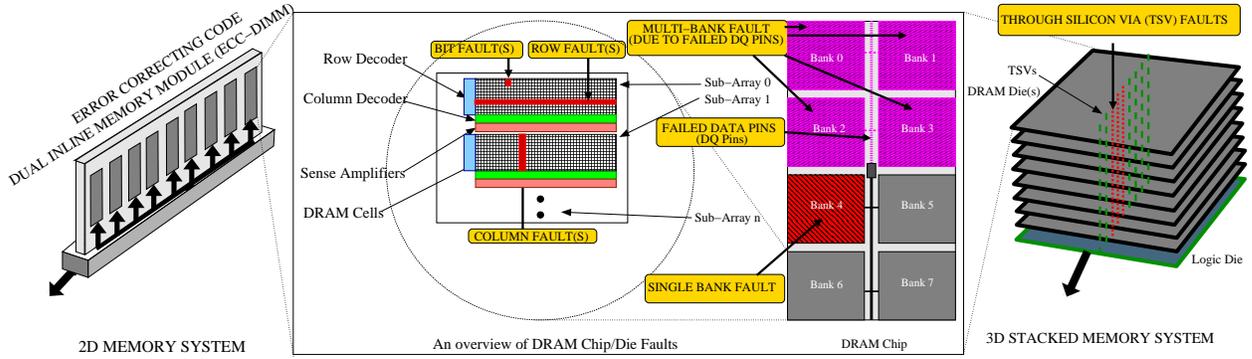,width=6.5 in}}
   \caption{Granularity of faults that occur in a DRAM
   Chip/Die. Faults can be at granularities of bit, column, row,
   bank(s), TSVs and I/O links for stacked memory systems. Common wiring
   faults within a chip can cause multiple banks to fail.}
  \label{fig:faults}
\end{figure*}

A memory DIMM consists of multiple DRAM chips. A DRAM chip is
organized into banks, where all banks share a common data bus. These
banks are composed of rows and columns and are divided into
sub-arrays. The banks contain row and column decoders that activate
the wordlines or select bitlines associated with the memory request.
Faults at the DIMM level can affect all DRAM chips within a
DIMM. However, the faults in individual chips are largely independent
of each other. In this dissertation, the definitions for the chip faults
follow that of Sridharan et. al.~\cite{failures:sridharan12} and are
represented in Figure~\ref{fig:faults}. Note that banks are
operated almost independently and share only wiring such as data,
address and command buses~\cite{DRAM_BANK:1996,DRAM_BANK_SHARING:1999}. Bank and rank faults occur mainly from faulty
data or address or command buses.

\subsection{Transposing Faults onto 3D Stacked Memories}

Layout of an individual die in 3D stacked memory systems shows that
its internal organization is very similar to that of a chip in
conventional 2D memory systems~\cite{addrdram,TOM_MICRON:2013,
TIMHOLLIS_MICRON:2012, MPR:2011}. To a first order, this dissertation
transposes failure rates for all fault types except complete bank and
complete rank for current 2D memory system onto stacked memory
systems.  The key difference is the introduction of TSVs for
connecting data and address lines~\cite{TSV_ISSCC:2009}. Due to this,
complete bank faults and complete rank faults in any 3D stacked memory are now
influenced by TSV faults.

\subsection{Stacked Memory: Organization and ECC Layout}

\begin{figure}[htb]
  \centering \centerline{\psfig{file=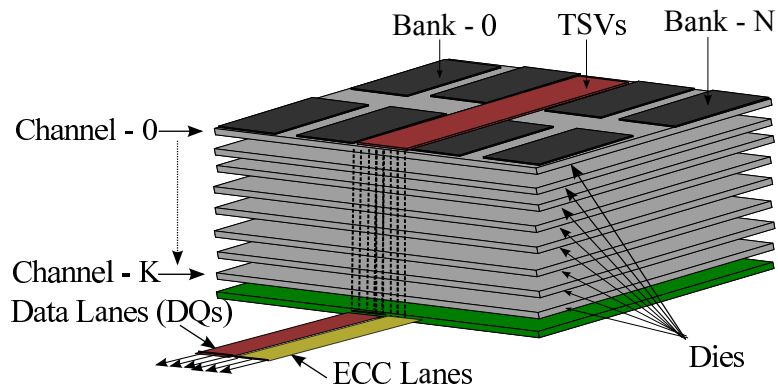,width=4in}}
  \caption{High Bandwidth Memory has a channel(s) per die and all
  banks in this channel are on the same die. HBM
  specification includes separate Data and ECC lanes} 
  \label{fig:3Dmem}
\end{figure}

There are several design prototypes of stacked memory, including the
High Bandwidth Memory (HBM)~\cite{HBM:2013}, Hybrid Memory Cube
(HMC)~\cite{HMC:2013,TOM_MICRON:2013} and Octopus from
Tezzaron~\cite{TEZ_OCTUPUS:2010}.  These standards differ in their
data organization and also share TSVs differently. However, these
stacked memory systems fundamentally have the same layout. This dissertation performs comprehensive analysis on an HBM like design. Subsequently, this dissertation also extends its analysis for HMC and Tezzaron designs.  Figure~\ref{fig:3Dmem} shows internal stack
organizations of HBM. Each channel may be fully contained in each DRAM
die in the stack.  A complete set of TSVs and buffers connect each
channel to the external interface.

The stacked memory consists of $D$ data dies and $E$ ECC dies
(depending on value of $D$ and the ECC implementation). ECC can be
stored in an additional space provided by $D$ dies or can be
distributed across $D+E$ dies. Similar to ECC-DIMMs, every data
request for a 512b data line also concurrently fetches its 64b ECC
metadata through dedicated ECC lanes~\cite{HBM:2013}. In this dissertation, an 8-die stack with one additional ECC die is used for ECC or metadata information. Such an organization has the same storage overhead as incurred in ECC
DIMMs (12.5\%).

\subsection{Data Striping in 3D Memory Systems}

The way data is striped in the memory system has a significant impact
not only on the power and performance but also the reliability of the overall
system. A conventional (2D) DIMM stripes a cache line across several
chips.  Similarly, a stacked memory system can place the cache line in
one of three ways:

\begin{itemize}
\item \textbf{Same Bank:} Within a single bank in a single channel.
\item \textbf{Across Banks:} Within a single die (channel) and striped across banks.
\item \textbf{Across Channels:} Within multiple dies (channels) and striped across one bank in each channel.
\end{itemize}

\subsection{Impact of Data Striping}

If one were to use an organization that places the entire cache line in the
same bank, then a failure of the bank would cause data loss of the
entire cache line.  To protect stacked DRAM from bank failures or
channel failures, one can stripe data across banks or channels.  In
such a case, each bank/channel would be responsible for only a portion
of the data for the cache line, and a correction mechanism (possibly
ECC scheme) can be used to fix the sub-line-granularity fault.  This
organization activates multiple banks/channels to
satisfy each memory request and reduces bank-level parallelism. Subsequently,
stacked DRAM consumes much higher power as it activates multiple banks.

\begin{figure}[htb]
\centering
   	\begin{subfigure}[b]{4in} 
 		\psfig{file=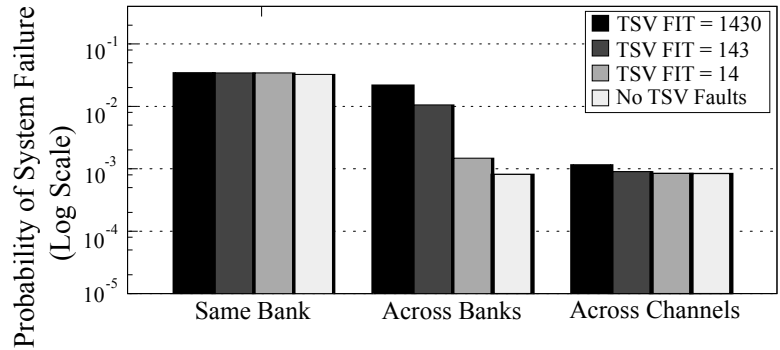,width=4in}
 		\caption{}
 		   \label{fig:Ng1} 
	\end{subfigure}
   	\begin{subfigure}[b]{5in} 
		\psfig{file=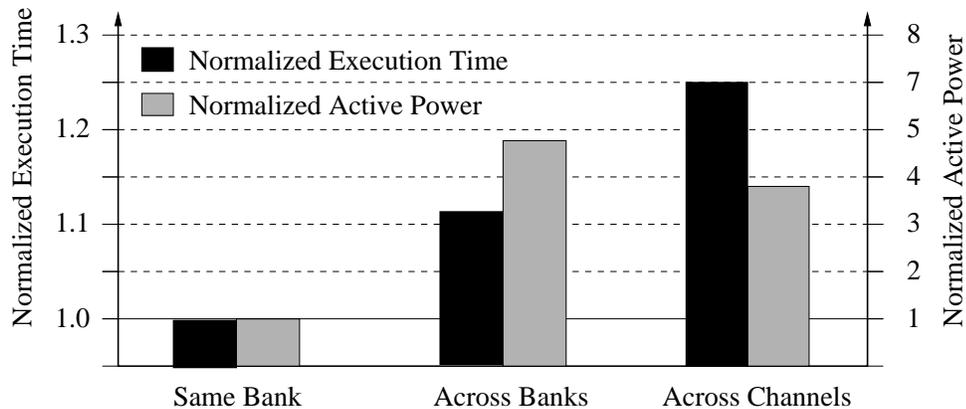,width=5in}
		   \label{fig:Ng2}
		   \caption{} 
	\end{subfigure}
	
\caption{Impact of data striping on Reliability, Power and Performance. (a) Striping data across banks or channels and using a strong 8-bit symbol based code (similar to Chipkill) gives higher reliability. (b) However, striping data across banks or channels comes at a significant price in performance (11\%-25\%) and power (3.8X-4.7X)} \label{fig:rpp}
\end{figure}

Figure~\ref{fig:rpp} compares the reliability for three data
mapping schemes for strong 8-bit symbol based ECC (similar to
ChipKill) for different TSV FIT rates (other parameters are described
in Section~\ref{sec:methodology3}).  System failure is the occurrence of
an uncorrectable fault within a seven-year
lifetime. Across-Channels configuration provides the highest
reliability.

Unfortunately, the reliability benefits of Across-Banks and
Across-Channels come at a significant price in terms of performance
and power. Figure~\ref{fig:rpp} shows that striping data Across-Banks
causes a slowdown of approximately 10\%, and Across-Channels causes a
slowdown of approximately 25\%.  Furthermore, Across-Channels and
Across-Banks consumes 3.8-4.7x more active power than the Same-Bank
mapping (Across-Channels takes longer to execute, consuming energy
over a longer time, hence the relative reduction in power compared to Across-Banks).

{\bf Goal:} A key goal of this chapter is to look at techniques that can enable performance and
power-efficient reliability by maintaining the data mapping of a Same-Bank
configuration. One of the requirements for stacked memories is protection against large
granularity faults. This chapter first describes the methodology before describing some solutions.

\section{Experimental Methodology}
\label{sec:methodology3}

\subsection{Fault Models and Failure Rates}

Real-world field data from Sridharan et
al.~\cite{failures:sridharan12} provides failure rates as
Failures In Time (FIT) for DRAM chips. As TSV failure data is not
publicly available, we perform a sensitivity study for TSV device
FITs. This study assumes 0.01 to 1 device failures in 7 years (translating to
Device FIT of 14 to 1,430) due to TSV faults.  Table~\ref{table:FIT3}
shows the failure rates per billion hours (FIT) and the failure
sensitivity for our evaluations (from~\cite{Citadel}).

\begin {table}[ht]
\begin{center}
\caption{Stacked Memory Failure Rates (8Gb Dies)}{
\begin{tabular}{|c|c|c|}
\hline
&\multicolumn{2}{c|}{Fault Rate (FIT)} \\ \cline{2-3} DRAM Die Failure
Mode & Transient & Permanent \\ \hline Single bit & 113.6 & 148.8\\
Single word & 11.2 & 2.4 \\ Single column & 2.6 & 10.5 \\ Single row &
0.8 & 32.8\\ Single bank & 6.4& 80\\ \hline
\multicolumn{3}{|c|}{TSV(Complete Bank/Channel)} \\
\hline \multicolumn{2}{|c}{TSV (Address and Data)}&
\multicolumn{1}{|c|}{Sweep:14 FIT - 1,430 FIT}\\\hline
\end{tabular}}
\label{table:FIT3}
\end{center}
\vspace{-0.2in}
\end{table}

\subsection{Simulation Infrastructure}
\subsubsection*{Reliability} 
To evaluate reliability of different schemes, this dissertation uses an industry-grade fault and repair simulator \textit{FaultSim} ~\cite{faultsim}. The scrub interval was configured for 12 hours. After intervals of 12
hours, correctable transient faults are removed due to the scrubbing
mechanism. FaultSim conducts Monte Carlo simulations for $10^5-10^6$
trials (more trails for schemes that show lower failure rates, to
improve accuracy) for lifetime of 7 years and report an average.

\subsubsection*{Performance} 
The baseline configuration is described in
Table~\ref{table:system_config3}.  The in-house system simulator uses 8 cores which share
an 8 MB LLC. The memory system uses 3D stacks with
eight 8 Gb dies for data and one additional die for ECC or metadata
in the case of Citadel. Virtual-to-physical translation uses a
first-touch policy with a 4KB page size. 

\begin {table}[ht]
\vspace{-0.2in}
\begin{center}
\caption{Baseline System Configuration for Citadel}{
\begin{tabular}{|c|c|}
\hline
				 \multicolumn{2}{|c|}{\textbf{Processors}} \\ \hline
                 Number of cores               & 8         \\ 
                 Processor clock speed         & 3.2 GHz    \\ \hline
                 \multicolumn{2}{|c|}{\textbf{Last-level Cache}} \\ \hline
                 L3 (shared)     & 8MB, 8-way, 24 cycles   \\
                 Associativity   & 8-way \\
                 Latency		 & 24 cycles \\ 
                 Cache-line size & 64Bytes    \\ \hline
                 \multicolumn{2}{|c|}{\textbf{DRAM 2x8GB 3D stacks}} \\ \hline
                 Memory bus speed             & 800MHz (DDR3 1.6GHz)\\
                 Memory channels              & 8/Stack         \\
                 Capacity per channel    & 1GB         \\
                 Banks per channel            & 8         \\
                 Row-buffer size			  & 2KB\\
                 Data TSVs	& 256/Channel\\
				 Addr TSVs	& 24/Channel\\
				 t$_{WTR}$-t$_{CAS}$-t$_{RCD}$-t$_{RP}$-t$_{RAS}$ & 7-9-9-9-36\\ \hline
\end{tabular}}
\label{table:system_config3}
\end{center}
\vspace{-0.2in}
\end{table}

For evaluations, this study used all 29 benchmarks from the SPECCPU 2006
\cite{spec} suite. This study also used memory-intensive benchmarks from the 
PARSEC \cite{PARSEC} suite, such as  \textit{black}, \textit{face},
\textit{ferret}, \textit{fluid}, \textit{freq}, \textit{stream} and
\textit{swapt}. From the BioBench \cite{BIOBENCH} suite, this study used
 \textit{tigr} and \textit{mummer}. A representative slice of 1
 billion instructions was generated using Pinpoints for simulation purposes \cite{pinpoints}.  

The evaluations executed the benchmarks in rate mode, in
which all eight cores execute the same benchmark. Timing
simulations were performed until all the benchmarks in the workload finish execution,
and measure the execution time as the average execution time of all
eight cores.

\subsubsection*{Power} The study also measured active (read, write, refresh and activation) power using the equations from the Micron Memory
System Power Technical Note for 8Gb chip~\cite{ddr3_power,micron_8Gb}. As per HBM, the refresh interval
is set to 32 ms ~\cite{HBM:2013,jedec_ddr4}.

\section{Citadel: An Overview}

\begin{figure}[htb]
  \centering
  \centerline{\psfig{file=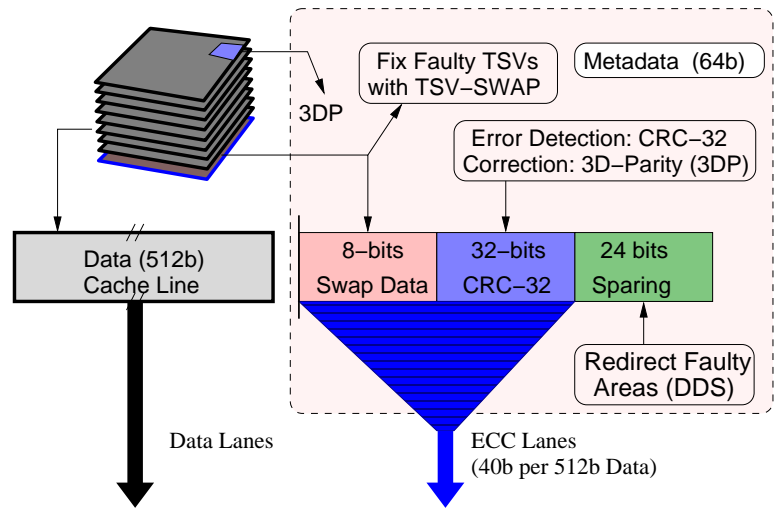,width=4in}}
   \caption{Overview of Citadel}
     \label{fig:overview}
\end{figure}

This dissertation proposes {\em Citadel}, a robust memory architecture that can
tolerate both small- and large-granularity faults effectively.
Figure~\ref{fig:overview} shows an overview of Citadel. HBM provisions
64 bits of ECC for every 64 Bytes, possibly in a separate ECC
die~\cite{HBM:2013}. Similarly, Citadel provisions each 64B cache line
with 64 bits of metadata.  However, Citadel uses the ECC die to store
different types of metadata information, each geared towards
tolerating different types of faults. Each 64B (512b) transaction
fetches 40bits of metadata over ECC lanes. The remaining 24 bits are
used to provision sparing of faulty blocks. Citadel consists of three
component schemes: {\em TSV-SWAP}, {\em Tri Dimensional Parity
(3DP)} and {\em Dynamic Dual-Granularity Sparing (DDS)}.

Citadel differentiates faults in memory elements from faults in TSVs.
The TSV-SWAP technique of Citadel can tolerate TSV faults by
dynamically identifying the faulty TSVs and decommissioning such TSVs.
The data of faulty TSVs is replicated in the metadata (up to 8
bits). TSV-SWAP protects against faulty data TSVs as well as
faulty address TSVs, which tend to be even more severe in
practice. Thus, TSV-Swap provides resilience to TSV faults at runtime,
without relying on manufacturer provided spare TSVs.

Citadel relies on CRC to detect data errors. Once an error is
detected, it is corrected using the 3DP scheme, which maintains parity
in three dimensions: across banks, across rows within one die, and
across rows of different dies. 3DP can not only tolerate
small-granularity failures such as bit and word failures as well as
large-granularity failures such as row and bank failures.  3DP uses
one of the data banks to implement bank-level parity (storage overhead
of 1.6\%).

Citadel employs data sparing to avoid frequent correction of faulty
data. This not only prevents the performance overheads of error
correction, but also makes the system more robust, as otherwise
permanent faults gets accumulated over time.  The DDS sparing scheme
of Citadel exploits the observation that a bank either has a few small
granularity faults (less than 4) or many (more than 1,000) faults; DDS
spares at either a row granularity or a bank granularity. DDS uses
three out of eight banks of the metadata die for sparing.

When combined, the three techniques of Citadel can tolerate TSV and multi-granularity granularity faults while consuming
a storage overhead similar to an ECC DIMM (14\% for Citadel versus 12.5\% for ECC DIMM) and allowing
the data of the cache line to be resident in the same bank.  The next
sections describe the three techniques in detail.

\section{Mitigating TSV Faults with TSV-SWAP}

Stacked memory systems use TSVs to connect data, address and
command links between the logic die and DRAM dies. Without loss of
generality, this section explains the working of
TSVs, fault models, and a solution that enables robust TSVs.

\subsection{TSV Organization within Stacked Memories}

The HBM system in this dissertation consists of 8 channels of 256 Data TSVs
(DTSV) with 24 address/command TSVs (ATSV). A memory request presents
an address and commands over external address/command
links. Internally, TSVs transfer the address and command information
for the channel to the corresponding die.  For a read request for one
cache line, the entire 2KB of data for the row (called a DRAM page) is
addressed and brought into the sense amplifiers. From the 2KB (16Kb)
page, 64B (512bits) of data are multiplexed and transferred via the
TSVs. Because there are only 256 DTSVs, each TSV will transfer data in
two DDR cycles. The DRAM row (2KB) contains data for 32 cache lines. Each
of these 32 cache lines is multiplexed to the same set of
TSVs. Furthermore, all banks within the same die share the TSVs, which
means a fault in the TSV causes multi-bank failures.

\begin{figure}[htb]
  \centering
  \centerline{\psfig{file=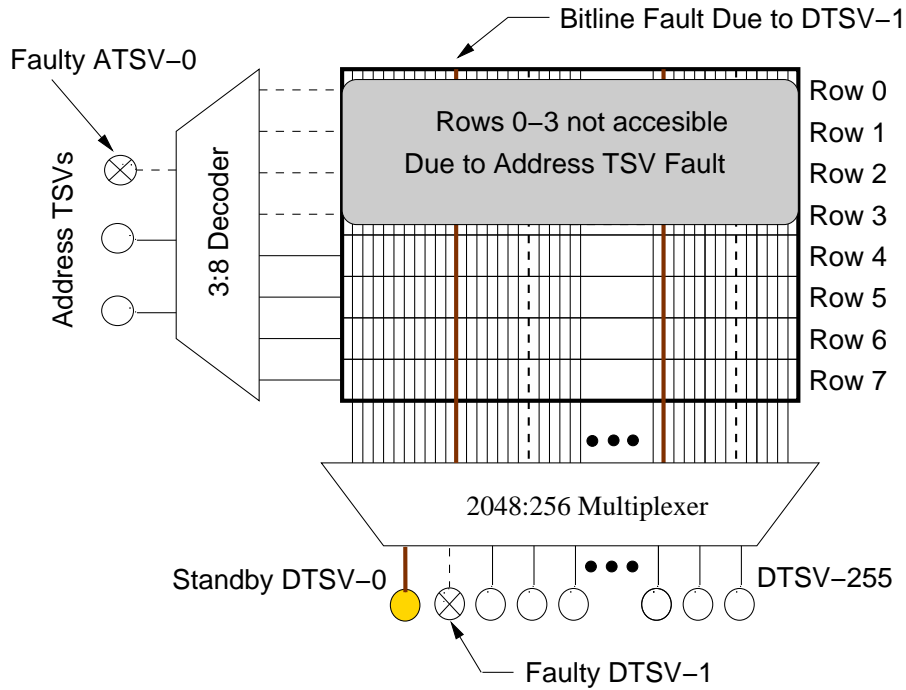,width=4.8in}}
  \caption{Faults in Data TSV (DTSV) and Address TSV (ATSV). TSV-SWAP
  creates stand-by TSVs from existing TSVs to tolerate TSV faults
  (such as DTSV-1 and ATSV-0).}  \label{fig:tsvfaults}
\end{figure}

\subsection{Severity of TSV Faults: DTSV vs. ATSV}
The vulnerability of the system to TSV faults depends on whether the
fault happens in DTSV or ATSV, as shown in Figure~\ref{fig:tsvfaults}.
Because the burst size for the HBM design is 2, each DTSV fault will
cause 2 bits to fail in every cache line.  For example, a failure of
DTSV-1 will cause bit[1] and bit[257] of each cache line to
fail. Faults in ATSV are even more severe; a single fault can make
half of the memory unreachable, because the decoder is unable to
address half of the memory space.  For example, a failure of ATSV-0
makes half of the rows (Row-0 to Row-3) unreachable.

\subsection{Efficient Runtime TSV Sparing with TSV-SWAP}
TSV faults at manufacturing time are typically mitigated by spare TSVs
provisioned for enhancing yield~\cite{tsv_red}.  Such spare TSVs may
or may not be available to the user to tolerate faulty TSVs that
happen at runtime. The proposal in this dissertation, {\em TSV-Swap} can mitigate TSV
faults at run-time without relying on manufacturer-provided spare TSVs
and distinguishes between the severity of faults in address and data
TSVs. TSV-SWAP differentiates between address and data TSVs with the help of a built in test logic. 
Instead of relying on spare TSVs, it creates a pool of stand-by
TSVs from the available DTSVs, and uses these stand-by TSVs to repair
the faulty DTSV and ATSV. TSV-SWAP consists of three steps and which are described as follows.

\subsubsection{Creating Stand-by TSVs} TSV-SWAP creates stand-by 
TSVs by duplicating the data of predefined TSV locations into the 8-bit
swap data provided by metadata in Citadel (see
Figure~\ref{fig:overview}). Such a design designates four TSVs as stand-by
TSVs from a pool of 256 DTSV (DTSV-0, DTSV-64, DTSV-128, and
DTSV-192). As each DTSV bursts two bits of data for each cache line, 8
bits from each cache line are replicated in the metadata (bit[0],
bit[64], ..., bit[448]).  The four stand-by TSVs which are created are
used to repair any faulty TSVs that occur at runtime.

\subsubsection{Detecting Faulty TSV}

Citadel computes a CRC-32 code using address and data information. A TSV
error will result in an incorrect checksum of the CRC-32 code. To differentiate between
TSV faults and data faults, TSV-SWAP employs two additional rows
(\textit{row1-fixed} and \textit{row2-fixed}) per die that stores a
fixed sequence of data. These rows are at locations where each bit of
addresses are the inverse of each other (for example, address
\textit{ 0x0000} and \textit{0xFFFF}). On detecting
a CRC mismatch, data from these fixed rows are read and compared
against the pre-decided sequence. If there is a mismatch between the compared values, the error is
highly likely (but not always) due to a TSV fault. The memory system
now invokes the BIST logic which checks for TSV faults.

\subsubsection{Redirecting Faulty TSV}
 
TSV-SWAP provisions both the DTSV and ATSV with a redirection circuit
that can replace a faulty TSV with one of the stand-by TSVs.  The
redirection circuit is simply a multiplexer and a register. On
detecting a TSV fault, the BIST circuitry enables the TSV redirection
circuit as a corrective action against the faulty TSV. The BIST circuitry then connects one of the stand-by
TSVs to replace the faulty DTSV or ATSV.
\subsection{Result: TSV-SWAP with ChipKill}
\begin{figure}[htb]
  \centering
  \centerline{\psfig{file=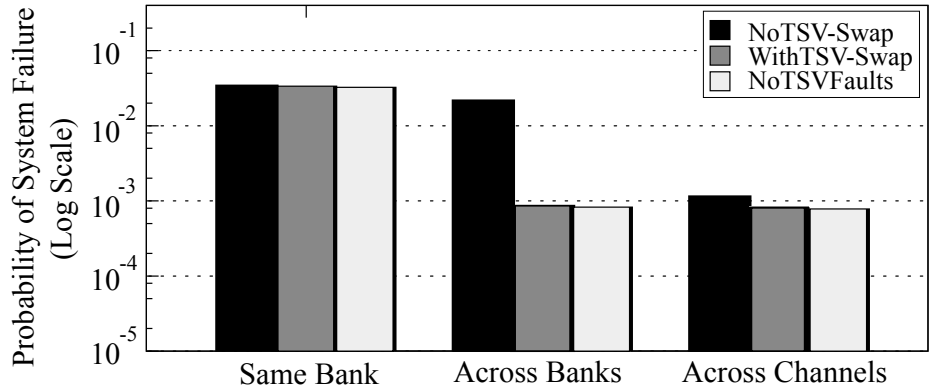,width=4.8in}}
   \caption{TSV-SWAP is effective at mitigating TSV faults and provides almost similar performance to an ideal ChipKill system}
  \label{fig:tsv_swap_chipkill}
\end{figure}

This dissertation analyzez the effectiveness of TSV-Swap at mitigating TSV
faults for a system employing ChipKill . Unfortunately, the FIT rate data for TSV faults is not
available publicly, so for this section, we assume a high TSV fault
rate (1430 FIT, corresponding to one TSV-caused die failure every
seven years) to assess the effectiveness of TSV-Swap at high TSV fault
rate. Figure~\ref{fig:tsv_swap_chipkill} shows the probability of
system failure for the three configurations (No TSV-Swap, With
TSV-Swap, and No TSV Faults) for the three data mappings. For all
systems, TSV-SWAP achieves a resilience similar to that of not having
any TSV faults, even with the assumed high failure rate for TSVs. One can therefore
conclude that TSV-SWAP is highly effective at mitigating TSV
failures.

\subsection{TSV-SWAP for Alternate Stacked Memory Organizations}\label{sec:alternate}
\begin{figure*}[htb]
  \centering
  \centerline{\psfig{file=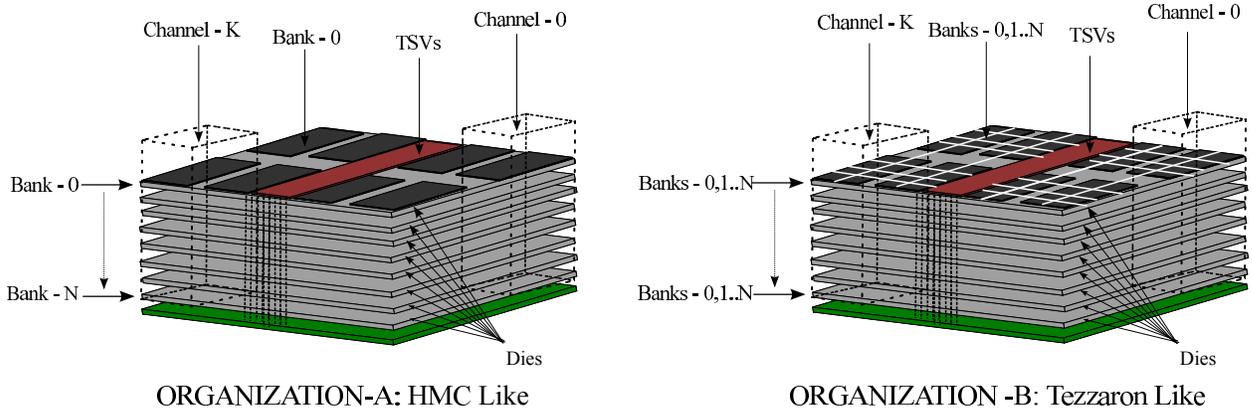,width=6.5in}}
   \caption{Alternate 3D stacked memory organizations. Organization-A has channels across dies (vertically) and all banks in this channel are in different dies and is similar to an HMC system. Organization-B has channels across dies (vertically) and is similar to a Tezzaron stacked memory system. Here, each die has a portion of all banks in that channel}
  \label{fig:3Dmemalt}
\end{figure*}

Until now, this chapter has evaluated TSV SWAP for an HBM-like organization. However, stacked memories can have alternate organizations in the placement of TSVs. Figure~\ref{fig:3Dmemalt} shows two alternate organizations of stacked memory systems that reorganize channels and banks by changing the placement of TSVs.

The first organization of stacked memory, Organization-A, is a HMC-like organization. In Organization-A, the channel(s) are organized vertically across dies. Every die contributes a single bank to each channel. The TSVs are distributed across dies and every channel in every die requires individual buffers. The second organization, Organization-B of stacked memory is a Tezzaron-like organization. In Organization-B, the channel(s) that are organized vertically across dies. However, every die holds a portion of multiple banks for a channel. This increase the number of address and data TSVs per channel per die.

\begin{figure}[htb]
  \centering
  \centerline{\psfig{file=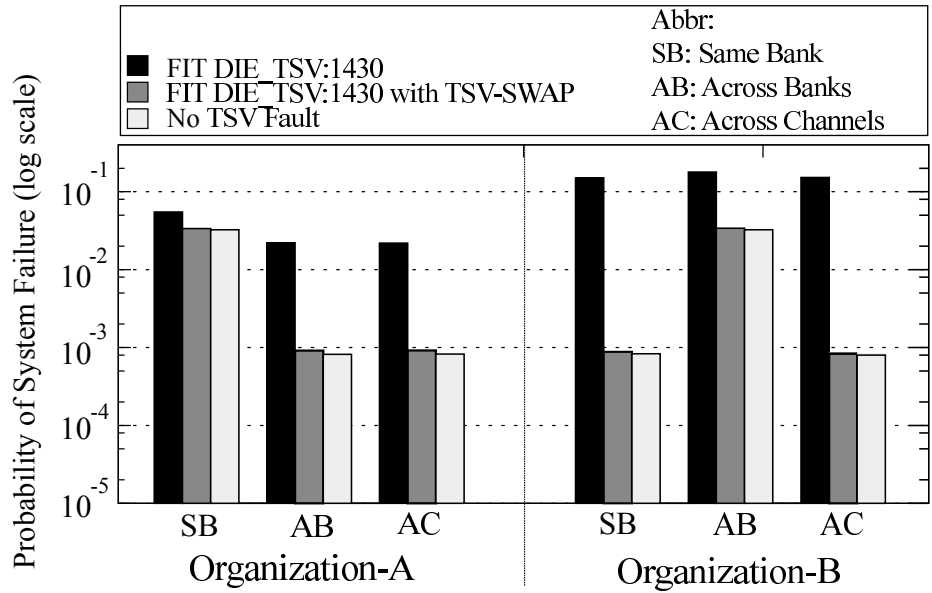,width=4.8in}}
   \caption{Organization-A (HMC-like) and Organization-B (Tezzaron-like) can be more sensitive to TSV faults when compared to a HBM like organization. TSV Swap mitigates almost all TSV faults}
  \label{fig:tsv_swap_chipkill2}
\end{figure}

Figure~\ref{fig:tsv_swap_chipkill2} shows the probability of system failure for two alternate stacked memory configurations. The evaluations in our study show that Tezzaron like designs are more prone to TSV faults due to higher density of TSVs for data and address. As every physical bank is further divided into several logical banks, placing data across these logical banks has the same effect as placing data in the same bank. Due to this the across bank data placement has the lowest reliability in a Tezzaron like design. An HMC like design has similar trend to that of a HBM like configuration. Even for alternate organizations, TSV-SWAP achieves a resilience similar to that of not having any TSV faults.

\subsection{Reducing the Complexity of TSV-SWAP}\label{sec:complex}
\begin{figure}[htb]
\vspace{0.2in}
  \centering
  \centerline{\psfig{file=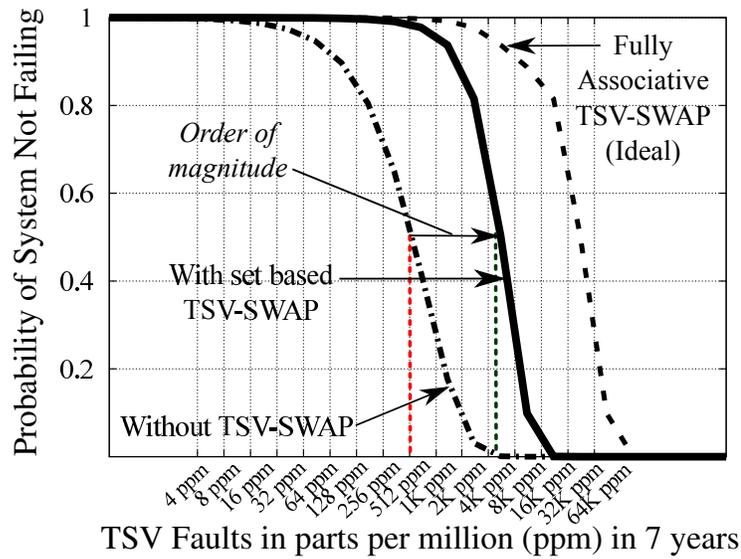,width=3.8in}}
   \caption{Set based TSV-SWAP can handle 10x more TSV failures before it causes system failure. In this study, a set consists of 70 TSVs, 63 data+\textit{1 Data Swap-TSV}+6 address/command TSVs }
  \label{fig:tsv_bucket_and_balls}
\end{figure}

Architecting any Data TSV to swap between address, command and other data TSVs increases the complexity of the swap logic. To overcome this, TSV-SWAP uses a set structure for swapping TSVs. In this structure, a set of TSVs (address/control+data) co-located with a fixed Standby Data-TSV (S-TSV). Only one TSV encountering a fault can be swapped with its \textit{S-TSV} in a set. In the analysis conducted for out study, a set consists of 70 TSVs, 63 data+\textit{1 Data Swap-TSV}+6 address/command TSVs. Our study performs a bucket and balls analysis to determine the probability of system failure for such set group. Figure~\ref{fig:tsv_bucket_and_balls} shows that such set based TSV-SWAP can handle 10x more TSV failures when compared to a system that does not employ TSV-SWAP. An ideal fully associative (complex) TSV-SWAP circuitry provides 10x higher reliability when compared to a Set-based scheme.

\section{Effectiveness of TSV SWAP for memory system employing Single Error Correction and Double Error Detection (SECDED)}\label{sec:TSV-SEC}
Until now we have assumed a system that employs a symbol based error correcting code like ChipKill. These symbol based codes can correct large granularity faults and single bit errors. Fortunately single or multiple random bit errors can be corrected using
Bose-Chaudhuri-Hocquenghem (BCH) codes, including Hamming Codes
~\cite{lin-costello-book}. Hamming codes have a bit storage overhead
of \textit{log$_{2}$(Size of the Code Word)+1} (including additional
error detection).  They provide single error correction, double error
detection (SECDED) computed over an 8 Byte codeword requires 8
additional bits for every 64 bits.  Decoding and encoding complexity,
check bit overhead and latency increase with the strength of the ECC.

\begin{figure}[htb]
\vspace{-0.25in}
  \centering
  \centerline{\psfig{file=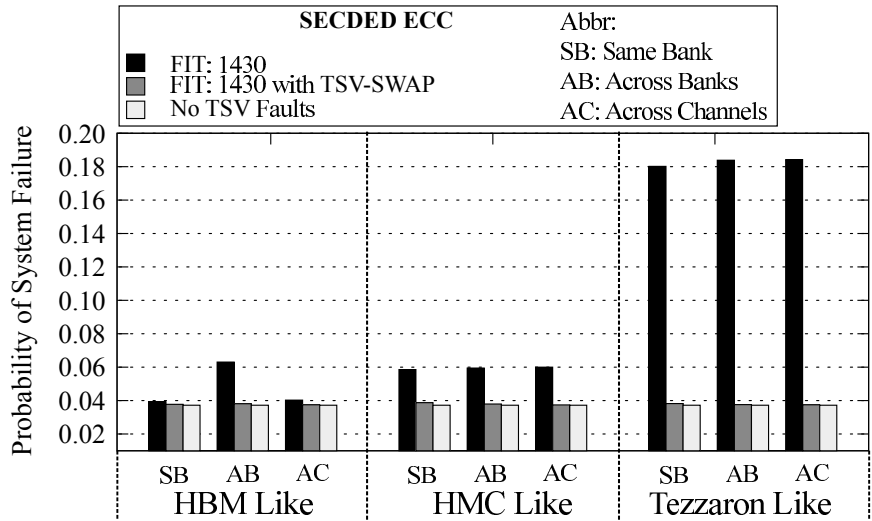,width=4.5in}}
    \vspace{-0.05in}
   \caption{TSV-SWAP is effective at mitigating TSV faults and provides almost similar performance to an ideal system employing SECDED codes}
  \label{fig:tsv_swap_secded}
  \vspace{-0.1in}
\end{figure}

Figure~\ref{fig:tsv_swap_secded} shows the effectiveness of TSV SWAP for a system that employs SECDED based ECC. SECDED provides lower reliability when compared to ChipKill, however TSV SWAP enables SECDED to overcome errors due to TSV faults and provides reliaibity close to an ideal system that employs SECDED protection.

Figure~\ref{fig:tsv_swap_secded} shows that Tezzaron like designs are more vulnerable to TSV faults due to their higher density of TSVs. Furthermore, since TSVs may cause large granularity failures, SECDED is ineffective against them. Subsequent sections assume that the system employs symbol based ECC (Chipkill like) and TSV faults are mitigated with TSV SWAP.

\section{Tri Dimensional Parity (3DP)}
The second component of Citadel targets efficient error detection and error
correction of data values. Several error detection codes such as SECDED, Checksums and CRC-32 or CRC64 are used in commercial systems\cite{CRC:1961,koopman, HP:2011}. Of these, CRC-32 tends to have a reasonable detection coverage and storage efficiency. Citadel provisions each line with a 32-bit
cyclic redundancy code (CRC-32), which is highly
effective\footnote{The probability of overlapping CRC-32 checksum is
$\frac{1}{2^{32}}\approx10^{-10}$. For false negative, the failed
element should have an overlapped CRC-32. The probability that an
element fails is less than 10$^{-6}$. Thus, the effective probability
of an overlapping CRC-32 is negligibly small ($\ll 10^{-16}$).} at
detecting data errors~\cite{CRC:1961,GABEISCA:2013}. Citadel uses a
novel scheme, called {\em Tri Dimensional Parity (3DP)}, to correct
data errors at multiple granularities. In 3DP, even if one dimension
encounters two faults, they are highly unlikely to fall into
the same block in the other two dimensions. On detecting an error, the
memory contents are read and the error gets corrected using
parity.\footnote{Error correction may take 700 milliseconds, however
given that correction is invoked once every few months, this results
in negligible performance overheads.}

\subsection{Design of Dimension 1}
\begin{figure}[htb]
  \psfig{file=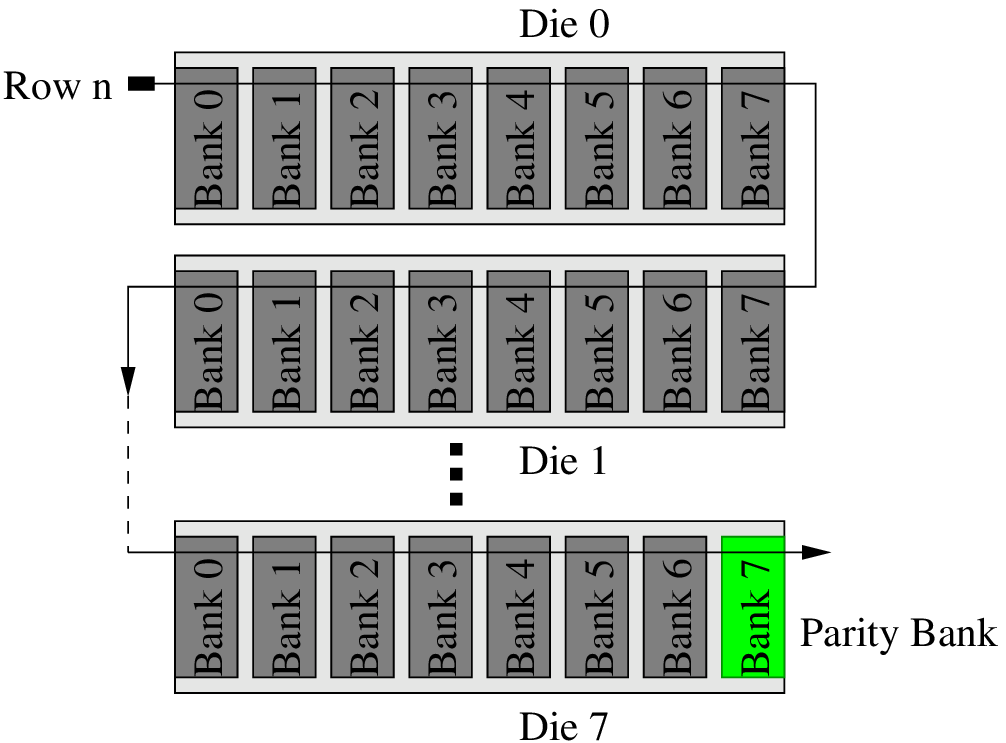,width=3.5in}
  \psfig{file=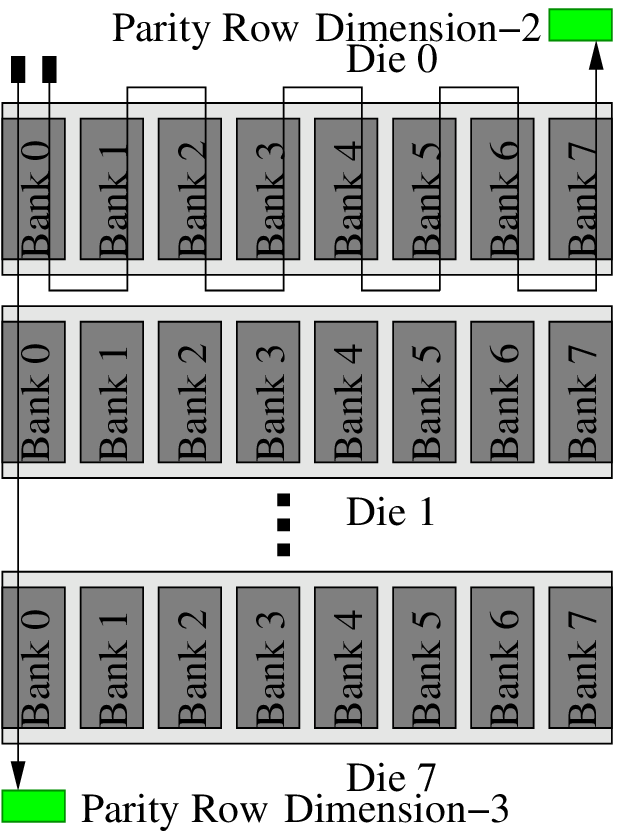,width=2.2in}
   \caption{Dimension 1 stripes parity across a single row in every bank for all dies and generates a row in the parity bank. Dimension 2 stripes parity across all row in every bank within a die and generates a parity row. Dimension 3 stripes parity across all rows in single bank across dies and generates a parity row.}
  \label{fig:dim123}
\end{figure}

Figure~\ref{fig:dim123} shows the design of Dimension 1. It computes the
parity for a row in every bank across dies as specified in
equation~\eqref{eqn_dim1}. This requires dedicating a range of single bank addresses as a
parity bank for the entire stack (1.6\% overhead, for our 8 channel
system, with 8 banks for each channel).\footnote{Parity bank is an abstraction, such a bank can have addresses across multiple physical banks in a stack. This can be done by swapping 2 bits (one lower bank bit and one higher channel bit) while addressing the parity bank. This prevents one physical bank from becoming a bottleneck.} A parity bank helps
mitigate single-bank faults. However, a one-dimensional parity
(1DP) scheme is intolerant to multiple faults. Even if
a single-bit failure occurs after a single-bank failure, it results in data loss. 

\begin{multline} \label{eqn_dim1}
ParityBank[row_{n}]=Die_{0}.Bank_{0}[row_{n}] \oplus
Die_{0}.Bank_{1}[row_{n}] \oplus \cdots \oplus
Die_{7}.Bank_{6}[row_{n}] 
\end{multline}

\subsection{Design of Dimensions 2 and 3}
Figure~\ref{fig:dim123} shows the design of Dimensions 2 and 3. In
Dimension 2, parity is taken across all rows in all banks within a
die. Equation~\eqref{eqn_dim2} shows the computation \textit{Parity
Row} in Dimension 2 for Die 0. Because there are 9 dies (including the
metadata die), the storage overhead is 9$\times$
the size of a DRAM row for each dimension.

\begin{multline} \label{eqn_dim2}
ParityRowDim2_{Die0}=[Bank_{0}[row_{0}] \oplus Bank_{0}[row_{1}]
\oplus \cdots \oplus Bank_{7}[row_{n}]]_{Die0}
\end{multline}

Dimension 3 computes parity across dies for all rows in a single
bank. Equation~\eqref{eqn_dim3} shows the computation for
\textit{Parity Row} in Dimension 3 for Bank 0. Because there are 8 banks
per die, the storage overhead of is 8$\times$size of DRAM row.  While
Dimension 1 is designed to tolerate bank failures, Dimensions 2 and 3
prevent independent row, word and bit failures. When used together,
3DP can correct multiple errors that occur at the same time within a
stack.

\begin{multline} 
\label{eqn_dim3}
ParityRowDim3_{Bank0}=[Die_{0}[row_{0}] \oplus Die_{0}[row_{1}] \oplus
\cdots \oplus Die_{7}[row_{n}]]_{Bank0}
\end{multline}
      
\subsection{Reducing Overheads for Parity Update}   
 
Citadel avoids the performance overheads of updating the parity for
Dimensions 2 and 3 by keeping the parity information on-chip. The size
of the row buffer of the stacked DRAM we simulate is
2KB~\cite{HMC:2013,HBM:2013}. Thus, maintaining Dimensions 2 and 3
would require a storage overhead of 34 KB (9 rows for Dimension 2 and
8 rows for Dimension 3), which can be kept at the memory controller.
Thus, updating the parity for Dimensions 2 and 3 can be done on-chip
with negligible timing and power overheads.

\begin{figure}[htb]
  \centering
  \centerline{\psfig{file=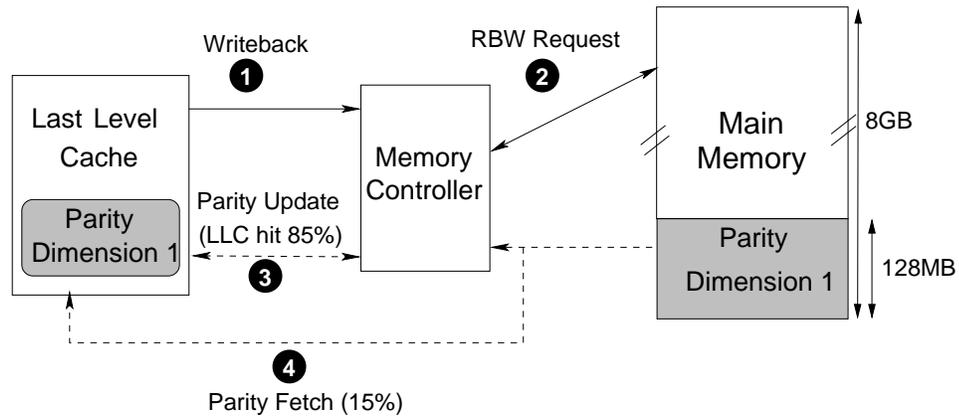,width=5in}}
  \caption{Memory System employing on-demand parity caching for Dimension 1 within
the LLC (Figure not to scale)}
  \label{fig:sys} \end{figure}

The total size of parity for Dimension 1 is equal to 1 Gb (128 MB)
which would be impractical to duplicate at the memory controller
side. To reduce the parity update overheads for Dimension 1, Citadel employs
parity caching within the on-chip LLC.  For Dimension 1, every parity
cache line is responsible for 63 data lines from 63 different
banks. Thus, accesses to parity lines are expected to have a very high
temporal locality. Figure~\ref{fig:sys} shows the operation of a
system that implements on-demand parity caching within the LLC for a
writeback request to a data line (action \circled{\textcolor{white}{1}}). 

To update the parity information, the old data of the written line is XORed with the new data. 
The memory controller performs such a Read Before Write (RBW)
request to obtain the old information of the line (action \circled{\textcolor{white}{2}}). 
As the row was recently opened, RBW tends to be a row-buffer hit. The XOR
forms a parity update. The memory controller then checks the LLC for
the parity line associated for the address for which writeback is
being made. In the common case (85\% of the time, on average) the parity
line is found in the LLC and the parity is updated with the XOR value
(action \circled{\textcolor{white}{3}}). In the uncommon case that the parity information for
Dimension 1 is not found in the LLC, then parity information is
fetched from the memory (action \circled{\textcolor{white}{4}}), installed in the LLC, and the parity
information is updated.

 \begin{figure}[htb]
  \centering
  \centerline{\psfig{file=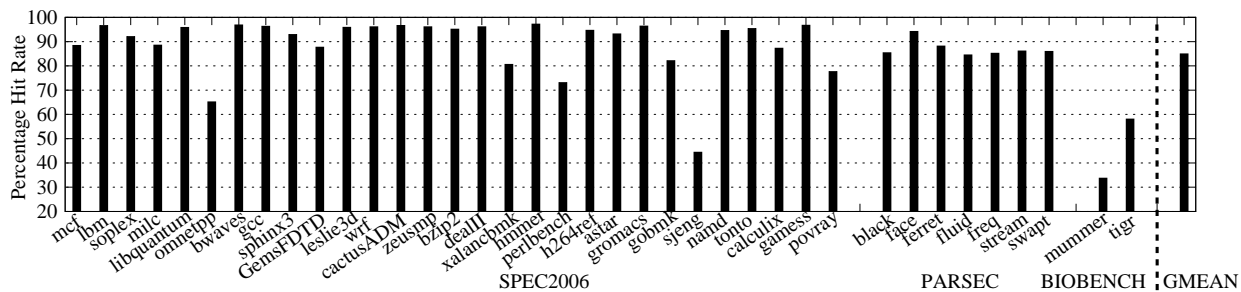,width=6.5in}}
  \caption{Hit rate for parity caching of Dimension 1}
  \label{fig:hitrate3} 
  \end{figure}
  
Figure~\ref{fig:hitrate3} shows the LLC hit-rate for parity update
requests. On average, the hit rate is 85\%, showing that parity caching is quite
effective. The BIOBENCH workloads mostly perform read operations, with
writes sparsely distribute between a large number of writes.  Hence,
read requests tend to evict parity lines. However, since the
frequency of writes for BIOBENCH is less, a low hit rate for parity update results in
negligible performance loss.

\subsection{Error Detection and Correction using 3DP}
On every read request, 3DP works in two phases. The first phase
consists of fast error checking using CRC-32 code. For most requests,
this phase will report no errors. However in the rare case of a
reported error (once in a few months), the second phase is activated
and the whole memory is read. 3DP then isolates the fault(s) using all
three dimensions of parity across the stack. If it is a small
granularity bit, word or row fault, then dimensions 2 and 3 parity can
fix such errors. However, large granularity faults such as column and
bank faults are corrected using dimension 1 parity. In the event of
simultaneous multi-granularity faults, dimensions 2 and 3 parity help
isolate small granularity faults and dimension 1 parity helps isolate
the large granularity fault.

\subsection{Results for 3DP}   
The 3DP scheme allows the memory system to retain the cache line
within the same bank, and yet be able to correct bit, word, row,
column and bank failures. Our study compares the resilience, performance, and
power of the 3DP scheme to a theoretical scheme that employs an 8-bit
symbol-based coding with data striping. For a fair comparison, our study assumes that
TSV-Swap is enabled for both the 8-bit symbol based code and 3DP.
 
\subsubsection{Resilience} 
\begin{figure}[htb]
     \vspace{-0.1in}
  \centering
  \centerline{\psfig{file=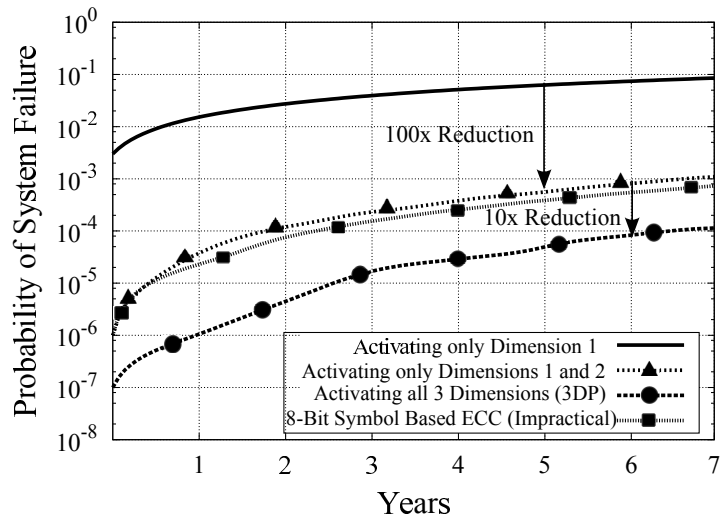,width=3.7in}} 
   \caption{3DP has 7x more resilient than an 8-bit
 symbol-based ECC for tolerating large-granularity failures
  in stacked memory. 3DP has 10x more resilience than 2DP}
   \label{fig:resilience}
   \vspace{-0.25in}
 \end{figure}

Figure~\ref{fig:resilience} compares
the multi-dimensional parity scheme with a very strong 8-bit symbol-
correcting code striped across channels. Enabling only a single
dimension of parity (at Bank Level) does not improve resilience
against multiple faults that occur concurrently. A single dimensional
parity scheme is unable to correct these faults. By enabling
all three dimensions, 3DP achieves a 1000x improvement in
resilience. Furthermore, 3DP achieves 7x stronger resilience than an
8-bit symbol-based ECC because it can handle higher number of multiple concurrent
faults.

\subsubsection{Performance} 

\begin{figure*}[htb]
     \vspace{-0.1in}
\centering
  \centerline{\psfig{file=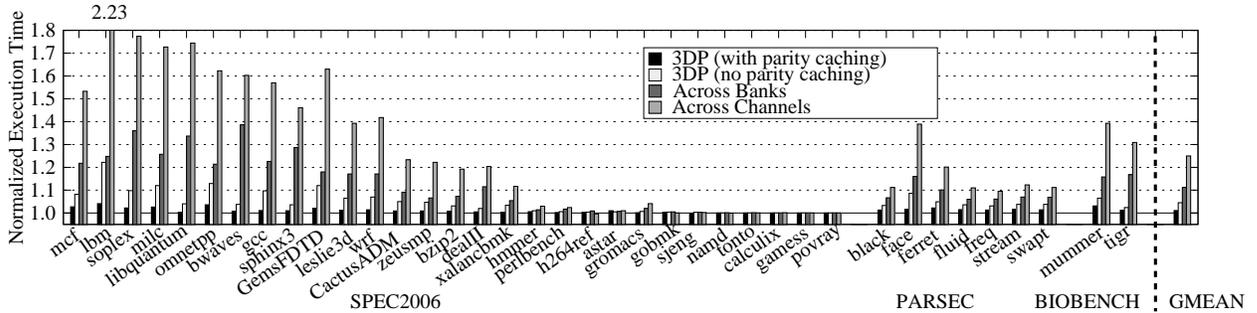,width=6.5in}}
  \caption{Normalized execution time: 3DP has negligible slow-down,
  whereas data striping causes 10-25\% slow-down.}
  \label{fig:perf3}
     \vspace{-0.1in}
  \end{figure*}

Figure~\ref{fig:perf3} compares the execution time of 3DP to the
organizations that stripe data either across a bank or a channel.  The
execution time is normalized to a baseline that retains the cache line
within the same bank and pays no overhead for error correction.  The
3DP scheme with caching has performance within 1\% of the baseline,
3DP without caching degrades performance by 4.5\%.  Thus, parity
caching is highly effective at mitigating the performance impact of
parity updates.  Alternative schemes, that rely on striping the data
in different banks or channels, degrade performance by as much as 10\%
to 25\%, on average due to the loss of bank/channel level
parallelism. Thus, 3DP not only improves the resilience of stacked
memory compared to data striping, but also helps brings the
performance impact of fault tolerance to a negligible level.

  \begin{figure*}[htb]
\centering
  \centerline{\psfig{file=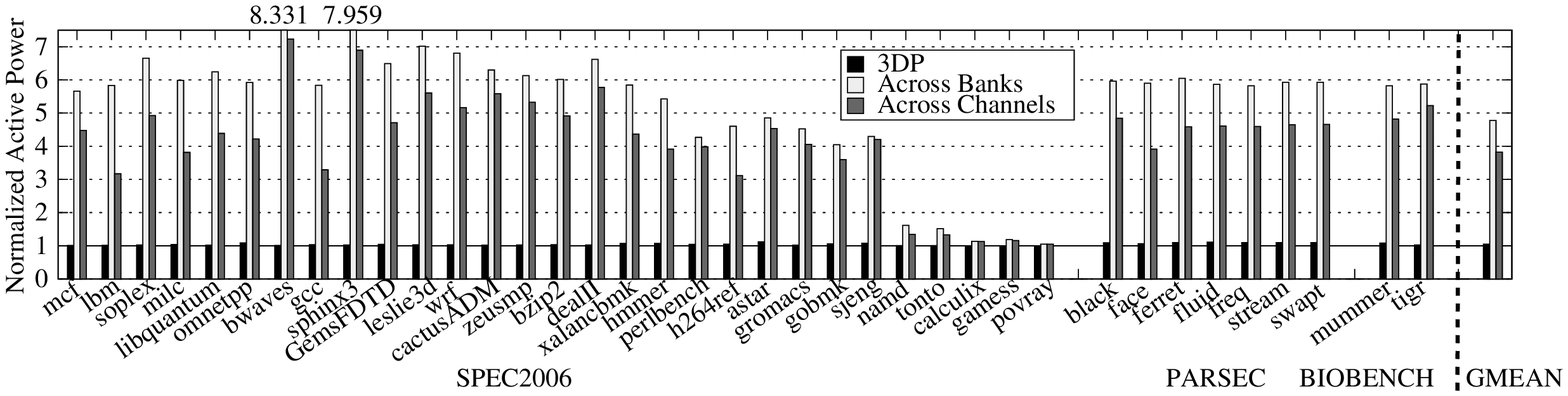,width=6.5in}}
  \caption{Active power consumption: 3DP has negligible power overheads, whereas data striping has 3-5x greater overhead.}
  \label{fig:pwr} 
     \vspace{-0.25in}
  \end{figure*}
  
\subsubsection{Power} 

Accessing multiple banks or channels to satisfy every memory request
also has the disadvantage that it consumes significantly higher
power. 3DP design allows Citadel to place the entire
cache line in one bank, and thus activate only one bank per read
request. This not only reduces the activation power but also improves
memory level parallelism, compared to the Across-Bank and
Across-Channel configuration. Figure~\ref{fig:pwr} shows the active
power for 3DP, Across-Bank, and Across-Channel configuration,
normalized to the fault-free baseline that places the cache line in
the same bank.  On average, 3DP increases active power by only 4\%,
whereas Across-Bank and Across-Channel configurations increase active
power by almost 3X-5X of higher bank/channel activations and row
conflicts.

\subsubsection{Additional Memory Traffic}\label{sec:traffic}
\begin{figure*}[htb]
 \centering
  \centerline{\psfig{file=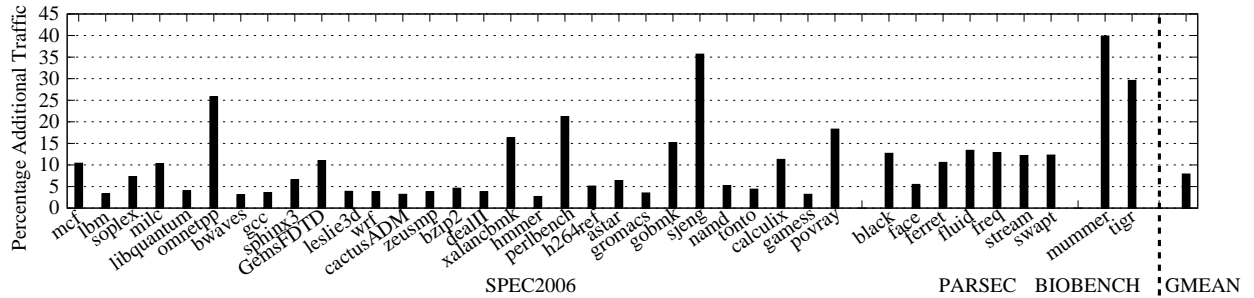,width=6.5in}}
  \caption{Percentage of additional memory traffic: 3DP with Dimension-1 caching, on an average incurs only 8\% with a maximum of 40\% for workloads with low LLC dimension parity hit rate}
  \label{fig:traffic3} 
  \end{figure*}

3DP updates dimension-1 parity for every write and accesses this from memory. Due to this there is additional traffic on every write. To overcome this, 3DP uses parity caching of dimension-1 parity. Figure~\ref{fig:traffic3} shows the additional traffic after caching dimension-1 parity. On an average, dimension-1 caching helps in reducing the average additional memory traffic to 8\%. The additional memory traffic is correlated to the hit rate of dimension-1 parity in last level cache. For instance, omnetpp and sjeng have low dimension-1 parity hit rates and therefore have upto 35\% higher traffic. Since writes in BIOBENCH are sparsely distributed, additional memory traffic does not have a significant impact on its performance.

\section{Dynamic Dual-granularity Sparing (DDS)}
The 3DP scheme performs error correction by recomputing the data based
on parity information. However, this can be a time-consuming process
(recomputing parity and isolating the fault in each
dimension). Fortunately, faults do not occur frequently, so employing
a slow correction mechanism is a viable option.  However, if the
faults are permanent then the correction scheme will be invoked
frequently and cause unacceptable performance degradation.  Citadel
avoids this by using dynamic sparing, whereby a data item once
corrected is redirected to an alternate location.  The key question in
designing a data-sparing scheme is the granularity of sparing.
Sparing at row granularity would be storage efficient, however it
would be fairly complex to tolerate bank failures, as the redirection
structures associated with row sparing would require several tens of
thousands of entries. One can implement sparing at a bank granularity,
but it suffers significant under-utilization of spare area. Thus, uniform
sparing is either complex or inefficient.  To address this dichotomy,
Citadel is provisioned with {\em Dynamic Dual-granularity Sparing
(DDS)}.  This section presents the key observation that motivates DDS.

\subsection{Key Observation: Failures Tend to be Bimodal}
   
\begin{figure}[htb]
    \centering
   \centerline{\psfig{file=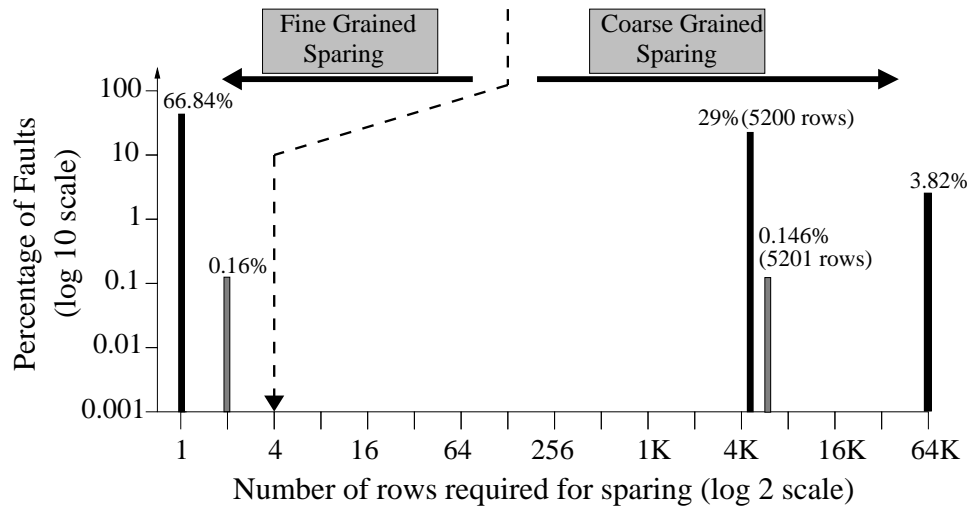,width=5in}}
   \caption{Permanent fault affects either very few (less than 4) rows or large number of 
             ($>$ 1000) rows.} 
\label{fig:fault_buckets}
\end{figure}

Only for the analysis in this section, all faults
that are smaller than or equal to a row fault are classified as causing a row
failure. These faults will consume one entry for a row-sparing
architecture.  A large-granularity fault would consume many entries of
row sparing. Figure~\ref{fig:fault_buckets} shows the distribution of
the number of rows that are used by a faulty bank, on average based on monte-carlo simulations using FaultSim \cite{faultsim}.

The number of failures show a bimodal distribution.  The
smaller-granularity faults do not occur in many multiples. In fact, in
all simulations for this study, no more than two rows per bank were affected by a
small-granularity fault within a scrubbing interval. However, there
are two peaks; one at 5,200 rows (most likely due to sub-arrays) and another at
65K rows (size of a bank).  A row-sparing architecture would be not
effective at tolerating 65K spare rows for a failed bank, because the
sparing associated table would become impractically large to
build and search on every access. Therefore, DDS implements two
granularities of sparing: either a row or a bank.

\subsection{Budgeting Spare Rows and Spare Banks}

DDS partitions faults into small- and large-granularity faults, then
replaces small-granularity faults with rows and large-granularity
faults with a bank.  Based on the data shown in
Figure~\ref{fig:fault_buckets} we deem any bank having more than four
faulty rows as a bank failure and spare that bank. Given that a bank
can have at most four row failures before the bank gets spared, the
number of spare rows required would be equal to four times the
number of banks (64 banks will have 256 spare rows).

The number of spare banks depends on the bank failure rate. Table~\ref{table:faultyBanks} shows the distribution of faulty banks for a system that has at least one failed bank (more than four row
faults), derived using monte-carlo simulations with FaultSim\cite{faultsim}. Even under our conservative definition of bank failure, Citadel needs at most two spare banks to handle 99.96\% of the systems that
have a bank failure, so DDS employs two spare banks.

\begin {table}[ht]
\centering
\caption{Num. Failed Banks (for system with $\ge$1 bank fail)}{
\begin{tabular}{|c|c|c|c|}
\hline
Number of Faulty Banks & 1  &  2  & 3+ \\ \hline\hline
Probability      & 66.98\% & 32.98\% & 0.04\% \\ \hline
\end{tabular}}
\label{table:faultyBanks}
\end{table}

\subsection{Design of Dynamic Dual-granularity Sparing}

DDS has two components; the spare area and the
redirection table. Because Citadel employs two granularities of sparing
it has two redirection tables; one at row granularity and the
other one at a bank granularity.

\subsubsection{Spare Area}

The metadata die in \textit{Citadel} has 8 banks. TSV and 3DP use 5
banks within the metadata die for storing CRC-32 and TSV-SWAP related
information. DDS uses the three remaining banks for sparing. These 3
banks are partitioned into coarse-granularity sparing banks
(\textit{spare bank-0} and \textit{spare bank-1}) and a fine granularity
bank (\textit{spare bank-2}) for row-based sparing.

\subsubsection{Row Remap Table (RRT)}DDS uses RRT to 
associate faulty row addresses with spare row addresses. Each RRT
entry contains a valid bit (1), the source row ID (16 bits) and a
destination row ID (16 bits). Each fault is tagged with a faulty row address and
its corresponding spare address.  Because DDS supports at most 4 spare rows
for each bank, each bank has 4 entries in RRT. The overhead of RRT for
our 8 die (8 banks per die) system is approximately 1 KB and the RRT
is stored on-chip. A memory access will check the 4 RRT entries of the
given bank for a valid row ID match. On a valid match, the spare
row is accessed.

\subsubsection{Bank Remap Table (BRT)}

If all four spare rows dedicated to a bank get exhausted, and a new
fault appears, then the fault is treated like a large-granularity
(bank) failure and coarse-granularity sparing is invoked. The data
from the failed bank is repaired and relocated to the spare bank. A
two-entry Bank Remap Table (BRT) provides redirection for faulty
banks.  Each BRT entry contains a valid bit, the ID of the failed bank
(6 bit ID), and ID of the spare bank (1 bit spare bank ID, to select one of two spare
banks). The BRT is located on chip, and is probed on every memory
access for a match, prior to looking up the RRT. On a BRT hit, the
spare bank is accessed.

\section{Single Error Correction (SEC) to mitigate correction latency}\label{sec:common}
Several studies have shown that soft errors ($\alpha$ particle strikes), scaling errors and retention errors are usually manifested as single bit errors. Unfortunately, Citadel, in a worst case, can take upto 700 milliseconds to correct such errors. This dissertation also proposes using a Single Error Correction Code (SEC) to optimize Citadel for the common case of single bit errors. Single Bit Error Correction using Hamming Code for a 512 bit cache line requires 10 additional bits.

This section proposes two techniques to implement SEC in Citadel without using additional area;
\begin{itemize}
\item First, do not use an additional bank for sparing small granularity failures. Instead, use the LLC to store values from these failed rows persistently. This will reduce the capacity of LLC by only 256KB (3.3\% for an 8MB LLC).
\item Second, since SEC uses 10 bits, one need space to store these additional 10 bits. To do this, one can employ the unused additional bank (previously used to spare small granularity faults) to store 8 bits per cache line (1 bank every 64 banks). To accommodate two additional bits, one can downgrade CRC-32 (32bits) into CRC-30 (30 bits).\footnote{CRC-30 is already used in CDMA technology} These additional bits is used to store the 9th and 10th bits for SEC.
\end{itemize}

\subsection{Quantifying the effect of using CRC-30 checksum against a CRC-32 checksum}
The chance of a CRC-32 checksum overlapping is $\frac{1}{2^{32}}(\approx10^{-10}$). The baseline design uses CRC-32 to maximize the bandwidth used on the ECC lanes. In the SEC based optimization for soft errors, the CRC-30 checksum will have an overlapping probability of $\frac{1}{2^{30}}(\approx10^{-9}$). Since the probability that an
element fails is much less than 10$^{-6}$. The CRC-30 checksum has a detection probability of $\ll 10^{-15}$ as compared to CRC-32 checksum with  a detection probability of $\ll 10^{-16}$.

\subsection{Operation}
For every access, we will update these 10 bits every cache line using the ECC lanes. SEC based correction works in three steps. On detecting a CRC error, the stacked memory system uses SEC to correct errors. Unfortunately, in case of multi-bit errors, SEC may not be able to correct the error. To avoid this, on correcting an error using SEC, Citadel re-computes the CRC again. In the common case of single bit errors, this will usually result in a CRC match. On a CRC match, Citadel infers that the error is corrected. In case of a CRC mismatch, Citadel denotes this as a multi-bit error and employs the longer latency error correcting of 3DP.

\section{Overall Results}
This section explains the impact of tying together TSV Swap, 3DP and DDS and explains the overheads in implementing Citadel.
\subsection{Tying it together}
\begin{figure}[htb]
\vspace{-0.1in}
  \centering
  \centerline{\psfig{file=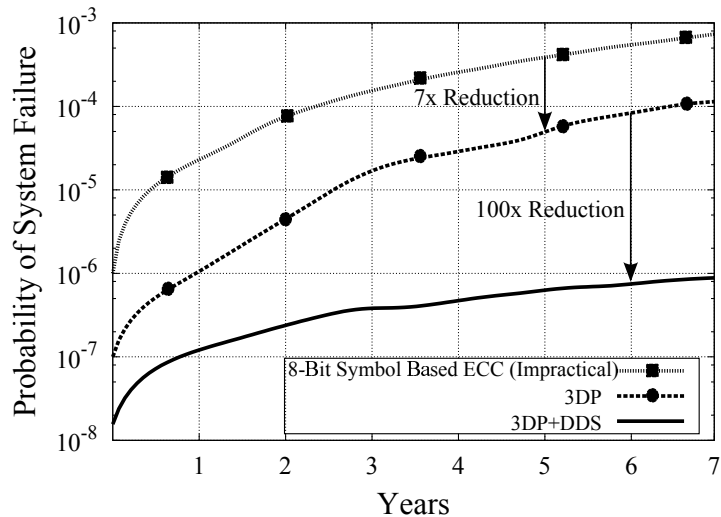,width=3.7in}}
  \vspace{-0.05in}
\caption{Resilience: The schemes 3DP and DDS together provides 700x higher
  resilience as compared to symbol-based codes}
\label{fig:alltogether} 
\end{figure}

Figure~\ref{fig:alltogether} compares the effectiveness of 3DP with
DDS to an 8-bit symbol correcting code. For all systems, we assume
that TSV-SWAP is enabled.  DDS when applied with 3DP delivers a 700x
improvement in resilience compared to the baseline strong 8-bit symbol-based
ECC code. DDS removes 99.995\% of all transient faults and 99.996\% of
all the permanent faults with a 12-hour scrubbing interval and thus
prevents the accumulation of faults.  Therefore, DDS can protect
against multiple faults if they occur during different scrub
intervals. Overall, these results show that Citadel can provide a reliability
 improvement of almost three orders of magnitude. It does so without
requiring the system to stripe data for a cache line across banks.

\subsection{Quantitative Comparison to Prior Work: 6EC7ED and RAID-5}
\begin{figure}[htb]
  \centering
  \centerline{\psfig{file=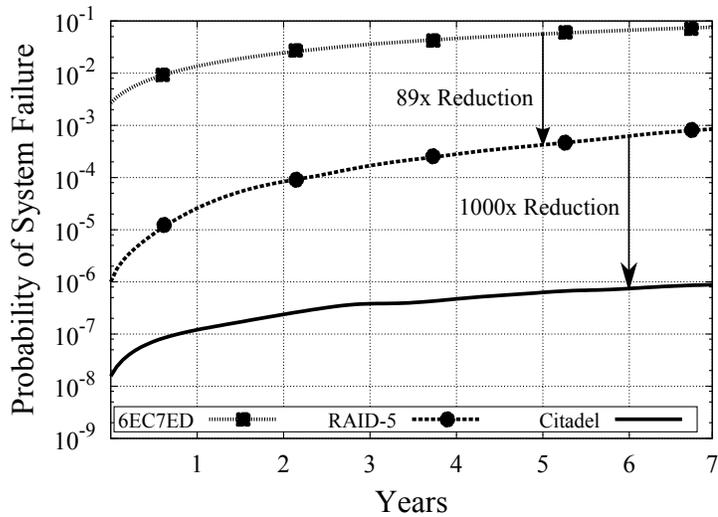,width=3.7in}}
  \caption{Comparing resilience of Citadel to strong ECC codes
  (6EC7ED) and RAID-5.}  \label{fig:related3} 
  \end{figure}

Citadel uses parity for error correction, as do other schemes such as
RAID ~\cite{RAID5:1995}. BCH codes can be used to provide protection for multiple-bit
errors (e.g. 6 or more
bits)~\cite{HP:2011}\cite{wilkerson:isca10}. Unfortunately, strong BCH
codes cannot handle large-granularity faults without significant
overheads. Figure~\ref{fig:related3} compares the resilience of Citadel with a
strong ECC scheme (6EC7ED) and with RAID-5.  Because these schemes are not
resilient to TSV faults, this study assumes a memory system with no TSV faults. Even
after discounting for TSV faults, these schemes end up having orders of
magnitude higher failure rates than Citadel. A RAID-5 scheme
provides 89x improvement in resilience compared to 6EC7ED. Citadel provides
1000x more resilience than a RAID-5 scheme.

\subsection{Storage Overhead of Citadel}
Citadel relies on having an extra die for storing metadata for the
eight data dies (12.5\% overhead).  In addition, bank-level parity
requires dedicating one of the data bank for storing parity (1.6\%
overhead, one bank out of 64 banks).  For 3DP, Citadel keeps parity for
Dimensions 2 and 3 on-chip (34 KB overhead), and the redirection
tables of DDS incur about 1KB overhead, for a total SRAM overhead of
only 35KB.  Thus, Citadel provides 700x better reliability while
requiring a storage overhead of 14\% which is similar to the overhead of ECC DIMM (12.5\%).

\section{Summary}
Memory stacking introduces new multi-bit failure modes, exacerbating
the large-granularity faults identified by DRAM field studies.
Typical approaches tolerate only random-bit
failures and tolerating large-granularity failures (such as tolerating
chip failures using ChipKill) typically relies on striping data to
multiple chips. Transposing such data striping to stacked memory
systems causes significant slowdown and 3-5x power overheads. This
dissertation proposes {\em Citadel} to tolerate large-granularity faults
efficiently, and makes the following contributions:

\begin{enumerate}

\item TSV-SWAP, which mitigates TSV faults at run-time,
without relying on manufacturer-provided spare TSVs. It remains
effective even at high TSV failure rates.

\item Tri-Dimensional Parity (3DP) which can correct a wide
variety of multi-granularity faults.

\item Dynamic Dual-granularity sparing (DDS) which can spare faulty
data blocks either at a row granularity or at a bank granularity to
avoid the accumulation of permanent faults and frequent error correction.

\end{enumerate}

Evaluations with real-world fault data for DRAM chips shows that
combining these three schemes is highly effective for tolerating high
rate of TSV failures and memory failures. This chapter shows that 3DP improves
reliability of stacked memory by 7x, and when combined with DDS by 700x,
compared to a symbol-based code that stripes data across
banks or channels.  Citadel provides high reliability while
maintaining high performance and low power, requiring a storage
overhead close to ECC DIMMs (14\% vs. 12.5\%).


\chapter{Efficient Mitigation of High Rates of Transient Failures}
Current memory systems tend to be power and performance constrained. However, by relaxing some operations in the memory system, one can improve performance and reduce power. For instance, by reducing the refresh-rate of DRAM cells, we can improve performance and reduce refresh power. While, such an optimization opens up avenues for new memory architectures, it also leads to errors that are transient in nature. This chapter discusses low-cost techniques to mitigate transient failures.

New memory technologies can also exhibit transient failures as they scale. To highlight the efficacy of the solutions to mitigate transient failures, this chapter identifies Spin Transfer Torque RAM (STTRAM) as a candidate. STTRAM is a promising technology for building large on-chip caches.  Scaling STTRAM to small feature sizes encounters major impediments such as retention failure. For example, reducing the thermal stability factor of STTRAM cells from 60 to 30 leads to a bit-failure rate as high as 1.9$\times$10$^{-6}$ during a 20ms period.  Unfortunately, conventional means of tolerating retention time failures, such as using  DRAM-style refresh are ineffective for STTRAM, because the failure behavior of retention-time failures in STTRAM resembles transient faults caused by particle strikes. Typically STTRAM failures are tolerated with periodic scrubbing and by provisioning each line with Error Correction Code (ECC).  However, for tolerating a desired error rate, the cache needs ECC-5 (five bit error correction) with each line, incurring unacceptably high storage and latency overheads.  Ideally, we want to tolerate retention failures in STTRAM without relying on multi-bit ECC. This dissertation proposes {\em SuDoku}, a strong reliable architecture that tolerates high-rate retention failures while using a single bit error-correction code (ECC-1) per line. SuDoku provisions each line with a strong error detection code and relies on a region-based RAID-4 to perform correction of multi-bit errors.

\section{Introduction}

Spin-Transfer Torque Random-Access Memory (STTRAM) has emerged as a promising technology that can enable large on-chip caches as it offers 3X-4X as high density as SRAM ~\cite{ITJ:2013}.  STTRAM cells store data in the form of the orientation of a soft ferromagnetic material which changes state with passage of current.  The ability of STTRAM to retain the stored data is dictated by a metric called as {\em Thermal Stability Factor ($\Delta$)}.  While demonstrations of STTRAM have shown that the cells can retain their state for several years, such designs typically use a $\Delta \ge 60$ and require larger cell area, higher energy per write, and long write latencies~\cite{sachin-stt}\cite{ITJ:2013}.  Recent proposals have suggested relaxing non-volatility~\cite{sudhanva-stt}\cite{STTRAM:MICRO2011} of STTRAM for caching applications to achieve larger capacity, lower write energy, and shorter write latency.  Furthermore, relaxing the thermal stability is one of the attractive means to provide dimensional scaling to STTRAM and scale it to smaller technology nodes. Unfortunately, retention time emerges as one of the main limitation to scale STTRAM to small feature sizes~\cite{ITJ:2013}.

This chapter targets STTRAM technology with a $\Delta=30$, as this regime of operation has been of interest to studies in both industry~\cite{ITJ:2013} and academia~~\cite{sudhanva-stt,STTRAM:MICRO2011,sachin-stt}.  The retention time of an STTRAM cell at 300K can be approximated as $1ns\cdot e^{\Delta}$~\cite{Rizzo:STTRAM,sudhanva-stt}.  Therefore, for an STTRAM cell with  $\Delta=30$ we can be expected to have a {\em Mean Time to Failure (MTTF)} of approximately three hours.  This range of retention time may seem more than sufficient for an on-chip cache, however, an LLC contains several millions of cells and the overall reliability of the LLC gets dictated by the failure\footnote{We use terms of failure, fault, and error interchangeably.} rate of all those cells. Furthermore, the retention failure occurs in STTRAM as a result of random thermal noise, which means even though the MTTF of a cell is fairly high, the cell will still fail within a short interval with a non-negligible probability. For example, even with an MTTF of three hours, the failure rate of a cell would be approximately 1.9$\times 10^{-6}$ in a period of 20ms.\footnote{We pick a period of 20ms for our analysis, as this period represents a duration in which a large LLC can be scrubbed while incurring an overhead of not more than a few percent~\cite{ITJ:2013}.  Analysis with other scrub interval is presented in Section~\ref{scrubanalysis}}  Therefore, for our baseline 64MB cache, we can expect on average 1020 bits to fail during each period of 20ms. To enable scaling of STTRAM  to small feature sizes, efficiently mitigating such high rate of retention failures is important.

Retention failures is a problem not only in STTRAM but also in other technologies~\cite{chang2013technology}. For example, DRAM systems rely on periodic refreshes to maintain data integrity.  Prior studies~\cite{sudhanva-stt,STTRAM:MICRO2011} on relaxing retention time of STTRAM have advocated a DRAM-style refresh for STTRAM, whereby periodically each line is read into a buffer and written back.  However, the retention failure occurs in STTRAM and DRAM quite differently.  While in DRAM, retention failure occurs as a result of charge leakage, in STTRAM, it occurs because of a random thermal noise that flips the direction of the magnetic celland. Therefore, unlike the DRAM, in which we can maintain data integrity simply by restoring the charge before it leaks below a certain threshold, we cannot restore the vaule of a STTRAM cell by simply reading and rewriting it. Moreover, the probability that a bit flips because of thermal noise within a given time window does not depend on the duration since the last access. Therefore, DRAM-style refresh is ineffective for STTRAM~\cite{ITJ:2013}.

In essence, retention failures in STTRAM are akin to transient errors in charge-based memories caused by external high energy particles. Prior techniques~\cite{chris:isca08,duwe2016rescuing,Prashant_ISCA:2013,cidra} that are highly effective at handling permanent faults also become inapplicable for such errors as they are transient, and any given cell is liable to incur a bit flip at a certain time interval. Such techniques typically rely on either disabling a faulty bit, or repairing a known bad bit with other data bits. In our case, all bits would be expected to experience a bit failure within several hours of operation, and such schemes would end up decommissioning almost all the bits in the cache.

\begin{figure}[htb]
\centering
    \centerline{\psfig{file=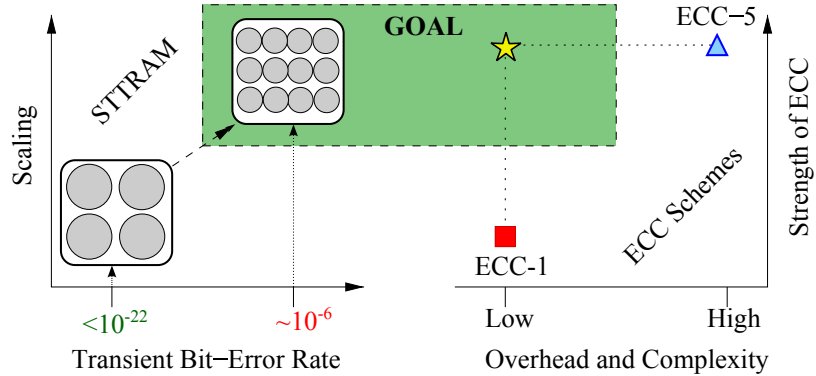,width=0.7\columnwidth}}
  \caption{Challenges for scaling STTRAM. We want to tolerate high error rates (1.9$\times 10^{-6}$), while retaining ECC-1 per line and avoiding the overheads of ECC-5. }
  \label{fig:intro4}
  \vspace{-0.1in}
\end{figure}

A practical solution to mitigate retention failures in STTRAM is employing periodic scrubbing and equipping each line with Error Correction Code (ECC)~\cite{sudhanva-stt,ITJ:2013,sachin-stt}, which should be strength enough to tolerate all the bit failures that occur between consecutive scrub operations. We use a 64MB STTRAM for our studies and employ a scrub interval of 20ms.  We seek a target FIT\footnote{Failures-In-Time, is the number of failures in one billion hours.} rate of one for the cache, which translates to one uncorrectable failure in one billion hours.  For tolerating a bit error rate (BER) of 1.9$\times10^{-6}$ within the scrub interval, each line would need to provisioned with ECC-5 (correcting five errors per line) with each cache line.  The storage overhead of such a design would be 50 bits per 64-byte line (10\%). Furthermore, it would require encoders and decoders for multi-bit ECC which can incur latencies of several tens of cycles.  As shown in Figure~\ref{fig:intro4}, ideally, we would like to tolerate high rate of retention failures by using only ECC-1 and avoiding the overheads of strong ECC.

This dissertation proposes {\em SuDoku},\footnote{SuDoku is a logic-based number placement puzzle, where the data value in a blank cell is constructed using the available data values in row, column, and smaller grid. Our design shares a similar spirit in data recovery for faults, hence the name.}  a resilient cache architecture that efficiently tolerates high error rates without incurring the storage and latency overhead of strong error correction. SuDoku is based on the insight that even at a BER of 1.9$\times 10^{-6}$, only two bits in every Million bits will be faulty. Therefore, 99.9999\% of the cachelines will either have no faulty bits or have only one faulty bit.  SuDoku handles the common case by provisioning each line with a single error correction code (ECC-1).  To handle the uncommon case of multi-bit faults, SuDoku appends each line with cyclic redundancy code (CRC-21) ~\cite{CRC:1961}, a strong detection code that detects up to five faults. For correcting multi-bit failures SuDoku relies on a region-based RAID-4 scheme, whereby each group of 512 lines is provisioned with one dedicated parity line. If SuDoku detects an uncorrectable fault in any line of the group, the parity line associated with the group is used to reconstruct the data.  The likelihood of invoking such a RAID-4-based correction is small (on average, only one line in seven 64MB caches will have a multi-bit error during the scrub interval of 20ms). We refer to this base SuDoku as {\em SuDoku-X} (Section~\ref{sudokux}). SuDoku-X leads to uncorrectable failure when a region of RAID-4 encounters two or more lines with multi-bit failures each, which occurs every 137 seconds on average that is inffered as the MTTF of SuDoku-X.

The dominant failure mode of SuDoku-X occurs when a region encounters two lines, each with exactly two bit-failures.  Traditionally, RAID-4 was unable to perform correction if two units of the region are deemed faulty.  We leverage the insight that we can still faithfully correct 2-bit failures with ECC-1, if we can identify the position of one of the faulty bits and flip that bit. We call this scheme to repair lines with 2-bit faults with ECC-1 as {\em Sequential Data Resurrection (SDR)}.  To perform SDR, we first scan the region of RAID-4 and correct all the lines with single bit faults.  Then, we compute the parity across all the lines in the group and compare it against the stored parity, to identify the bit positions with a mismatch.  For the faulty lines, we sequentially flip each identified position of faulty-bit sequentially and perform ECC-1 based correction.  If after performing the ECC-1 correction, the CRC associated with the line indicates no error, the line is deemed to be corrected successfully.  We refer to this design with SDR as {\em SuDoku-Y} (Section~\ref{sudokuy}).  SuDoku-Y has an MTTF of 129 hours.

SuDoku-Y fails in two situations: First, when a region has two lines each with two faulty bits, whcih overlap, so parity is not be able to detect the their positions. Second, when the region has 3+ faulty lines each with more than two faulty bits, so ECC-1 is unable to correct by flipping one faulty bit. As a result, we use the concepts of skewed-hashing to significantly enhance the effectiveness of SuDoku.  Rather than restricting a line to participate in exactly one parity group, we use two different hash functions and let each line in the cache to map to two separate groups.   If the faulty lines for a given region are uncorrectable under the first hash, SuDoku tries to repair each of the faulty lines using the group formed under the second hash function.  We refer to this design of SuDoku with skewed hashing as {\em SuDoku-Z} (Section~\ref{sudokuz}).

The MTTF of SuDoku-Z is 924 billion hours, which performe 1.8$\times$10$^{6}$ times as reliable as ECC-5 with an MTTF of 2.85 billion hours (0.351 FIT). SuDoku achieves the high reliability without relying the storage and latency overheads of ECC-5.  Unlike ECC-5, which requires 50 bits per line, SuDoku-Z requires 31 bits per line (10 bits for ECC-1 and 21 bits for CRC).  SuDoku also incurs a 128KB overhead for storing parity information for our 64MB cache. The latency overheads of error correction with SuDoku are incurred rarely and do not have any measurable impact on system performance ( $<$0.01\%).  Note that while SuDoku is designed to tolerate high rates of transient faults, it will be effective for tolerating permanent faults too, without the need to know which bits are faulty.

\section{Background and Motivation}
\subsection{Challenges in Scaling STTRAM}
STTRAM provides higher density than SRAM does, and enables large on-chip caches.  An STTRAM cell uses a Magnetic Tunnel Junction (MTJ) layer to store binary data. The MTJ layer can be polarized in two directions. The direction of polarization determines data inside the cell. Once the MTJ layer is polarized, it is susceptible to temperature variations and transient failures. The BER ($p_{cell}$) indicates the robustness of the MTJ-layer to temperature variations. $p_{cell}$ follows a Poisson distribution and depends on the thermal stability factor ($\Delta$) of the MTJ layer. Equation~\ref{eqn:retention} shows the impact of $\Delta$ on $p_{cell}$ for a given time period (t$_{s}$), where $f_{0}$ is the thermal attempt frequency and is nearly 1GHz. For a $\Delta$ of 60 and even after a time period (t$_{s}$) of 10 years, we get tolerable p$_{cell}$ of only 10$^{-19}$ ~\cite{ITJ:2013}.
\begin{equation}\label{eqn:retention}
\begin{aligned}
p_{cell}(t_{s}) = 1 - e^{-\lambda \cdot t_{s}} \text{\textcolor{white}{aaa} \Big( where } \lambda=\frac{f_{0}}{e^{\Delta}} \implies \frac{10^{9}}{e^{\Delta}} \text{ \Big)}
\end{aligned}
\end{equation}

Technology scaling increases memory density by reducing the feature size of STTRAM cells. As cells scale, their $\Delta$ must remain unaffected to maintain a low $p_{cell}$. However, $\Delta$ is proportional to the volume of the free layer (\(V_{f}\)). Thus, as STTRAM scales, maintaining \(V_{f}\) becomes more challenging.~\footnote{DRAM also faces similar constraints, as its cell volume (i.e., capacitance) must be kept constant to maintain the retention-time. To scale DRAM to below sub-20nm nodes, the volume of DRAM cells was reduced by 2x and new memory standards (DDR4 and LPDDR4) now dictate refreshing DRAM cells at 2x the rate.}  While reducing the feature size, if the \(V_{f}\) decreases by 2x, then $\Delta$ also reduces by 2x and increases $p_{cell}$. Table~\ref{table:pcell} shows that as $\Delta$ reduces from 60 to 30, the BER (for a duration of 20ms) increases nearly 13 orders of magnitude to 1.9$\times 10^{-6}$. Retention failures are one of the biggest obstacles of scaling STTRAM~\cite{ITJ:2013,jog2012cache}.
\begin {table}[htb]
\begin{center}
\caption{Thermal Stability vs Error Rate (20ms period)}
\resizebox{0.6\columnwidth}{!}{
\begin{tabular}{| c || c | c | c |}
\hline
Thermal Stability ($\Delta$)		& 60	& 45 &  30 \\ \hline
Bit-Error Rate ($p_{cell}$)		& \cellcolor{Green}10$^{-19}$ 	& \cellcolor{Green} 10$^{-12}$	& \cellcolor{Red} 1.9$\times 10^{-6}$ \\ \hline
\end{tabular}}
\label{table:pcell}
\end{center}
\end{table}

\subsection{Ineffectiveness of DRAM-Style Refresh}
Prior work~\cite{sudhanva-stt,STTRAM:MICRO2011} has suggested tolerating retention failures in STTRAM by applying DRAM-style refresh, whereby the cells are read and rewritten at periodic intervals. Unfortunately, the failure model of STTRAM is quite different from DRAM.  Retention failure in DRAM cell occurs because of gradual loss of charge. In addition, as long as we can restore the charge on the cell before the charge falls below a certain threshold, we can maintain data integrity, which implicitly assumes that the $p_{cell}$ of all the cells is ``0'' within the refresh interval.  However, as shown in Equation~\ref{eqn:retention}, retention-related failures in STTRAM follow a Poisson distribution, and a cell failure occurs abruptly and not gradually.  Even for a short time interval (20ms), the likelihood of cell failure is fairly significant (1.9$\times 10^{-6}$). If a cell flips within the refresh interval, simply reading the same value and rewriting it to the cell will not tolerate failures. Therefore, DRAM-style refreshing is ineffective for STTRAM based memory systems~\cite{ITJ:2013}.

\subsection{Solutions for Handling Permanent-Faults}
Several recent studies~\cite{chris:isca08,chisti:micro09,Alameldeen:2011,qureshi2013operating} have looked at enabling SRAM caches at low voltages by tolerating bit failures. However, these studies are aimed at handling only permanent faults and they rely on precisely knowing which bit fails at low voltage.  Handling a permanent fault rate of 1.9$\times 10^{-6}$ is relatively easy, as only 0.1\% of the lines are expected to have any faulty bit, so we can simply disable these lines.  Unfortunately, the retention failures of STTRAM are akin to transient failures caused by particle strike. Thus, we do not have a prior knowledge of which bit will fail.  Given a $p_{cell}$ of 1.9$\times 10^{-6}$ during 20ms, within a few hours, almost all the cells in the cache would encounter retention failure at least once.  Therefore, prior schemes  that rely on disabling a faulty cells would end up disabling the entire cache. In general, efficiently handling a high rate of transient errors is a more difficult problem than handling permanent faults.

\subsection{Effective Solution: Scrubbing and ECC}
A practical solution to mitigate retention failures in STTRAM is employing periodic scrubbing and equipping each line with a strong enough ECC~\cite{sudhanva-stt,ITJ:2013,sachin-stt} to tolerate all the bit failures that occur between consecutive scrub operations. We use a 64MB STTRAM for our studies and employ a scrub interval of 20ms.  We seek a target FIT rate of atmost one for our cache design, translating to atmost one uncorrectable failure in one billion hours of operation.\footnote{Typically, a Chipkill protected DRAM memory  has a FIT rate slightly exceeding 1 FIT, so our target FIT ensures that the reliability of the overall system  is not dominated by the cache.} To reach the target FIT rate, we need to equip each line with ECC-5, as shown in Table~\ref{table:strength}. Unfortunately, provisioning each cache line with ECC-5 incurs significant overheads in terms of latency and storage, which is 50 bits per line (10\%) for ECC-5. and the encoders and decoders for multi-bit ECC incur latencies of several tens of cycles~\cite{chris:isca08,morphableecc}. Ideally, we would like to use a simple ECC-1 with only 2\% area-overheads and still be able to tolerate five or more faulty-bits in a cacheline.

\subsection{Insight: Optimize for Common Case}
As shown in Table~\ref{table:strength} the likelihood of multi-bit error is very uncommon.  For example, even if each line was provisioned with ECC-1, only one line out of two 64MB caches would fail.  Therefore, on average, a single 64MB cache (1 million lines), is expected to have a line with multi-bit fault every seven scrub intervals.  Unfortunately, we do not know which line would encounter the multi-bit failures. Moreover the lines with multi-bit faults will change between scrub intervals.  As we lack the knowledge of which line will encounter failures, the prior work on tolerating STTRAM failures~\cite{sachin-stt,ITJ:2013} naively allocated uniform amount of error correction entries with each line, and thus incur significant ECC overheads.  Our insight for reducing the overhead of tolerating high error rates is to give lines enough ECC entries to tolerate the common case (ECC-1). We equip each line with strong detection code (CRC-21) to detect the episode of multi-bit failures and rely on an alternate low-cost mechanism (region-based RAID-4 in our case) to perform correction.

\begin {table*}[htbp]
\begin{center}
\caption{FIT Rate of 64MB Cache for various ECC schemes (BER of $1.9\times 10^{-6}$ in scrub interval of 20ms)}{
\resizebox{\columnwidth}{!}
{
\begin{tabular}{| c || c | c | c | c | c |}
\hline
ECC code per line& \textbf{ECC-1}  &  \textbf{ECC-2} &  \textbf{ECC-3}&  \textbf{ECC-4}&  \textbf{ECC-5} \\  \hline \hline

Probability of line-failure in 20ms   & 4.8$\times$10$^{-7}$  &  1.7$\times$10$^{-10}$  &   4.4$\times$10$^{-14}$          &    9.8$\times$10$^{-18}$             &     1.9$\times$10$^{-21}$             \\ \hline

Probability of cache-failure in 20ms   &  4$\times$10$^{-1}$    &  1.7$\times$10$^{-4}$   &   4.6$\times$10$^{-8}$  &    1$\times$10$^{-11}$  & 2$\times$10$^{-15}$ \\ \hline

\textbf{Cache FIT-Rate}  & $>10^{11}$   &      3.11$\times$10$^{10}$      &   8.3$\times 10^{6}$  &     184           &   0.351                      \\ \hline

\end{tabular}}
}
\label{table:strength}
\end{center}
\end{table*}

\section{Sudoku-X: Base Design}~\label{sudokux}

This dissertation propose {\em SuDoku}, a resilient architecture that tolerates high rate of transient failures at low cost.  Before discussing our enhancements of SuDoku, we first explain the basic design, which we call {\em SuDoku-X}.  Our solution is based on the insight that even at a BER of 1.9$\times$10$^{-6}$, only two in every Million bits will be faulty. Therefore, 99.9999\% of cache lines will either have zero or one faulty bit.  SuDoku handles the common case by provisioning each line with an ECC-1, and provides an alternate means to provide correcting multibit errors.  The cache employs periodic scrubbing.  Unless specified otherwise,  we use a scrub interval of 20ms for our studies and a BER of 1.9$\times 10^{-6}$ within the scrub interval  (sensitivity to these parameters is provided in Section~\ref{sec:results}). We use a 64MB cache with a 64-byte lines.

\begin{figure}[htb]
\centering
    \centerline{\psfig{file=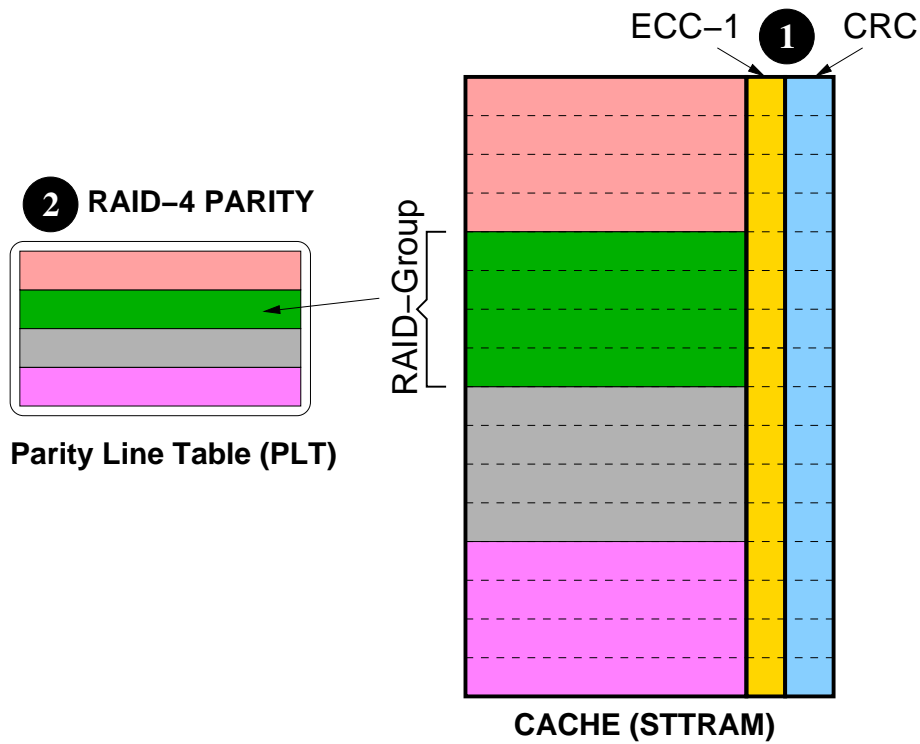,width=0.8\columnwidth}}
  \caption{The Organization of SuDoku-X. Each line is equipped with an ECC-1 and CRC-21. RAID-4 corrects multi-bit failures. The PLT stores the parities.}
  \label{fig:sudoku-x}
\end{figure}

\subsection{ The Organization}

Even at a high BER, only a few lines would encounter multi-bit failures. For example, only one line within two scrub intervals of the 64MB cache would be expected to have multi-bit failures at a BER of 1.9$\times 10^{-6}$.  SuDoku-X provides two levels of protection -- one to handle single-bit failure (common case) and another to handle multi-bit failures (low cost).  Figure~\ref{fig:sudoku-x} shows the organization of the cache with SuDoku-X.  Each line is equipped with an ECC-1 to handle the case of one bit error locally and quickly. Each line is also provisioned with a strong error detection code, CRC-21, which detects up to five errors in a line (all even number of errors beyond five bits can still be detected with high probability, but not guaranteed).  The CRC-21 requires a storage overhead of 21 bits per line.  When the line encounters a multi-bit error and the CRC detects it, we need an alternate mechanism to provide correction.  To achieve correction of multi-bit errors at low cost, we use a scheme based on the concept of RAID~\cite{gibson1992redundant,RAID5:1995}, more specifically RAID-4.

RAID-4 corrects multi-bit failures without requiring the complex encoders and decoders for strong ECCs. However, the limitation of RAID-4 is that it can correct only one faulty line within the protected region.  In our case, we expect several lines (two on average) with multi-bit failures in ten scrub intervals.  If we have a RAID-4 across all the lines, we will be unable to correct them.  Therefore, we partion the cache into several equal-sized regions, called {\em RAID-Group}, each provisioned by a parity line for RAID-Group. In other words, this line maintains the parity information for all the lines in the RAID-Group. For example, in Figure~\ref{fig:sudoku-x} the cache contains 16 lines, which are split into four RAID-Groups of four lines each.  The parity for each Raid-Group is maintained in a separate structure called the {\em Parity Line Table (PLT)}.  We use a default size of 1024 lines for the RAID-Group, so the PLT is much smaller than the cache ($0.1\%$).  As each write to the cache must also update the PLT, one can provide sufficient bandwidth to the PLT, either by making it heavily banked or in SRAM or both.  A line with multi-bit error can be repaired using the respective parity line stored in the PLT and the RAID-4 scheme.\footnote{In general, RAID-5 is more popular than RAID-4, as it can stripe the parity information across all the diss and avoid the parity disk becoming the  bottleneck.  In our case, only a single line can be accessed from a bank at the given time, so these lines are not independent read/write units (their concurrency is limited by the number of banks in the cache). As long as we have the same number of banks in the PLT as there are in the cache, we can avoid parity updates of the PLT becoming the bottleneck.}

\subsection{Error-Free Operation}
This section explains read and write operations to the cache that implements SuDoku-X.  Then, the next section will explain the correction scheme. 

For a read operation, the cache reads the ECC-1 and CRC-21 along with the dataline.  The cache controller  first checks if the line is faulty using its CRC, which requires checking the syndrome for CRC, so can be performed within one cycle. As long as the syndrome is ``0'', the line is deemed to be non-faulty, and data can be sent to the processor.  Note that PLT is not accessed for a read operation. 

For a write operation, the cache controller must update data in the stored cache line, as well as the associated parity information in the PLT.  These updates can be performed as two sequential read-modify-write operations. STTRAM usually employ a read-modify-write scheme to reduce the number of bit-flips and reducing write power and latency ~\cite{cho2009flip, loh2013memory}.  The first read-modify-write is to the dataline in the cache. As part of this operation, the controller identifies the position of the bits get modified due to the write.  The second read-modify-write is to the respective parity line in the PLT, and flips the bits corresponding to the locations for which data bits had changed.

\subsection{Performing Error Correction}

This section examines how SuDoku-X performs correction.  When a line is accessed, the CRC associated with the line will detect possible errors.  The repair depends on whether the line encountered a single-bit or multi-bit fault.

\subsubsection{ Repairing Single Bit Faults}  As the most common case of faults is a single-bit failure, once we get a CRC error, we first try the ECC-1 based repair for the line.  If the ECC-1 performed a correction, we recompute the CRC using the corrected data value. If the CRC does not detect any error, we deem the line to be corrected successfully.  This corrected line is sent to the processor, and written back to the cache.

\subsubsection{ Repairing Multi-Bit Faults}  If a line encounters a multi-bit failure, then even after undergoing ECC-1, the CRC still deems the line to be faulty. For correcting multi-bit failures SuDoku relies on a RAID-4 scheme, whereby each group of 1024 lines is provisioned with one dedicated parity line. To perform this correction, we first read all the lines in the RAID-group (and fix an single bit failures that are encountered).  Then, we compute data for the faulty line by computing the parity of over the parity line and all the lines in the RAID-Group, except for the line being repaired. The likelihood of invoking such a RAID-4 based correction is small (on average, only two lines have multi-bit error after ten scrub intervals of 20ms).

\begin{figure}[htb]
\vspace{-0.1in}
\centering
    \centerline{\psfig{file=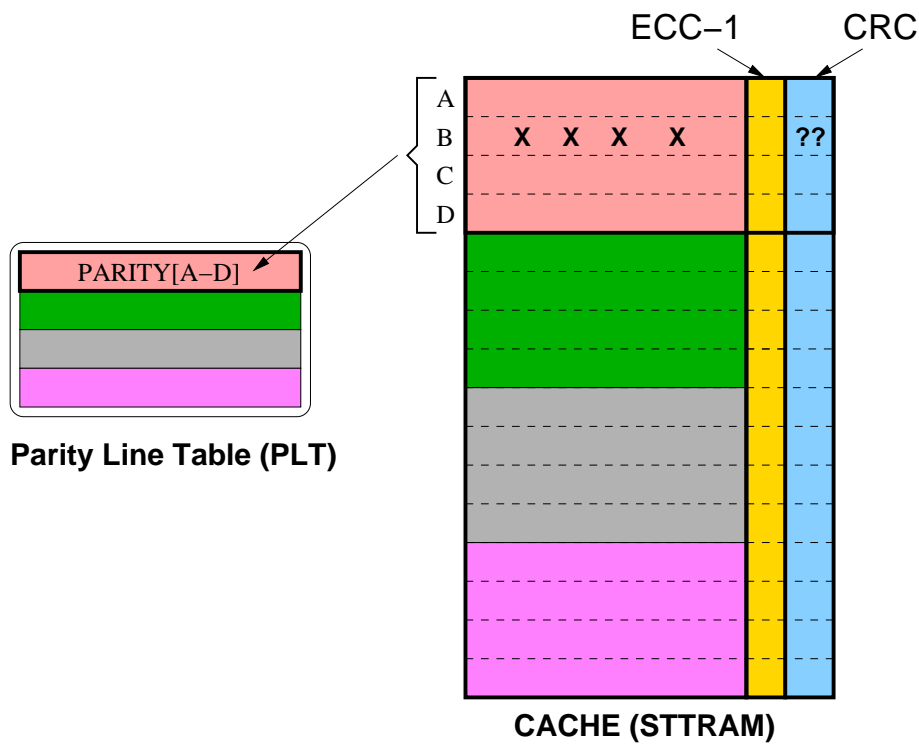,width=0.8\columnwidth}}
  \caption{Correction of multibit failures with SuDoku-X. Line B encounters a multibit error, which is detected by CRC. The data for B is repaired by exoring the data for lines A,C,D and the respective parity line from the PLT.  }
  \label{fig:sudoku-x-correct}
\end{figure}

This paragraph explains the repair of multi-bit failures with SuDoku-X using an example.  Figure~\ref{fig:sudoku-x-correct} shows a cache with 16 lines, each RAID-Group of which contains four lines.  Lines A-D form a RAID-Group and the parity line for this group is the top-most line in the PLT.  Line B encounters a four-bit failure, which is detected by the CRC.  Even after undergoing an ECC-1-based correction, the CRC still indicates error.  As a result, we reconstruct the data for B by computing the parity of lines A, C, D, and the parity line.  If a line encounters any single-bit error, then such an error is first corrected before participating in the RAID based correction. This way, we can repair the multi-bit failure in Line B, without requiring any storage or circuitry for multi-bit ECC.

\subsection{Considerations on Size of RAID-Group}

We use a default size of 1024 lines for the RAID-Group.  The size of the RAID-Group determines the storage overhead for storing the parity lines, the latency for performing error correction using RAID-4, and the overall reliability of the scheme.  With a RAID-Group of 1024 lines, the RAID-4 based correction would incur a storage overhead of 64KB for a cache of 64MB (the 64MB cache has 1K groups of 1024 lines). This storage overhead is sufficiently small to be stored in SRAM. Furthermore, the latency overhead of repairing using RAID-4 (1024 line reads) is incurred infrequently -- on average two lines over ten scrub intervals of 20ms.  This repair latency (of approximately eight micro second per repair) is usually encountered only once in every ten scrub intervals. Even if we encounter all of the repairs on demand read, the overall latency impact would be less than 0.001\% (eight microsecond every ten scrub intervals of 20ms).  

\subsection{SDC Rate of SuDoku-X}

The Silent Data Corruption (SDC)  of SuDoku-X is dictated by the error detection capability CRC-21, which detects all errors up to five bits per line. For 6+ bit errors, CRC-21 has a small misdetection probability of 2$^{-21}$. Unfortunately, with SuDoku-X, a line with 5 bit-error can get miscorrected to a line with six-bit error by the ECC-1,  and subsequently the CRC-21 can let this event go undetected. In fact, this is the dominant source of SDC in SuDoku-X. Table~\ref{table:sudokux} lists the SDC Rate (over a billion hour period) for cases when the line has either five errors or 6+ errors.  Note that, if the line has four or fewer errors, SuDoku-X will not result in SDC, as CRC-21 is guaranteed to detect. The total  SDC FIT-Rate of SuDoku-X is 0.0009, about three orders of magnitude lower than that of our target goal that is  one FIT.  Thus, SDC FIT-Rate of SuDoku-X is not a concern.

\begin {table}[htp]
\begin{center}
\caption{SDC Rates of Cache with SuDoku-X}{
\resizebox{0.7\columnwidth}{!}{
\begin{tabular}{| c | c | c | }
\hline
\textbf{Vulnerability} & 5 Faults/Line & 6+ Faults/Line \\ \hline \hline
Event  (per Billion Hours) & 1840 & 0.4  \\ \cline{1-3}
CRC-21: Prob. of Misdetection & 2$^{-21}$ & 2$^{-21}$ \\ \hline \hline
\textbf{SDC Rate (per Billlion Hours)} & \textbf{8.8$\times$10$^{-4}$} & \textbf{1.7$\times$10$^{-7}$} \\ \hline
\end{tabular}
}
}
\label{table:sudokux}
\end{center}
\end{table} 

\subsection{Limitations of SuDoku-X: DUE Rate}
SuDoku-X leads to uncorrectable errors when a single RAID-Group encounters two or more lines, each with multi-bit failures, causing an episode of Detected Unrecoverable Error (DUE).  Even though, there are only a few lines with multi-bit failure exist (on average, a line with multi-bit failure occurs every 200 ms) and we have a large number of RAID-Groups (1K), it is just a matter of time before we encounter a RAID-Group with multiple lines with multi-bit failure. On average, the situation of getting multiple faulty lines (with multi-bit faults) within the same RAID-Group happens once every 137 seconds, which is equivalent to a DUE FIT rate of 22 Billion. The total FIT Rate of SuDoku-X is dominated by the DUE rate, and results in an MTTF of SuDoku-X of only 137 seconds.  We discuss extension that can increase the effectiveness of SuDoku significantly in the next Section.

\section{SuDoku-Y: Data Resurrection}~\label{sudokuy}

The dominant failure mode of SuDoku-X is when two lines have two errors each and they map to the same RAID-Group.  Out of all cases of failures with SuDoku-X, the case of exactly two faulty lines (with multi-bit errors) accounts for 99.98\% of the cases. Furthermore, even in the case of multi-bit failures, the case of two bit failures dominates the rest (99.98\% versus 0.02\% for all the rest).  In this section, we focus on how to correct faulty data in such scenarios without any extra storage overhead. We leverage the insight that we can still faithfully correct two-bit failures with ECC-1, if we can identify the position of one of the faulty-bits and flip that bit. This technique is called {\em Sequential Data Resurrection (SDR)}, and SuDoku-X is equipped with SDR as {\em SuDoku-Y}.

\subsection{Overview of SDR}

In general, RAID-4 schemes can only tolerate one failure.  Data for the failed disk is recreated by computing the parity of all other disks.  However, if two disks fail, we cannot repair data using RAID-4.  We leverage the insight that unlike disk failures, in our design we are handling bit errors, and when we state that the line has uncorrectable faults, still the overwhelming number of bits in the line are still fault free (for example, in the typical case of two-bit error, 510 bits are still error-free).  For a line with two faults, if we can identify the position of one of the faulty bit, then we can perform correction for the line by flipping that bit and employing the ECC-1. The correction that is performed can be checked with the CRC associated with the line to make sure that the correction is indeed successful.  In case of SuDoku-X, when there are multiple lines with multi-bit failures, the parity of the RAID-Group be used to identify the location of faulty bits because such bits can lead to parity mismatch (in the common case).  We can use the bit positions of the mismatch to correct lines with two-bit errors using ECC-1. This can be done by flipping each of the bits in the mismatch positions one by one and then performing the correction with ECC-1 and checking with CRC. If the CRC does not match, we check with the next bit position.  We try this until all the bit positions for which the parity mismatched are exhausted.  Note that if we can correct even N-1 faulty lines out of the N faulty lines of a  RAID-Group using SDR, we can correct the final uncorrectable line using the RAID-4 based correction.  We analyze the effectiveness of SDR for the case of two faulty lines in the group, with two faults each.

\subsection{Operation of SDR for Two Faulty Lines}

\begin{figure*}[htp!]
\centering 
    \centerline{\psfig{file=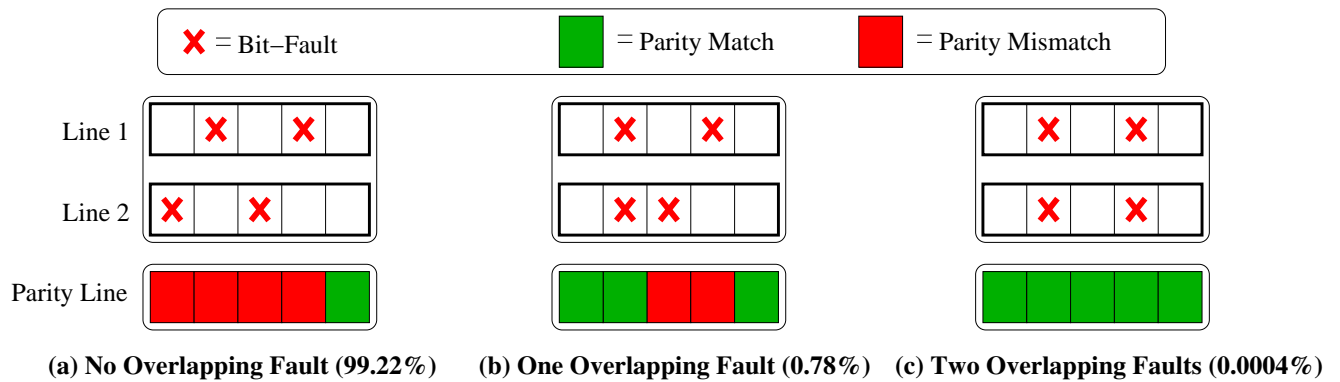,width=7in}}
  \caption{Three scenarios for Selective Data Resurrection using SuDoku.  In general ECC-1 cannot repair lines with two faults. However, if we know the position of one of the faults (from the Parity Line) we can correct using ECC-1 by flipping one of the faulty bit (the CRC of line can validate if the correction was indeed successful). }
  \label{fig:sdr}
\end{figure*}

If a region has two faulty lines, each with two faulty bits, then only a maximum of four locations will encounter a mismatch in the parity line.  The parity is computed by correcting all lines with 1-bit error in the group, and by using the original (uncorrected data values) for both the faulty lines with two-bit errors.   Figure~\ref{fig:sdr}  illustrates a scenario in which two lines (Line 1 and Line 2) that are part of the same RAID-Group encounter two faults.  For simplicity, lets assume none of the other lines in the group encounter any faults.  In the rest, we explain the operation of SDR for three possible scenarios that can occur.

\textbf{Case 1: No overlapping faults (99.22\% probability)}\newline
Figure~\ref{fig:sdr}(a) shows the scenario in which no overlapping faults in two lines occurs. In this case the parity line generates four mismatch locations that can be faulty in each line. The SDR then fetches Line-1 and sequentially tries to flip only the bits in the locations that are mismatch in the parity line and invoke ECC-1. If the bit-flipped was indeed faulty, then ECC-1 can correct the remaining faulty-bit. The CRC will indicate that the cacheline is non-faulty. As Line-1 is no longer faulty, Line-2 can be corrected using RAID-4.

\textbf{Case 2: One overlapping fault (0.78\% probability)}\newline
Figure~\ref{fig:sdr}(b) shows the scenario in which one overlapping fault occurs. In this case, the parity line will have only two mismatches. SDR fetches Line-1 and sequentially tries to flip only the bits in the locations that are mismatch in the parity line and invoke ECC-1. If the flipped bit was indeed faulty, then ECC-1 can correct the remaining faulty-bit. CRC-21 will indicate that the cacheline is non-faulty.  Note that even if the location of one faulty-bit was unknown, we were still able to correct both the faulty bits. The CRC will indicate that the cacheline is non-faulty. As Line-1 is no longer faulty, Line-2 can be corrected using RAID-4.

\textbf{Case 3: Both faults overlap (0.0004\% probability)}\newline
Figure~\ref{fig:sdr}(c) shows the scenario in which both faults overlap. In this case, the parity line will not have any mismatch and SDR cannot be applied. The likelihood that two faulty bits of one line (512 bits) will  overlap exactly with the two faulty bits of another line is quite low ($\frac{2}{512} \cdot \frac{1}{511}$ = 0.0004\%).

The latency of SDR-based correction is only a few cycles of try and error on mismatch position (4-6), so it is in the regime of few tens of nanosecond. However, this latency is incurred once every 137 seconds (the MTTF of SuDoku-X), so the overall latency impact remains negligible.

\subsection{Effectiveness of SDR in Other Cases}

\begin{figure}[htb]
\centering 
    \centerline{\psfig{file=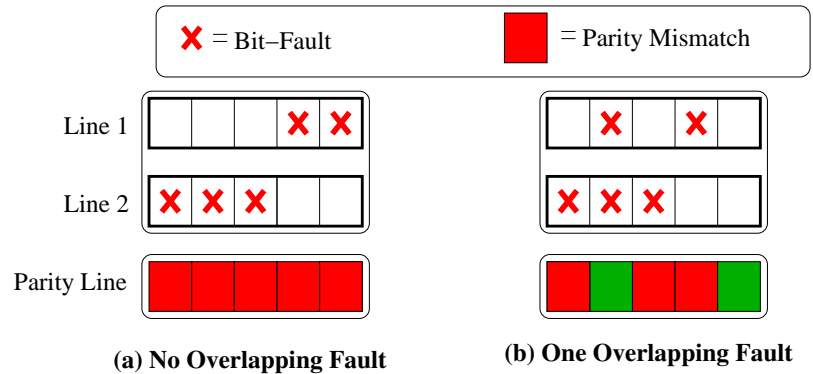,width=0.7\columnwidth}}
  \caption{SDR can repair a line with 3-bit fault if it does not have less than 1 bit of overlap with a line with 2-bit fault. }
  \label{fig:sdr23}
\end{figure}

SDR is highly effective in the case of two faulty lines, each with two faulty bits, as it can repair both lines in 99.9996\% of the cases.  However, these are not the only scenarios where SDR is effective.  It may seem that SDR is unable to repair if one of the line has 3 or more faulty bits.  However, the most common case of this is when there are two faulty lines in the group, one of them has two faulty bits and the other three or more faulty bits, as shown in Figure~\ref{fig:sdr23} (if two faulty bits overlap then SDR cannot repair). If we can repair Line 1 using SDR, then we can repair Line 2 using RAID-4.

Similarly, we have so far only analyzed cases where the RAID-Group has only two faulty lines. But SDR is useful in other cases too. For example, if the are three faulty lines with two bit failures each (the most common case of three lines with multi-bit failure), we can repair each of the faulty lines using SDR. Just that in this scenario, we would have six possible position of mismatch and each line will sequentially undergo repair through these six possible locations.  The effectiveness of SDR even in this case is 99.9\%.  We do not perform SDR if there are more than six mismatches.

\subsection{SDC Rate of SuDoku-Y}~\label{SDCY}
Correction is invoked under SuDoku-Y only when SuDoku-X encounters muliple lines with multi-bit failures.  Once such correction is invoked, it is extremely unlikely for SuDoku-Y to encounter Silent Data Corruption, as it would mean each of the miscorrected line goes undetected by the CRC-21 and these miscorrected lines also go undetected in the Parity Line of the RAID-Group (about $10^{-25}$ probability).  The dominant scenario for SuDoku-Y to cause SDC is identical to that of SuDoku-X (one line in the group has 5+ errors and the CRC-21 is unable to detect).  As per Table~\ref{table:sudokux}, the total SDC rate of SuDoku-Y is also 0.0009, about three orders of magnitude lower than our target goal of one FIT.  Thus, SDC rate of SuDoku-Y is not a concern. 

\subsection{Limitations of SuDoku-Y: DUE Rate}
SuDoku-Y encounters DUE when a SDR fails to perform correction. This occurs in two scenarios. First, when there are multiple faulty lines and at least two of them have three or more errors. Second, when two faulty bits overlap. As SuDoku-Y fixes most multi-line failures, it has an MTTF of 129 hours (3390X higher than SuDoku-X) and provides a DUE FIT of 6.5 Million. The next section describes a scheme that reduces the FIT-Rate to be less than 0.001.

\section{SuDoku-Z: Skewed-Hash For RAID}~\label{sudokuz}

SuDoku-Y fails when a RAID-Group contains multiple faulty lines each with more than two-bit error.  In such a case, SDR is unable to correct the faulty lines, as these lines have three or more faults, so identifying one faulty bit position will not enable the line repair using the per-line ECC-1.  We leverage the concepts of skewed-hashing~\cite{sardashti2014skewed} and multi-hash Bloom Filters~\cite{bloom1970space} to enhance the effectiveness of SuDoku.  The SuDoku designs described thus far restrict a given cache line to map to exactly one RAID-Group. Rather than restricting a line to participate in exactly one RAID-Group, we use two hash functions (Hash-1 and Hash-2) to let each lines participate in two different RAID-Groups. If the faulty lines for a given RAID-Group is deemed uncorrectable under Hash-1 (the case of failure for SuDoku-Y), then this design tries to repair each of the uncorrectable lines using the RAID-Groups formed under Hash-2.  This design of SuDoku with skewed hashing is called {\em SuDoku-Z}.\footnote{Although we implement SuDoku-Z along with SuDoku-Y, we can implement SuDoku-Z alone too. Such a design will not be as effective because of the high DUE rate, causing a FIT rate of 935.}

\subsection{Organization}

\begin{figure}[htb]
\centering
    \centerline{\psfig{file=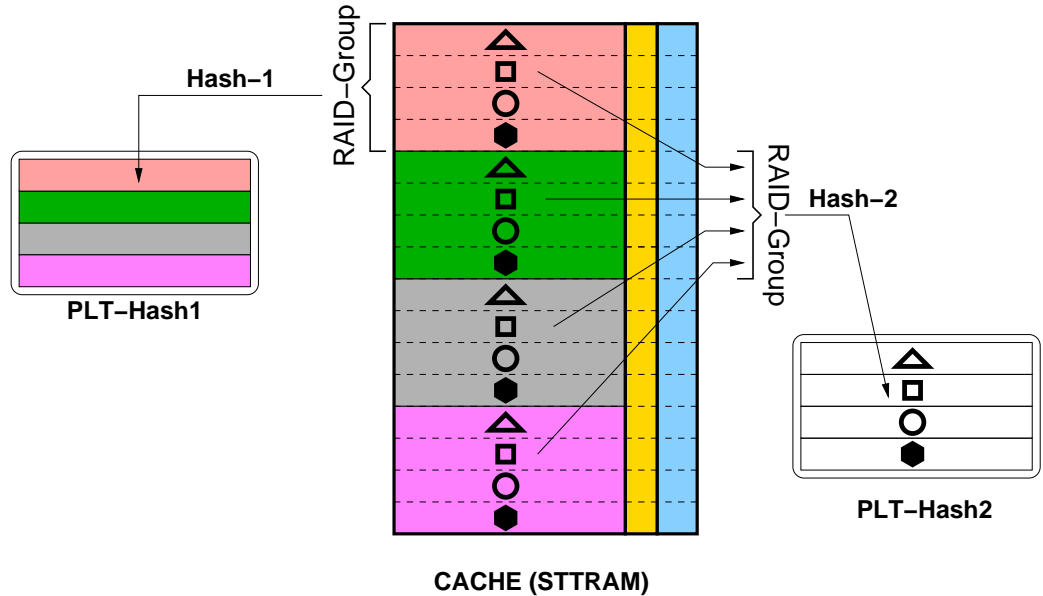,width=0.9\columnwidth}}
  \caption{Organization of SuDoku-Z using two Hash functions: Hash-1 and Hash-2. A newly added structure (PLT-Hash2) stores parity lines of RAID-Groups formed under Hash-2. SuDoku-Z performs correction with Hash-2 only if correction fails under Hash-1. }
  \label{fig:sudoku-z}
\end{figure}

Figure~\ref{fig:sudoku-z} shows the organization of SuDoku-Z.  SuDoku-Z contains two Parity Line Tables (PLT-Hash1 and PLT-Hash2).  The lines are mapped to the two PLT using two Hash functions, Hash-1 and Hash-2.  PLT-Hash1 stores the parity lines of the RAID-Groups formed under Hash-1 (identical to SuDoku-Y).  Similarly, PLT-Hash2 stores the parity lines of the RAID-Groups formed under Hash-2 (newly added for SuDoku-Z).  The Hash functions are selected such that the lines that are mapped to a given RAID-Group under Hash-1 are guaranteed to map to different RAID-Groups under Hash-2.  This avoids the same set of lines from making a RAID-Group fail under both Hash-1 and Hash-2.  For example, if the size of the RAID-Group is 512 lines, we can form Hash-1 by masking out the 9 least significant bits (CacheLineAddr[7:0]) of the cache line address, and Hash-2 by masking out the next nine least significant bits (CacheLineAddr[15:8]) of the cache line address. To keep parity information on both the PLT updated,  each write into the STTRAM cache must now update both PLT-1 and PLT-2.

Figure~\ref{fig:sudoku-z} shows an example of  a cache with 16 lines implementing SuDoku-Z.  The size of the RAID-Group is four lines.  Under Hash-1, the consecutive four lines (same color) form a RAID-Group, and their parity is stored in PLT-Hash1.  Under Hash-2, every fourth line (same symbol) form a RAID-Group, and their parity is stored in PLT-Hash2. Note that if a line shares a RAID-Group under Hash-1 with another line, then it does not share a RAID-Group with that lines under Hash-2.  This helps with correction of the faulty lines -- if a set of faulty lines are unrepairable under Hash-1, then those lines are guaranteed to map to different RAID-Groups under Hash-2, and can undergo a correction of those RAID-Group by applying SuDoku-Y on them. skewed-hashing of RAID for SuDoku-Z  is highly effective because the likelihood of a faulty line mapping into two different RAID-Groups that are both uncorrectable is extremely small.

\subsection{Repairing Faulty Lines}

The correction of SuDoku-Z is invoked only if the SuDoku-Y-based correction fails for the RAID-Group formed using Hash-1. Note that this is a relatively infrequent event, and occurs once every 129 hours on average.  When this occurs, we first identify the set of lines that are unrepairable under Hash-1.  Then, for each such line, we try to  repair using the RAID-Group under Hash-2. For a line to be deemed uncorrectable under SuDoku-Z, it will have to be unrepairable under both Hash-1 and Hash-2.  In fact, if there are N unrepairable lines in a RAID-Group under Hash-1, and we are able to repair say N-1 lines under Hash-2, then we can repair the remaining uncorrectable line by using Hash-1 and the corrected values for all the remaining faulty lines.  As the RAID-Group will have only one faulty line, the RAID-4 based correction will be able to correct the line.  Thus, for SuDoku-Z to fail, we must have two lines that are uncorrectable under both Hash-1 and Hash-2. As we will see this is an extremely unlikely scenario.

\begin{figure}[htb]
\centering
    \centerline{\psfig{file=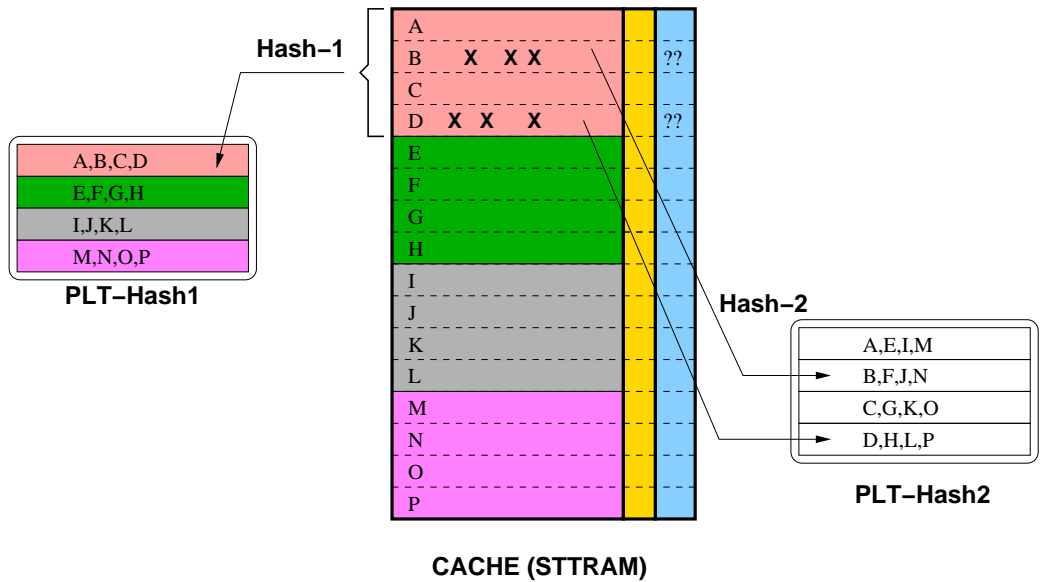,width=0.9\columnwidth}}
  \caption{Correction with SuDoku-Z. Lines B and D have and uncorrectable failure under Hash-1.  Under Hash-2, they map to different RAID-Group and can be corrected.   }
  \label{fig:sudoku-z-correct}
\end{figure}

We explain the correction of SuDoku-Z with an example.  Figure~\ref{fig:sudoku-z-correct} shows a cache with 16 lines (A-P) which use two hash functions, Hash-1 and Hash-2, including two lines (B and D) with three faulty bits that reside in the same RAID-Group under Hash-1. Correction under Hash-1 fails.  By design, B and D map to different RAID-Groups under Hash-2.  B can perform correction under the RAID-Group under Hash-2 that is formed with lines B,F,J and N.  If this RAID-Group can be corrected using SuDoku-Y then line B can be repaired. Similarly, can Line D can perform correction under the RAID-Group under Hash-2 that is formed with lines D,H,L and P. If this RAID-Group can be corrected using SuDoku-Y then line D can be repaired. In fact, even if one of the line can be repaired (say only Line D can be repaired), then we can use the corrected value of that line to repair the other line (using corrected Line D under Hash-1 to repair Line B). SuDoku-Z fails only if both lines are deemed uncorrectable under both Hash-1 and Hash-2.

The analysis can be extended to the case when there are more than two uncorrectable lines in a RAID-Group.   For example, consider there are N faulty lines in a RAID-Group formed under Hash-1.  Then , we will try correction for all N lines under Hash-2. As long as at least (N-1) lines can be corrected using Hash-2, we will be able to repair all N lines. 

\subsection{SDC Rate of SuDoku-Z}
Correction is invoked under SuDoku-Z only when SuDoku-Y encounters an uncorrectable error.  The likelihood that this correction will yield an undetected error is negligible (miscorrected lines  go undetected by CRC-21 and Parity Lines match under both Hash-1 and Hash-2).  The dominant cause of SDC for SuDoku-Z is identical to that of SuDoku-X (one line has 5+ errors and the CRC-21 is unable to detect).  From Table~\ref{table:sudokux}, the SDC Fit-Rate of SuDoku-Z is also 0.0009,  three orders of magnitude less than our target of 1 FIT.

SuDoku-Z encounters a DUE when the faulty line cannot be corrected using both Hash-1 and Hash-2. Given the likelihood of a group failing is quite small (nearly 4.3$\times 10^{-11}$), the likelihood that a line fails under both Hash-1 and Hash-2 is extremely small, and for the system to fail, we will need two of such lines. The DUE FIT-Rate of SuDoku-Z is 2$\times 10^{-7}$ (32 trillion times smaller than SuDoku-Y). 

\begin{figure}[htb]
\centering
\vspace{-0.15in}
   \centerline{\psfig{file=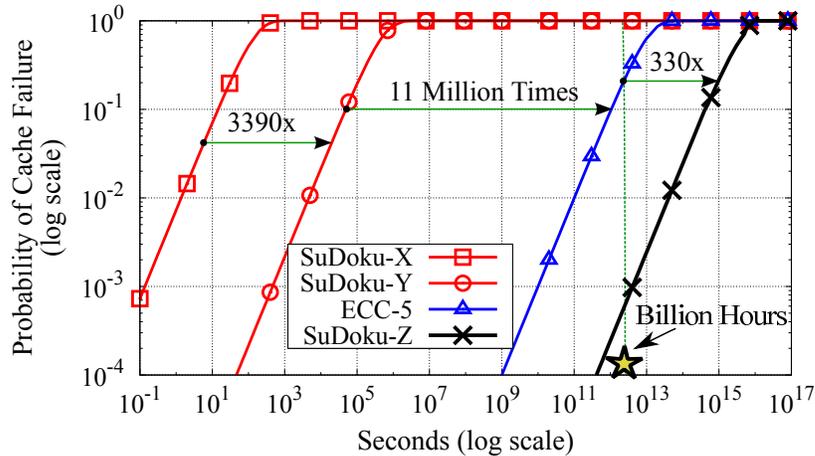,width=0.7\columnwidth}}
     \vspace{-0.15in}
  \caption{The probability of cache failure (DUE+SDC) with SuDoku-X, SuDoku-Y, SuDoku-Z, and ECC-5. Note, SuDoku-Z has 330x as high MTTF as ECC-5.}
  \label{fig:mttf-z}
  \vspace{-0.3in}
\end{figure}

As the DUE FIT-Rate of SuDoku-Z is 4500x as small as the SDC FIT-Rate, the total FIT-Rate of SuDoku-Z is determined by its SDC Rate. Thus, SuDoku-Z has a total FIT-Rate of 0.0009, two orders of magnitude lower than ECC-5. As shown in Figure~\ref{fig:mttf-z}, the Mean Time to Failure (MTTF) of SuDoku-Z is about 330x as high as that of ECC-5. Note that SuDoku-Z provides this level of resilience without requiring the storage and latency overheads of ECC-5. 

\section{Results and Analysis}~\label{sec:results}
Thus far, we have performed analysis using a 64MB STTRAM Cache, which employs a scrub interval of 20ms, and encounters a BER of 1.9$\times$10$^{-6}$.  We perform sensitivity studies to these parameters. We also provide performance and power evaluations. For all our reliability evaluations, we use analytical models to report FIT-Rates.

\subsection{Impact of Scrub Interval}~\label{scrubanalysis}
We used a scrub interval of 20ms, which is in line with the
recommended scrub period for a 64MB STTRAM cache to keep
the cache bandwidth overheads to within a few percent~\cite{ITJ:2013}. Reducing the scrub interval reduces the BER (almost linearly), however it increases the time the cache is busy doing scrub operations.  Table~\ref{table:scrubres} shows the impact
of varying the scrub interval from 10ms to 80ms on the FIT-Rate of ECC-4, ECC-5 and SuDoku-Z.  Note that ECC-4 is insufficient at providing FIT of one even at 5ms scrub interval, whereas SuDoku-Z can provide one FIT even at 80ms.

\begin {table}[ht]
\begin{center} 
\caption{FIT-Rate vs. Scrub Intervals (default: 20ms)}{
\resizebox{0.65\columnwidth}{!}{
\begin{tabular}{| c | c | c | c | c |} \hline
Scrub  &  BER & FIT-Rate  & FIT-Rate & FIT-Rate \\ 
Interval &  per scrub & ECC-4 &  ECC-5 & SuDoku-Z \\ \hline \hline
5ms& 4.7$\times$10$^{-7}$ &   7.2   &0.0003 & 4x$10^{-6}$ \\ \hline
10ms& 9.4$\times$10$^{-7}$ &  115   & 0.011 & 6x$10^{-5}$ \\ \hline
{\bf 20ms} & 1.9$\times$10$^{-6}$ &   1.8K  & 0.351 & 0.0009 \\ \hline
40ms& 3.8$\times$10$^{-6}$ &  3K   & 11.2 & 0.014	\\ \hline
80ms& 7.5$\times$10$^{-6}$ &   471K  & 359 & 0.224	\\ \hline
\end{tabular}}}
\label{table:scrubres}
\vspace{-0.5in}
\end{center}
\end{table}

\subsection{Impact of RAID-Group Size}
We use a RAID-Group size of 1024 lines. The size of the RAID-Group determines the DUE Rate of SuDoku-Z.  However, the FIT-Rate of SuDoku is dominated by the SDC rate (due to CRC-21). Therefore, even if RAID-group sizes range from 64 to 1024, their FIT-Rate remains at 0.0009. However, the size of the RAID-Group impacts correction latency, which we discuss next.

\subsection{Analysis of Correction Latency} 
Lines with 1 error can be corrected with the per-line ECC-1 at low latency. However, for lines with multi-bit error, a RAID based correction is invoked. For our fault rate, the system encounters a line with multi-bit error, on average, once every 200ms. Such lines would invoke SuDoku-X and require reading all the 1024 lines in the RAID-Group, incurring a latency of at-most 10$\mu$s (9ns per line). Fortunately, incurring 10$\mu$s overhead for correction once every 200ms would cause a degradation of less than 0.01\%.  The correction latency of SuDoku-Y (20$\mu$s) and SuDoku-Z (80$\mu$s) is longer, however, these are incurred every 137 seconds and 129 hours, respectively, so the performance impact from such corrections remains negligible (<0.00001\%).

\subsection{Impact on Performance}
The performance impact of SuDoku comes from two aspects.  First, the
increased delay incurred due to CRC decoding (one cycle). Second, the
latency incurred in performing error correction for multi-bit
errors. As corrections are performed infrequently,  the
impact on performance is negligibly small. To
assess the performance impact of SuDoku, we integrate
the STTRAM cache into USIMM~\cite{usimm}. USIMM is a cycle-accurate memory system simulator that enforces strict timings as per the JEDEC DDR3 protocol specifications. Table~\ref{table:system_config4} lists the key parameters for the Baseline System.

\begin {table}[ht]
\begin{center} 
\caption{Baseline System Configuration}{
\resizebox{0.75\columnwidth}{!}{
\begin{tabular}{|c|c|}
\hline
                 Processor             & 4-wide, OoO-core; 8 cores       \\ \hline
                 STTRAM Last Level Cache (Shared)   & 64MB, 8-Way, 64B lines \\ 
                 STTRAM Latencies & Read: 9ns, Write: 18ns \\ \hline
                 Memory bus speed             & 800MHz    \\
                 DDR3 Memory channels         & 2 Channels @ 8GB Each \\ \hline

\end{tabular}}
}
\label{table:system_config4}
\end{center}
\end{table}

We choose all benchmarks in the
SPEC2006 suite~\cite{spec} and PARSEC~\cite{PARSEC} , BioBench (BIO)~\cite{BIOBENCH} and  commercial (COMM) benchmarks from
the MSC suite~\cite{msc}. For SPEC2006, we generate a representative slice of
one billion instructions using Pinpoints~\cite{pinpoints}. We directly use the traced workloads present in the MSC suite. We also form four MIXED workloads by randomly selecting benchmarks. We perform timing simulation until all the benchmarks in the workload finish execution, and measure the average execution time.

\begin{figure*}[htpb]
\centering
   \centerline{\psfig{file=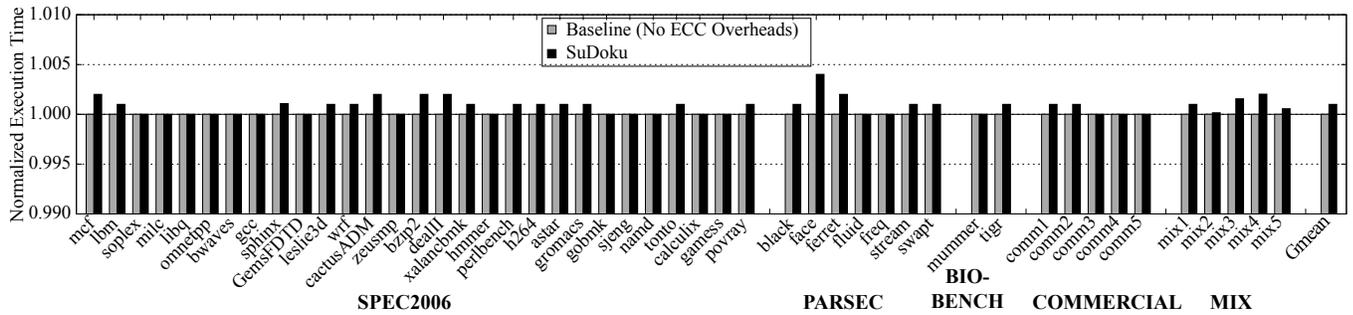,width=7in}}
  \caption{The Execution Time of SuDoku-Z normalized to an Idealized
    cache that does not encounter any error (and hence pays no
    overhead for error correction).  On average, SuDoku incurs a
    slowdown of 0.15\%.}
  \label{fig:exec}
  \vspace{-0.2in}
\end{figure*}

Figure~\ref{fig:exec} shows the execution time for SuDoku-Z as compared
to an the idealized cache that does not encounter any error (and thus
pays no overhead for error correction).  Since SuDoku-X requires a
single cycle to check the ECC-1 and CRC-21 syndrome for every request,
additional latency overhead is small. The overall
impact of this latency overhead is negligibly small, 0.1\% on an
average. Furthermore, the common-case fault is also a single-bit
fault, so the high-latency of RAID-based correction is incurred infrequently (10$\mu$s overhead once every 200ms).

\subsection{Impact on Energy and Power}

SuDoku-Z consumes additional energy as due to the parity updates in the
PLT on each write access to the cache.  We use the parameters shown in
Table~\ref{table:power} for our energy evaluations.

\begin {table}[ht]
\begin{center} 
\vspace{-0.1 in}
\caption{Characteristics of STTRAM and SRAM \cite{oscar}}{
\resizebox{0.6\columnwidth}{!}{
\begin{tabular}{| c | c | c |} \hline
Characteristic & STTRAM & SRAM\\ \hline  \hline
Write energy per access (nJ)& 0.35 & 0.11  \\  \hline
Read energy per access (nJ)& 0.13 & 0.05  \\  \hline
Static power per cell (nW) & 0.07 & 4.02	\\ \hline
\end{tabular}}}
\label{table:power}
\end{center}
\vspace{-0.2in}
\end{table}

\begin{figure}[htb]
\centering
   \centerline{\psfig{file=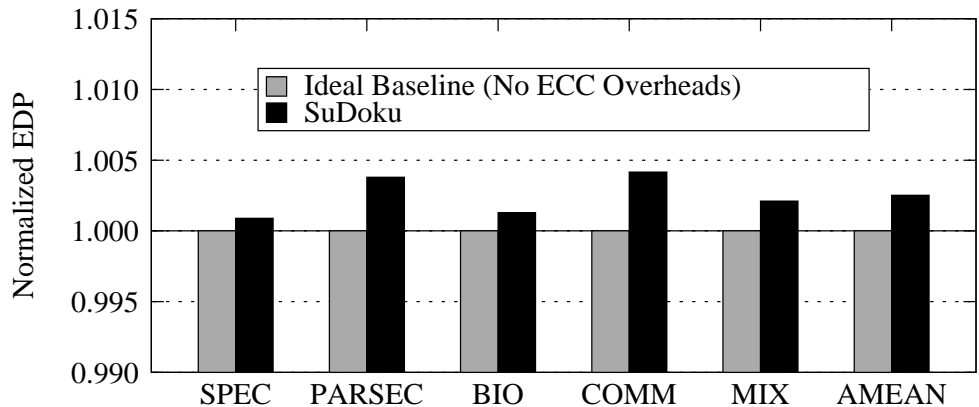,width=0.85\columnwidth}}
  \caption{The Energy Delay Product of a System with SuDoku-Z normalized to an
    error-free baseline. SuDoku requires negligible energy to update PLTs.  }
  \label{fig:edp}
\end{figure}

We compute the overall system energy and the compared the Energy Delay Product (EDP)
of SuDoku-Z with an idealized baseline that does not encounter any error, therefore it does not pay any energy overheads for error correction. Figure~\ref{fig:edp} shows the System-EDP for SuDoku-Z normalized
to the idealized baseline.  On average, the updates of SuDoku-Z cause an overall System-EDP to increase by at most 0.4\%.

\subsection{Storage Overheads of SuDoku-Z}
SuDoku-Z requires 10 bits of ECC-1 and 21 bits of CRC-21 for every 512-bit cachelines. 
Furthermore, it also uses two PLTs, each 64KB for the 64MB cache. 
Therefore, the amortized cost of these two PLTs is 1 bits per cachelines.
Therefore, SuDoku requires a total storage overheard of 32 bits per
cacheline, which is much less than the 50 bits per line incurred for
ECC-5. Furthermore, the PLT structures are sufficiently small that they
can be kept in a small 128KB SRAM structure beside the 64MB STTRAM cache.

\section{Summary}
As STTRAM cells are scaled to small feature sizes, the volume of the cell reduces, which makes the cell susceptible to external thermal noise.  Retention failures is considered as a critical obstacle to the scalability of STTRAM. We investigated a regime where by the Thermal Stability Factor of STTRAM is reduced to 30, which results in a bit error rate of 1.9$\times$10$^{-6}$ over a period of 20ms.  Due to the transient nature of these retention failures, prior work on DRAM-style refresh as well as efficient means of handling permanent faults become ineffective.  An effective means of tolerating retention failure is to do periodic scrubbing and employ per-line ECC.  Unfortunately, to tolerate our target error rate, we would need to ECC-5 per line, which is costly in terms of both storage and latency. Ideally, we would like to tolerate a high rate of transient failures while avoiding the overheads of strong error correction. To that end, this chapter makes the following contributions:

\begin{enumerate}
\item We propose {\em SuDoku}, a design that can efficiently tolerate a high rate of transient errors while requiring low storage overhead.  SuDoku optimizes for the common case of 1 bit failure and provisions each line with only ECC-1. It provides a strong error detection code (CRC-21) with each line that can detect multibit failures.  We describe three flavors of SuDoku.

\item We propose {\em SuDoku-X}, a design that uses RAID-4 to perform correction of multibit failures. In particular, we use a region-based RAID-4, whereby a given number of lines form a group, (called {\em RAID-Group}), and there is a parity line associated with each RAID-Group. Correction of multibit errors is performing using the parity line and all the other lines in the RAID-Group.  

\item We propose {\em SuDoku-Y}, a design that can correct two faulty lines in the RAID-Group by using {\em Sequential Data Resurrection (SDR)}.  SDR uses the mismatch in parity to identify the faulty locations, and the bits in these locations are flipped one at a time in order to allow ECC-1 to correct a line with two errors.  SDR improves the MTTF of SuDoku-X by 3390 times.

\item We propose {\em SuDoku-Z}, a design that allows each line to participate in two different RAID-Groups, formed by using two different hash functions.  When a line is deemed uncorrectable under one hash function, its correction is performed using the second hash function. SuDoku-Z provides an MTTF of 1138 Billion hours.

\end{enumerate}

SuDoku-Z provides 330x times as high reliability as ECC-5 does, while incurring only two-thirds the storage overhead, and avoiding the latency overheads associated with encoding and decoding of ECC-5.  Our evaluations shows that SuDoku performs within 0.1\% of the performance of an idealized fault-free baseline.  While we analyze SuDoku only in the context of STTRAM and only to handle retention failures, it is a general scheme that can be applied to handle a high error rate of transient failures.  SuDoku can also be applied to handle permanent faults, and is especially useful in domains where it may not be practical to precisely identify the location of the faulty bits using  testing.   Exploring such extensions is a part of our future work.


\chapter{Conclusions and Future Work}
\section{Conclusion}
Technology scaling tends to reduce the reliability of memory systems. As we venture into the sub-20nm regime, DRAM based systems are finding it difficult to scale reliably. This is because the feature sizes of cells become extremely small and they tend to break frequently. Furthermore, even at runtime, DRAM based systems show multi-granularity faults. This is not only a problem for current memory systems, but even new memory technologies can exhibit different modes of failures. For instance, stacked memories may exhibit TSV and large-granularity failures and promising technologies like STTRAM tend to incur high rates of transient failures as it scales. Therefore memory reliability is a key concern to enable technology scaling for building high-density memory chips. To address these concerns, this dissertation suggests four broad designs and techniques over four chapters.

Chapter 3 proposes a cross-layer technique to enable seamless and robust technology scaling for DRAM-based memories as they venture into the sub-20nm regime. The technique called as ArchShield exposes the scaling-related faults in the DRAM chips and enables the architecture to maintain replicas while using simple ECC. ArchShield can handle error rates of up to 100ppm with less than 1\% performance and 2.5\% area overhead.

Chapter 4 proposes a strong runtime ECC scheme that protects against large granularity failures in DRAM chips. This technique, called as XED, exposes the ``hidden'' On-Die ECC and implements RAID-3 type correction. XED enables commodity DIMMs to implement Chipkill using 2x fewer chips while requiring no changes in the memory protocols. XED provides 172x higher reliability as compared to a system that conceals its On-Die ECC.

Chapter 5 proposes techniques to build reliable stacked memories. This proposal, called Citadel, enables reliable and efficient stacked-memories by fixing TSV failures and employing RAID-5 type correction. Citadel provides 700x higher reliability over a naive Chipkill based implementation while being performance and power efficient.

Chapter 6 discusses techniques to enable scalable new memory technologies like STTRAM. At small feature sizes, STTRAM cells retain data only for a few milliseconds and turn erroneous in a transient fashion. To mitigate these transient failures, one would have to use strong ECC which are complex, incurring large penalties. This dissertation proposes a scheme, called as SuDoku, that uses simple ECC for the common-case faults and uses RAID-4 based strong ECC for the uncommon case. SuDoku provides 2000x higher reliability while incurring negligible overheads as compared to a 6EC7ED based ECC scheme.

Broadly, to enable scalable memories, this dissertation advocates for cross-layer techniques at the architecture-level to provide 100x-1000x higher reliability while incurring minimal overheads in terms of area, performance, and power.

\section{Future Work}
While this dissertation investigates architecture techniques that enable reliable and scalable memory systems, the ideas on reliability can be extended to some future vectors.

\subsection{Morphable RAID}
Memory systems employ different levels of RAID and they incur different overheads for any RAID level. For instance, if the memory system uses RAID-1, it loses half its capacity but does not incur any bandwidth abilities. On the other hand, if the memory system uses RAID-5, it can get most of the capacity but will incur some additional bandwidth overheads. As workloads have different characteristics, some workloads may perform better in a particular level of RAID over the other. There is scope in exploring morphing different levels of RAID to suit the workload and optimize its performance and power while maintaining strong reliability.

\subsection{Advanced Cross-Layer Resilience Schemes}
The reliability of memory systems can be improved further if reliability schemes can be designed with greater coordination with the operating systems (OS), software, and algorithms layer in the system stack. For instance,  one can design advanced fault-tolerant algorithms by using the ECC information from the memory system. Instead of taking care of corner-case failures in the hardware, one can simply use the OS to remap or decommission pages. Even in the software layer, there is potential for creative data structures that can be made more resilient to faults by using the ECC from the memory system.

\subsection{Link Error Models and Bandwidth Throttling}
Memory links tend to exhibit errors if we scale their voltage and frequencies. The error model for link errors can help the academic community understand the scaling challenges for memory links. Furthermore, after understanding the error models, it would be easier for architects to tailor ECC codes to tackle link errors. This will also unlock new opportunities to employ dynamic bandwidth throttling of the memory links, thereby providing a boost in the memory bandwidth during critical time periods.

\subsection{Co-Architecting Secure and Reliable Systems}
Memory systems that are secure tend to have various apparatus such as MACs and Counters to enable encryption, ensure the integrity and provide security. Most times, some of these security apparatus can be re-purposed for reliability with very low overheads. For instance, the MAC, apart from checking for the integrity of data, can be used to detect faults in memory chips. Therefore, one can try to co-architect secure memory systems to have strong reliability while paying minimal overheads for reliability.

\subsection{Optimal Designs for Heterogeneous Memory Systems by using ECC}
With the advent of 3D Xpoint memories, it is conceivable to have memory designs that place such non-volatile memories (NVM) in the same channel as  DRAM. NVMs offer higher capacity than DRAM, but they are slow and have variations in their read and write latencies. By creatively using ECC, one can co-architect such heterogeneous memory systems to lock high-variation lines in NVM within its front-end DRAM (that is placed in the same channel). Such simple and low-cost techniques can help reduce the effective read latency of the entire memory system. 

In the similar spirit, one can also design reliability schemes that can potentially checkpoint large portions of DRAM within the NVM. Additionally, NVMs also exhibit unique faults (such as endurance-related faults), they present an opportunity to rethink techniques that can help improve the effective reliability of the entire memory system by tailoring reliability techniques for unique types of faults.

\subsection{Managing Error Correction for Quantum Accelerators}  
Quantum computer consists of quantum bits (qubits) and a control processor that acts as an interface between the programmer and the qubits. Qubits are extremely sensitive towards the noise and rely on continuous error correction to maintain the correct state. Current proposals of software managed error correction results in an extremely high instruction bandwidth as the instruction bandwidth scales proportionally to the number of qubits. While such a design may be reasonable for small-scale quantum computers, instruction bandwidth will become a critical bottleneck for scaling quantum computers. Typically a large portion of the instructions in the instruction stream of a typical quantum workload stems from error correction. One can look at techniques that help delegate the task of quantum error correction to the hardware.




\begin{singlespace}  
	\bibliography{references}
\end{singlespace}

\addcontentsline{toc}{chapter}{References}  

\chapter*{Vita}
\renewcommand{\thefootnote}{\fnsymbol{footnote}}
\addcontentsline{toc}{chapter}{Vita}  
\paragraph{}Prashant Nair, the son of Jayaprakash Thumparambil Nair and Mythili Jayaprakash Nair, was
born in Santacruz, Mumbai, India on February 13th, 1987. He received the Bachelor of Engineering degree
in Electronics from the University of Mumbai in 2009. Thereafter, he worked as an Engineer at the Reliance Infrastructure Ltd. in the Technology Automation Group in Navi Mumbai,
India. 

He enrolled in the Ph.D. program in 2011 at the Georgia Institute of Technology where he began working with his Ph.D. adviser Dr. Moinuddin K. Qureshi. He received the Master of Science
degree in Electrical and Computer Engineering in 2013.
While in graduate school, he served as a teaching assistant for four semesters. He also interned five times in many top-class industry labs including Intel, IBM, AMD, and Samsung. 

He has published papers in The
International Symposium on Computer Architecture (ISCA-40, ISCA-43, and ISCA-44),
The International Symposium on Microarchitecture (MICRO-47), The International Symposium
on High Performance Computer Architecture (HPCA-19, HPCA-21, HPCA-22), The International Conference on Architectural Support for Programming Languages and Operating Systems (ASPLOS-20), and The IEEE/IFIP International Conference on Dependable Systems and Networks (DSN-45).
\vspace{0.2in}

{\setlength{\parindent}{0cm}
\underline{Permanent Address}:\newline 
B1103, RNA Heights CHS Ltd., \newline
JV Link Road, Andheri  (East), \newline
Mumbai 400093, India \newline
}

{\setlength{\parindent}{0cm}
    This dissertation was typeset with \LaTeX ~\footnote[2]{ \LaTeX {\color{white} $\cdot$}is a document preparation system developed by Leslie Lamport as a special version of Donald Knuth’s \TeX {\color{white}$\cdot$} Program.} by the author.
}

\end{document}